\documentclass[10pt,letterpaper]{article}
\usepackage[T1]{fontenc}
\usepackage{jheppub}
\usepackage{comment,xcolor}
\usepackage{amsmath}
\usepackage{mathtools}
\usepackage{amssymb}
\usepackage{amsfonts}
\usepackage{amsthm}
\usepackage{mathrsfs}
\usepackage[all]{xy}
\usepackage{tensor}
\usepackage{footnote}
\usepackage{graphicx} 
\usepackage[justification=raggedright,singlelinecheck=false]{caption}
\usepackage{subcaption}
\usepackage{cleveref}
\usepackage{booktabs}

\newtheorem{innercustomgeneric}{\customgenericname}
\providecommand{\customgenericname}{}
\newcommand{\newcustomtheorem}[2]{%
  \newenvironment{#1}[1]
  {%
   \renewcommand\customgenericname{#2}%
   \renewcommand\theinnercustomgeneric{##1}%
   \innercustomgeneric
  }
  {\endinnercustomgeneric}
}

\newcustomtheorem{customthm}{Theorem}
\newcustomtheorem{customlemma}{Lemma}

\newtheorem{thm}{Theorem}

\newtheorem{defn}{Definition}

\newtheorem{alg}{Algorithm}

\newcommand{\dd}{\mathrm{d}}
\newcommand{\beq}{\begin{equation}}
\newcommand{\eeq}{\end{equation}}
\numberwithin{equation}{section}

\newcommand{\ket}[1]{| #1 \rangle}
\newcommand{\bra}[1]{\langle #1 |}
\newcommand{\braket}[2]{\left\langle #1 \mid #2 \right\rangle}

\def\al{\alpha}

\def\la{\lambda}
\def\ma{\mathrm{max}}
\def\mi{\mathrm{min}}

\def\eps{\varepsilon}

\def\A{\mathcal{A}}

\def\O{\mathcal{O}}

\def\h{\mathcal{H}}

\def\W{\mathcal{W}}

\def\sbh{S_\mathrm{BH}}
\newcommand*{\tr}{\mathrm{Tr}}
\newcommand*{\ttr}{\mathrm{tr}}

\def\id{\mathbf{1}}
\def\E{\mathbb{E}}
\def\nc{\mathrm{NC}}

\title{Beyond islands: A free probabilistic approach}
\author{Jinzhao Wang}
\affiliation{\small \it Institute for Theoretical Physics, ETH Z\"urich, 8093 Z\"urich, Switzerland}
\affiliation{\small \it Stanford Institute for Theoretical Physics, Stanford University, Stanford, CA 94305}
\emailAdd{jinzhao@stanford.edu}

\abstract{We give a free probabilistic proposal to compute the fine-grained radiation entropy for an arbitrary bulk radiation state, in the context of the Penington-Shenker-Stanford-Yang (PSSY) model where the gravitational path integral can be implemented with full control. We observe that the replica trick gravitational path integral is combinatorially matching the free multiplicative convolution between the spectra of the gravitational sector and the matter sector respectively.  The convolution formula computes the radiation entropy accurately even in cases when the island formula fails to apply. It also helps to justify this gravitational replica trick as a soluble Hausdorff moment problem. We then work out how the free convolution formula can be evaluated using free harmonic analysis, which also gives a new free probabilistic treatment of resolving the separable sample covariance matrix spectrum.

The free convolution formula suggests that the quantum information encoded in competing quantum extremal surfaces can be modelled as free random variables in a finite von Neumann algebra. Using the close tie between free probability and random matrix theory, we show that the PSSY model can be described as a random matrix model that is essentially a generalization of Page’s model. It is then manifest that the island formula is only applicable when the convolution factorizes in regimes characterized by the one-shot entropies. We further show that the convolution formula can be reorganized to a generalized entropy formula in terms of the relative entropy.}

\begin{document}
\maketitle

\section{Introduction}\label{sec:intro}

Recent advances in the black hole information puzzle~\cite{hawking1975particle,hawking1976breakdown}
feature a quantitative characterisation of the entropy of Hawking radiation that is consistent with unitarity~\cite{penington2020entanglement,almheiri2019entropy,almheiri2020page,penington2022replica,almheiri2020replica,almheiri2020entropy}. The radiation entropy is shown to follow the Page curve and it resolves the \emph{entropic information puzzle},\footnote{This does not concern other aspects of the information puzzle, such as the black hole microstates and the typical state firewall problem~\cite{marolf2013gauge}. cf. discussions in~\cite{almheiri2020entropy}. } which we refer to as the tension between an ever-increasing radiation entropy computed by Hawking, and the entropy following the Page curve which stops increasing after the Page time as demanded by unitarity \cite{page1993average,page1993information}.  The entropy is computed using the gravitational path integral (GPI) implementing the \emph{replica trick}~\cite{calabrese2004entanglement,lewkowycz2013generalized,faulkner2013quantum}. In semiclassical gravity, the GPI can be evaluated under the saddle-point approximation, and the key insight is that a wormhole saddle connecting the replicas is generically dominant for an old evaporating black hole. The resulting entropy of Hawking radiation is described by the island formula,
\begin{equation}\label{eq:island}
    S(\tilde R)=\mathrm{min\,ext}_I\left(\frac{\mathrm{Area}[\partial I]}{4G_N}+S(I\cup R)\right).
\end{equation}

This formula equates the radiation entropy to the generalized entropy evaluated at the island, $\partial I$, in the black hole spacetime. The surface that bounds the island is identified via an extremisation procedure and is known as the \emph{quantum extremal surface} (QES)~\cite{engelhardt2015quantum}. In particular, it identifies an island $I$ inside the black hole after the Page time, and this allows the resulting radiation entropy $S(\tilde R$ to stop increasing and hence resolving the paradox.

Some fineprints are needed to interpret this formula. The entropy that appears on the right is the entropy in the semiclassical theory. It is the von Neumann entropy of the quantum state reduced on the joint region $I\cup R$ and we shall refer it as the state in the \emph{effective description}, or simply as \emph{the bulk state} if we borrow the language of holographic duality. Then the island formula is believed to capture the radiation entropy, as would be obtained from the von Neumann entropy of a density operator computed directly in a complete theory of quantum gravity. We shall refer to such a density operator as the \emph{fundamental description} of the radiation, or as \emph{the boundary state} again using the analogy of AdS/CFT. We add a tilde (e.g. $\tilde R$) in the fundamental description to help distinguish it from the effective description. In general, we cannot deduce the fundamental description in full from the replica trick calculation and the best we can hope for is to extract the spectrum of its density operator.

However, the island formula is not guaranteed to hold in all circumstances. If one examines carefully its derivation~\cite{lewkowycz2013generalized,faulkner2013quantum,almheiri2020replica,penington2022replica}, an important step is to assume that the replica-symmetric (i.e. $\mathbb{Z}_n$-symmetric) saddles always dominate over the non-symmetric ones, such as the fully disconnected Hawking saddle before the Page time and the fully connected replica wormhole saddle after the Page time. When the non-replica-symmetric saddles aggregate a significant contribution comparable to the contributions from the replica-symmetric saddles, this assumption becomes invalid and \emph{replica symmetry breaking} kicks in. 

This is a fair simplification as far as demonstrating the qualitative features of the Page curve is concerned.  However, it is recently argued and shown that the island formula doesn't compute the radiation entropy accurately when the effect of replica symmetry breaking is significant, which could occur in presence of a non-flat entanglement spectrum between the radiation and the black hole~\cite{akers2021leading,wang2022refined}. In such cases, the quantum extremal surfaces and islands cannot be identified. This is well-expected because when the GPI is not dominated by a single geometry, so the entropy cannot be evaluated using a single semiclassical geometry when a superposition of distinct geometries is relevant.

The investigation of replica symmetry breaking in the context of black hole information problem is largely unexplored, mostly because the non-replica-symmetric saddles cannot be analysed with good control in the GPI. One exception is the toy model due to Penington-Shenker-Stanford-Yang (PSSY)~\cite{penington2022replica}. It consists of a black hole in JT gravity coupled with a static end of world (EOW) brane, that carries internal degrees of freedom, modelling the black hole interior. It is perhaps the simplest setup which mimics an evaporating black hole such that the entropic information puzzle is still manifest. The upshot of the simplification is that one can make sense of summing over all saddles with good control. It is thus possible to carry out detailed quantitative analysis probing replica-symmetry, which could be challenging in other models.

\subsection{Problem: The PSSY model with non-flat entanglement spectrum}\label{sec:pssy}

Our goal in this work is to explicitly compute in the PSSY model the spectrum of the radiation state in the fundamental description, given \emph{any bulk entanglement spectrum}, and thus the von Neumann or R\'enyi entropies of the radiation. Non-flat spectrum is relevant when the we apply quantum information processing to the radiation or when the reservoir is ruled by some interesting Hamiltonian. The bottom line is that it is a more general setting to study the entropic information puzzle that was not thoroughly investigated. We shall see that this slight generalization offers some insights on the replica trick and the information puzzle.\\

We start by briefly reviewing the PSSY model. The black hole is described in JT gravity, which is a 2D theory of gravity that couples to a dilaton field $\phi$.  Concretely, we work with the Euclidean JT action~\cite{harlow2020factorization,penington2022replica,almheiri2020replica},
\begin{equation} \label{eq:JTaction}
    I_{\mathrm{JT}}[g, \phi] = - \frac{S_0}{2\pi} \Bigl( \frac{1}{2} \int_{\mathcal{M}} \sqrt{g} R + \int_{\partial \mathcal{M}} \sqrt{h} K \Bigl)
    - \Bigl( \frac{1}{2} \int_{\mathcal{M}} \sqrt{g} (R+2) \phi + \int_{\partial \mathcal{M}} \sqrt{h} (K-1) \phi  \Bigl) \ ,
\end{equation}
where $S_0$ is the extremal entropy of the black hole solution, i.e., the black hole entropy in the zero temperature limit. The total action is appended by the addition term describing a static brane that sits at the end of the world,
\begin{equation}
    I_{\mathrm{JT}+EOW}[g, \phi]=I_{\mathrm{JT}}[g, \phi]+\mu \int_{\mathrm{EOW}} \dd s
\end{equation}
where $\mu$ denotes the tension of the EOW brane. 

\begin{figure}
\centering
\begin{subfigure}{.22\textwidth}
  \centering
  \includegraphics[width=.9\linewidth]{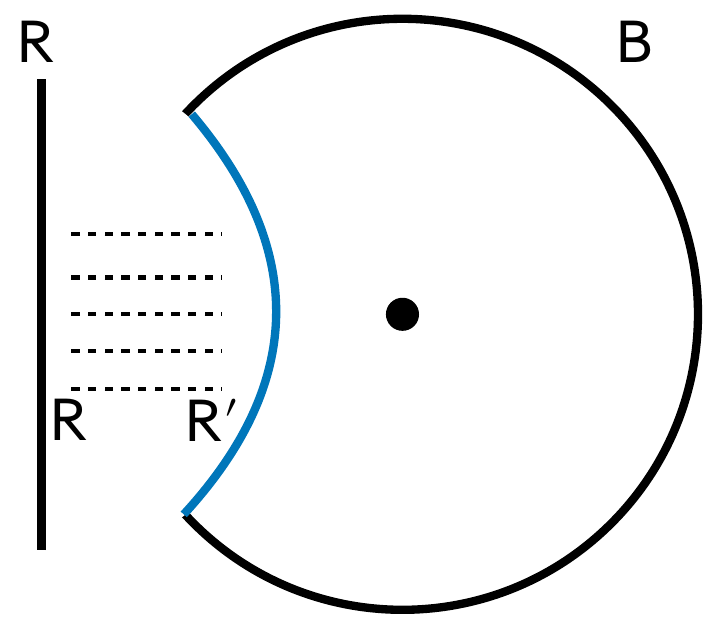}
  \caption{\centering}\label{fig:setup}
\end{subfigure}%
\begin{subfigure}{.22\textwidth}
  \centering
  \includegraphics[width=0.9\linewidth]{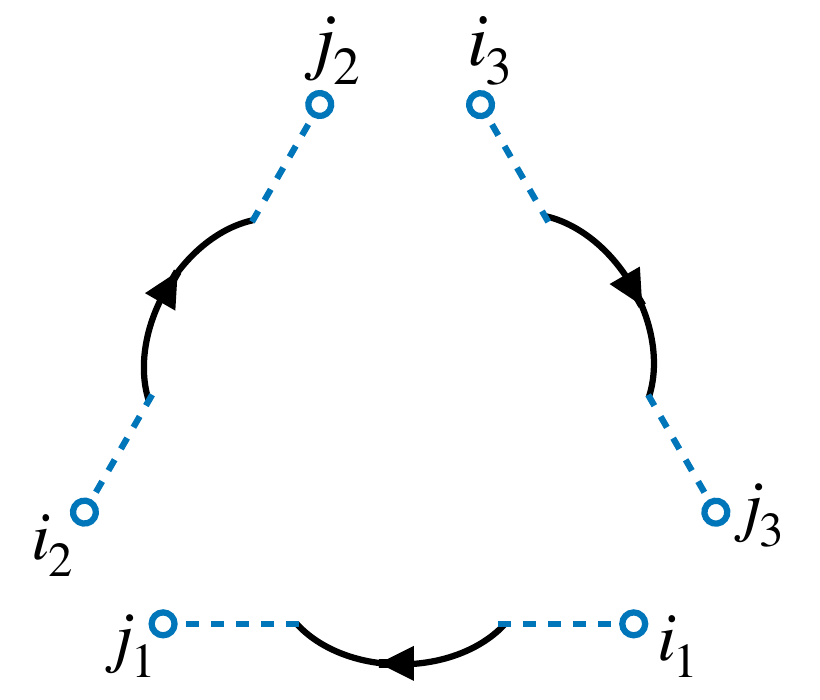}
  \caption{\centering}\label{fig:bc}
\end{subfigure}
\begin{subfigure}{.22\textwidth}
  \centering
  \includegraphics[width=.8\linewidth]{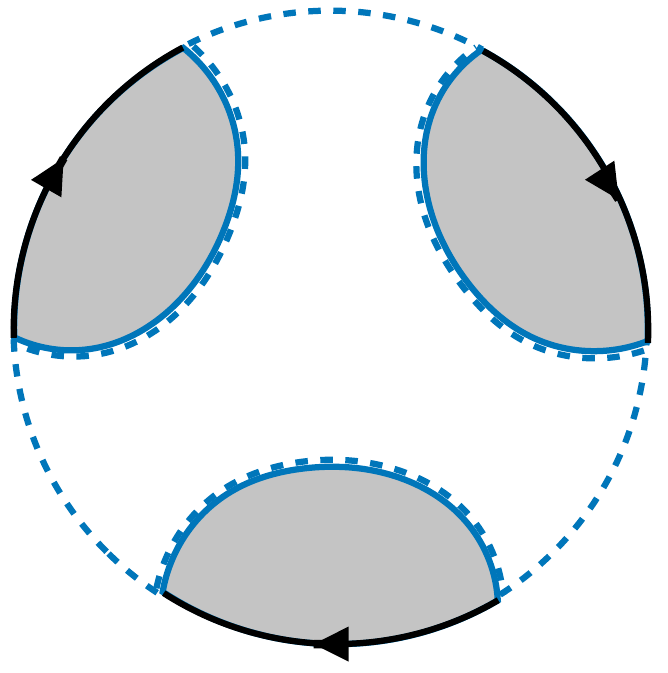}
  \caption{\centering}\label{fig:discs}
\end{subfigure}%
\begin{subfigure}{.25\textwidth}
  \centering
  \includegraphics[width=0.7\linewidth]{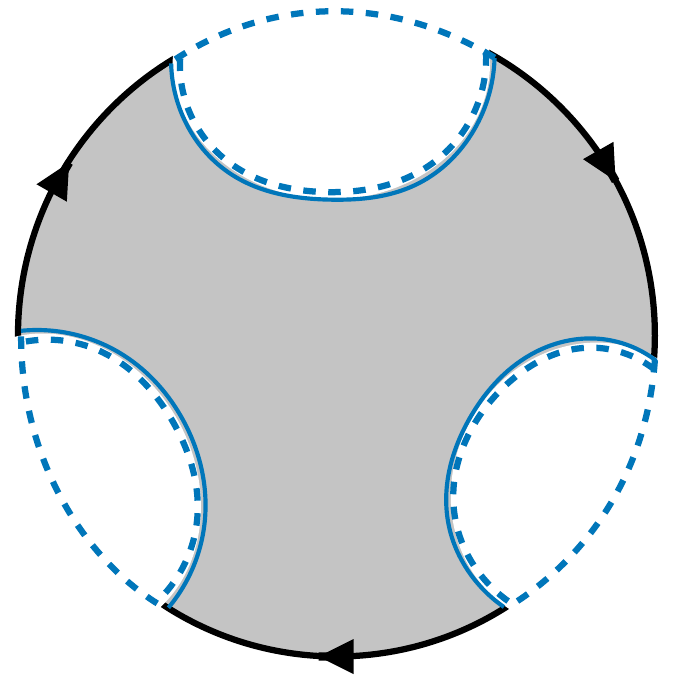}
  \caption{\centering}\label{fig:wormhole}
\end{subfigure}
\caption{\textbf{The PSSY model.} The setup is sketched in Fig.~\ref{fig:setup}, where the EOW brane-radiation bipartite system is labeled as $\mathsf{R'R}$ and the black hole-radiation system is labeled as $\mathsf{BR}$. We shall consider arbitrary pure bipartite state on $\mathsf{R'R}$ (effective description), denoted as the dashed line, and aim to compute the entanglement spectrum between $\mathsf{BR}$ (fundamental description). This calculation is achieved using the replica trick that concerns for instance three replicas of boundary conditions as depicted in Fig.~\ref{fig:bc}. The EOW brane indices are contracted by the $3$-swap operator $\eta_3=\delta_{j_1i_2}\delta_{j_2i_3}\delta_{j_3i_1}$. The dominant replica-symmetric saddles in the GPI are the fully disconnected (Hawking) saddle (Fig.~\ref{fig:discs}) and the fully connected wormhole saddle (Fig.~\ref{fig:wormhole}).}
\end{figure}

The EOW brane carries internal degrees of freedoms that are entangled with the Hawking radiation in an auxiliary $k$-dimensional reservoir system outside the spacetime (cf. Fig.~\ref{fig:setup}). Specifically, the scenario studied by PSSY concerns the \emph{maximally entangled bulk state} in the effective description:
\begin{equation}\label{eq:entangled}
    \ket{\rho}_{\mathsf{R'R}}:=\frac{1}{\sqrt{k}}\sum_{i=1}^{k} \ket{\psi_i}_{\mathsf R'}\ket{i}_{\mathsf R}
\end{equation}
where $\ket{\psi_i}_{\mathsf R'}$ is the brane state acting as the Hawking partner that purifies the $i^\mathrm{th}$ radiation mode $\ket{i}_{\mathsf R}$.\footnote{In PSSY equation (2.5), they write down the state $\ket{\tilde\rho}_{\mathsf{BR}}:=\frac{1}{\sqrt{k}}\sum_{i=1}^{k} \ket{\psi_i}_{\mathsf B}\ket{i}_{\mathsf R}$ to motivate the boundary conditions for the GPI. We put a tilde to indicate that, in our terms, $\ket{\tilde\rho}_{\mathsf{BR}}$ is the state in the fundamental description, where $\ket{\psi_i}_{\mathsf B}$ is the black hole state when the brane is in state $i$. They have overlap of $\O(e^{-\sbh})$ such that $S(R)$ follows the Page curve. Here we specifically refer to the state in the effective description. It is written in Schmidt decomposition where the brane states $\ket{\psi_i}_{\mathsf R'}$ are strictly orthogonal.} Here the Schmidt coefficients are chosen to be flat with the Schmidt rank $k$. The model does not describe a dynamical evaporation process. With the entanglement between the radiation and the black hole put in by hand, the PSSY model should be viewed as a snapshot of a physical black hole that is evaporating quasi-statically. To see the evolution of the radiation entropy and the Page curve, one simply tune up $k$. Regardless of its simplicity, the entropic information puzzle is still manifest as the radiation entropy seems to increase without bound as we tune up the bulk entanglement.

The island formula in the PSSY model is simply,
\begin{equation}\label{eq:island_pssy}
    S(\tilde R)=\min\{\sbh, \log k\}\ .
\end{equation}
where $\sbh$ is the Bekenstein-Hawking (BH) entropy.\footnote{Later in Section~\ref{sec:convolution}, we will provide details of how the BH entropy can be computed in the PSSY model.}

According to the formula, the Page curve has two regimes with $\log k< \sbh$ and $\log k> \sbh$. In the former case, the QES surface is the empty surface so we have no area term but only a bulk entropy term with contribution $\log k$; and in the latter case, the QES surface sits at the horizon and we have no bulk entropy contribution $S(I\cup R)=0$ but only a constant area term $\sbh$. The island formula therefore provides us a sketch of the Page curve for an eternal black hole as we tune up $k$, which increases with $k$ until it flattens out for $\log k> \sbh$. It’s very accurate in the microcanonical ensemble and subject to subleading $1/\beta$ correction near the Page transition in the canonical ensemble of inverse temperature $\beta$.

To see how this formula comes from, we need use the replica trick. The idea is to compute the moments, $\tr\tilde \rho_{\mathsf R}^n$, of a hypothetical radiation density matrix $\tilde \rho_{\mathsf R}$. Since we do not have such a density matrix a priori, we compute the moments by preparing replicas of the black hole-radiation system, and evaluate the expectation value of the $n$-swap operator $\langle \eta_n\rangle$ using the GPI. The rule is to integrate over all the configurations compatible with the \emph{boundary conditions} that prepares the state and imposes the observable $\eta_n$. Since the full path integral is challenging, we instead sum over all saddlepoint configurations to approximate the GPI. This quantity that the GPI computes can be thought of as a partition function $\tilde Z_n=\langle \eta_n\rangle$ with the replica trick boundary condition.

The boundary condition is illustrated for three replicas in Fig.~\ref{fig:bc}. It consists of three copies of a asymptotic boundary (solid line) of length $\beta$ that prepares the JT black hole at temperature $1/\beta$ and the end points are joined by the sources that create the EOW brane. They carry internal degrees of freedom indicated by the dangling dashed lines. As labeled, the indices indicate the matrix elements of the radiation density operator that we'd like to prepare. The dashed lines are connected cyclically corresponding to evaluating the expectation value $\langle\eta_3\rangle$. When the EOW branes form in a saddle geometry that fills into this boundary condition, the the dashed lines will extend along the branes as the internal degrees of freedom are physically residing on the branes. So they form in loops as depicted in Fig.~\ref{fig:discs}. We refer them as the \emph{EOW brane loops}. 

We work in the \emph{planar limit} which means that we ignore topologies without crossings or higher genus. This is legit when we work in the regime $\sbh,\log k\gg 1$ as the nonplanar geometries incur additional factors of order $\O(e^{-\sbh}), \O(1/k)$ in the on-shell action. Such planar saddles are organized by \emph{non-crossing} (NC) partitions. (cf. Definition~\ref{def:nc})

From each saddle, there are two contribution to the partition function $\tilde Z_n$: One from the \emph{gravitational sector}, denoted as the gravitational partition function $Z_n$ from an $n$-connected disk, i.e. a \emph{replica wormhole} connecting $n$ boundaries; and the other from the \emph{matter sector}, where a length-$n$ EOW brane loop gives a contribution of $\tr(k\rho_{\mathsf R})^n$, which always evaluates to $k$ for the reduced state $\rho_\mathsf{R}$ of \eqref{eq:entangled}. 

A quick calculation shows that when $\log k<\sbh$ ($\log k>\sbh$), the fully disconnected (connected) dominates yielding the island formula \eqref{eq:island_pssy}. Note that we've ignored the non-replica-symmetric saddles for the island formula, but they can be counted in and it would result in a correction of order $1/\sqrt{\beta}$ near the transition. Nonetheless, the leading order behaviour is captured by the island formula. \\


Consider now a pure bulk state with an arbitrary entanglement spectrum\footnote{For a generic mixed state, which can always be considered as a marginal of a purification, we have a tripartite correlated system among the black hole, radiation and the reference. The reason we restrict to pure bulk states is that we mostly care about the quantum correlation between the radiation and the black hole in the context of the information puzzle. Also, the Page curve concerns the entanglement entropy, and this is only an operationally meaningful measure of the entanglement between the radiation and the black hole. }
\begin{equation}\label{eq:bulkentangled}
    \ket{\rho}_{\mathsf{R'R}}:=\sum_{i=1}^{k} \sqrt{\lambda_i}\ket{\psi_i}_{\mathsf R'}\ket{i}_{\mathsf R}\ .
\end{equation}

Unlike in the original PSSY with the bulk state~\eqref{eq:entangled}, the EOW brane loops are now weighted by $c_i$ and an $n$-connected loop evaluates to $\tr(k\rho_{\mathsf R})^n$. We keep using $k$ to denote the \emph{Schmidt rank} of the bipartite state. It's shown explicitly in~\cite{wang2022refined} that a bulk state non-trivial entanglement spectrum, specifically a state in superposition of two branches,\footnote{This state has the Schmidt coefficients in \eqref{eq:bulkentangled} chosen as $\{\la_i\}_{i=1}^m=x, \{\la_i\}_{i=m+1}^k=y$ and $x\gg y, k\gg m$. } does lead to a leading order correction to the island formula.

This feature was first pointed out by Akers-Penington (AP) in the QES prescription in AdS/CFT~\cite{akers2021leading}. Heuristically, the non-uniform correlation in the matter sector enhances the effect of replica symmetry breaking. The non-flatness of the spectrum is gauged by the difference between the min-entropy and the max-entropy of the bulk state, which are the entropic measures invented to characterize information-processing tasks in the one-shot regime~\cite{renner2008security}.

We expect the failure of the island formula at the leading order if their difference is comparable to the area term (BH entropy). This deviation occurs in a large regime ``near'' the transition governed by the gap between the min-entropy and the max-entropy. In fact, the Page time, usually defined as the time when the bulk entanglement entropy is equal to the BH entropy, is not longer a relevant scale to characterize the transition. It is superseded by the one-shot versions indicating when the island formula seizes to apply. 

\begin{figure}
    \centering
    \includegraphics[width=.87\linewidth]{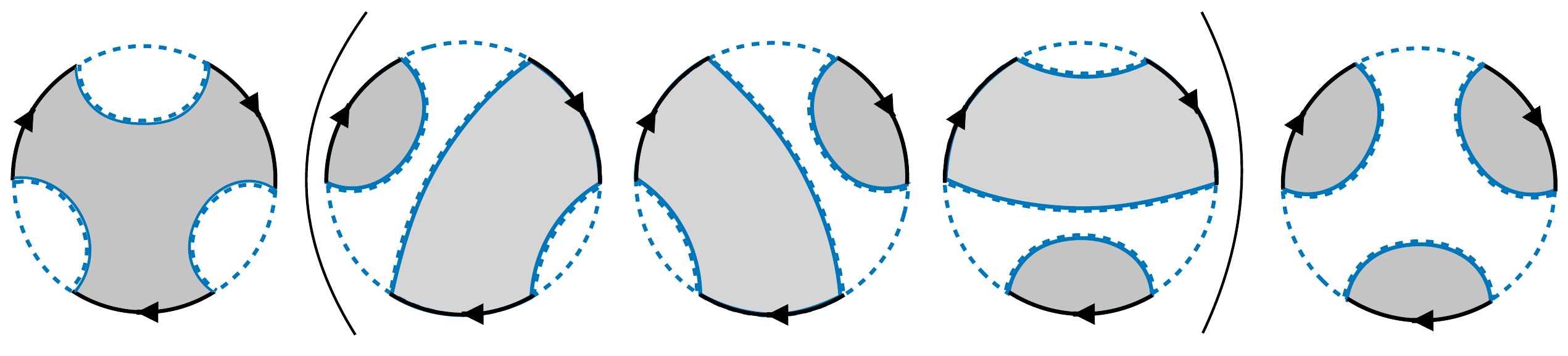}
    \caption{\textbf{Saddles for three replicas.} The figure shows all the relevant five saddles for three replicas. The middle three in the bracket are the non-replica-symmetric ones which are ignored in the island formula derivation but we include in our calculation. The partition function evaluates to $\tilde Z_3=Z_3\cdot 1^3+3Z_2Z_1\cdot k^2\tr\rho_{\mathsf R}^2+Z_1^3\cdot k^3\tr\rho_{\mathsf R}^3$ .}
    \label{fig:3-replicas}
\end{figure}

\begin{figure}
\centering
\begin{subfigure}{.4\textwidth}
  \centering
  \includegraphics[width=.8\linewidth]{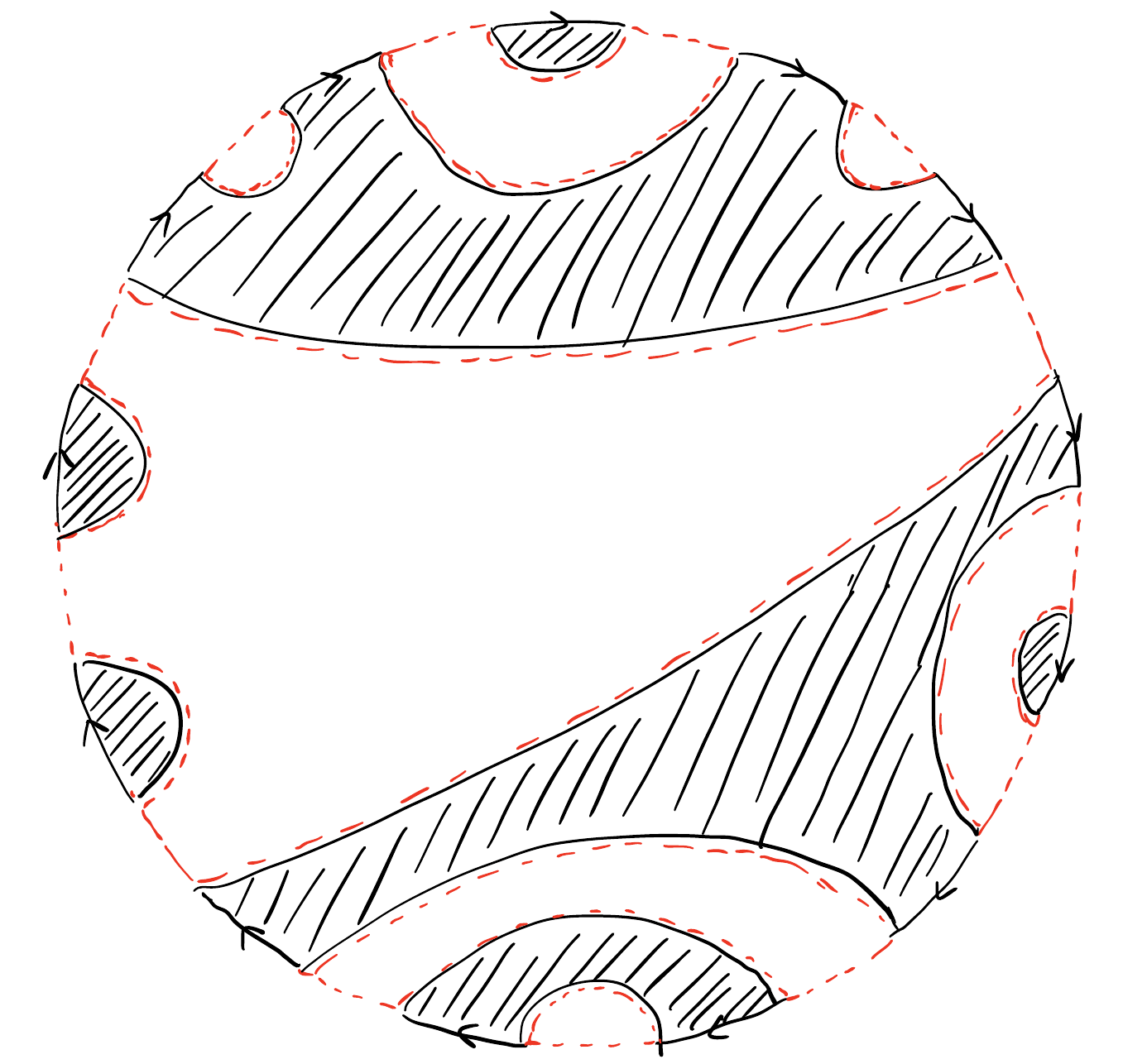}
  \caption{\centering A generic saddle}
\end{subfigure}%
\begin{subfigure}{.4\textwidth}
  \centering
  \includegraphics[width=0.8\linewidth]{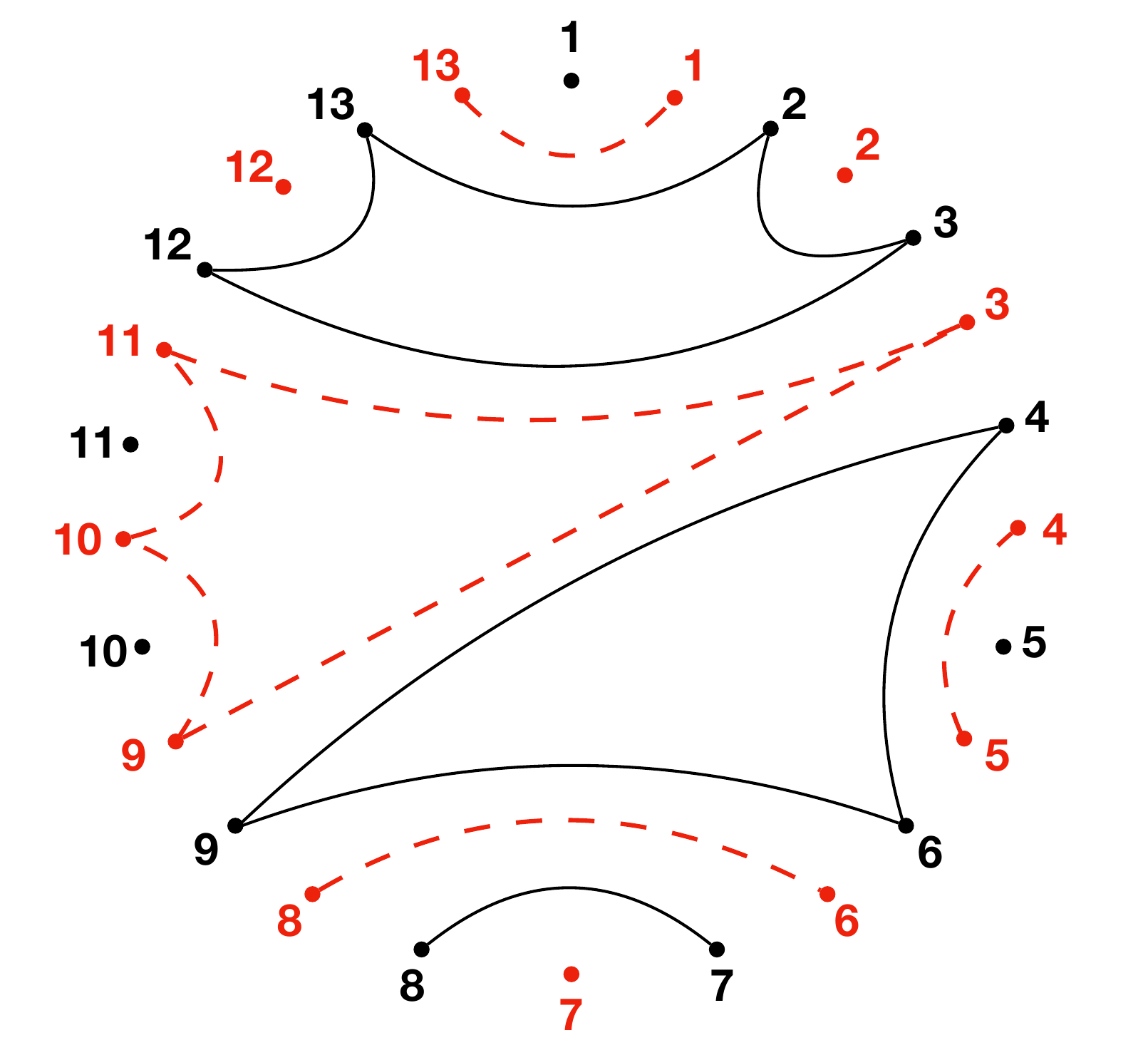}
  \caption{\centering A NC partition (black) and its Kreweras complement (red)}\label{fig:kreweras_b}
\end{subfigure}
    \caption{\textbf{The combinatorics of the saddles.} The saddles are fully characterized by the set of NC partitions. Specifically, any wormhole configuration corresponds to a NC partition and the configuration of the EOW brane loops corresponds to the Kreweras complement, which is precisely defined via a partition diagram that underlies the saddle configuration. Here we illustrate a generic example that the wormholes are depicted by the partition (in black) $\pi = (1)(2\ 3\ 12\ 13)(4\ 6\ 9)(5)(7\ 8)(10)(11)\in \mathrm{NC}_{13}$, and the EOW brane loops are depicted by its Kreweras complement (in red) $\bar\pi = (1\ 13)(2)(3\ 9\ 10\ 11)(4\ 5)(6\ 8)(7)(12)\in \mathrm{NC}_{13}$.}
    \label{fig:kreweras}
\end{figure}

We shall now attack the problem for an arbitrary bulk entanglement spectrum in \eqref{eq:bulkentangled}, and we need a replacement for the island formula. Implementing the replica trick gives the partition functions $\{\tilde Z_n\}_{n\in\mathbb{N}}$ which admit the following form,
\begin{equation}
    \tilde Z_n = \sum_{\pi\in \mathrm{NC}_n}\left(\prod_{V\in\pi} Z_{|V|}\cdot \left(\mathrm{matter\ sector}\right)\right).
\end{equation}
where the contributions are organized by the non-crossing partitions. Each saddle corresponds to a configuration of wormholes that corresponds to a NC partition $\pi$ (cf. Fig.~\ref{fig:kreweras}), and the total contribution from the gravitational sector is simply the multiplication of the contributions $\{Z_{|V|}\}_{V\in\pi}$ for each $|V|$-connected wormhole. See Fig.~\ref{fig:3-replicas} for the example of $\tilde Z_3$.

To sort out contributions from the matter sector, we observe that the EOW brane loops define another set of cycles/partitions corresponding to a unique NC partition $\bar\pi\in \mathrm{NC}_n$ dual to each $\pi\in \mathrm{NC}_n$. This is called the \emph{Kreweras complement} of $\pi$, defined as follows. The rule is to replicate another set of $[n]$ (say in red)  and interlace them with the original set $[n]$ (say in black); and then link up as many red dots as possible without crossing the black dots. It is evident that this procedure defines a unique NC partition $\bar\pi$. (See an illustration in Fig.~\ref{fig:kreweras_b} and a precise definition is provided in Section~\ref{sec:freeness}.).

We therefore have the matter contributions also organized,
\begin{equation}\label{eq:replicatrick}
    \tilde Z_n=\sum_{\pi\in \mathrm{NC}_n}\left(\prod_{V\in\pi}Z_{|V|}\cdot\prod_{\bar V\in\bar\pi}\tr \left[(k\rho_{\mathsf R})^{|\bar V|}\right]\right)\ .
\end{equation}
With the notion of Kreweras complement defined, this formula can be readily read off from the geometric diagrams of the saddles.

The main \emph{hypothesis} of the replica trick is that the $\tilde Z_n$'s (with appropriate normalization) are the moments of a density operator of the radiation in the fundamental description, i.e. $\tr\rho_{\tilde R}^n$. The replica trick GPI acts like an oracle, that takes as inputs the moments of the matter density operator and the moments of the black hole thermal spectrum, and then outputs a collection of moments. Mathematically, the replica trick should be understood as a \emph{moment problem}. Generally, there is no guarantee that this moment problem is well-posed, in the sense that the solution exists and it is unique. Here we have a rather explicit formula for its moments, we can in principle address this issue and hopefully work out the spectrum of the density operator. The only trouble is that it isn't obvious what the formula is actually doing, and the key is to examine it through the lens of free probability.

\subsection{Solution: The free probabilistic toolkit}

The key observation is that the combinatorial structure of how the moments are processed in~\eqref{eq:replicatrick} precisely matches with how the \emph{free multiplicative convolution} between two probability distributions works (cf. \eqref{eq:freemulti1}). To see this, we'd need a brief introduction of free probability theory. We shall follow Speicher's combinatorial treatment~\cite{nica2006lectures,mingo2017free,speicher2019lecture},\footnote{See also~\cite{novak2014three} for a pedagogical introduction.} in which the theory is built upon the key object of \emph{free cumulants}. We will give an introduction of non-commutative random variables, freeness and the free probabilistic toolkit in Section~\ref{sec:primer}.

In order to match the GPI precisely with the free multiplicative convolution, we have to deal with the issue that the black hole thermal spectral distribution, denoted as $\nu_b$ (cf. \eqref{eq:nub}), is \emph{not} a probability distribution, because the black hole density of states in the large $N$ limit is non-integrable and so is $\nu_b$.  Instead, the moments of $\nu_b$ should instead be viewed as the free cumulants of some other spectral distribution, known as the free compound Poisson distribution $\mu_{\nu_b}$. Note that $\mu_{\nu_b}$ is completely determined by $\nu_b$, which can be extracted from the GPI in the JT gravity+EOW brane theory and is parameterized by the temperature and the brane tension. Finally, we obtain the spectral distribution of the radiation as a free multiplicative convolution, which we sometimes simply refer as the \emph{free convolution},
\beq\label{eq:mainresult}
\nu_{\tilde r}=\nu_r\boxtimes \mu_{\nu_b}\ .
\eeq

Mathematically, this result says that the replica trick GPI implements the free multiplicative convolution for the two input (bulk) distributions effectively describing the black hole $\nu_b$ and the radiation $\nu_r$.  The convolution formula computes the radiation entropy accurately even in cases when the island formula fails to apply.  The result in turn justifies that our moment problem is indeed well-posed. We shall argue the existence and uniqueness of the resulting distribution, satisfying the premise of the replica trick. This is our main result and the details will be provided in Section~\ref{sec:convolution}.

\begin{figure}
    \centering
    \includegraphics[width=.6\linewidth]{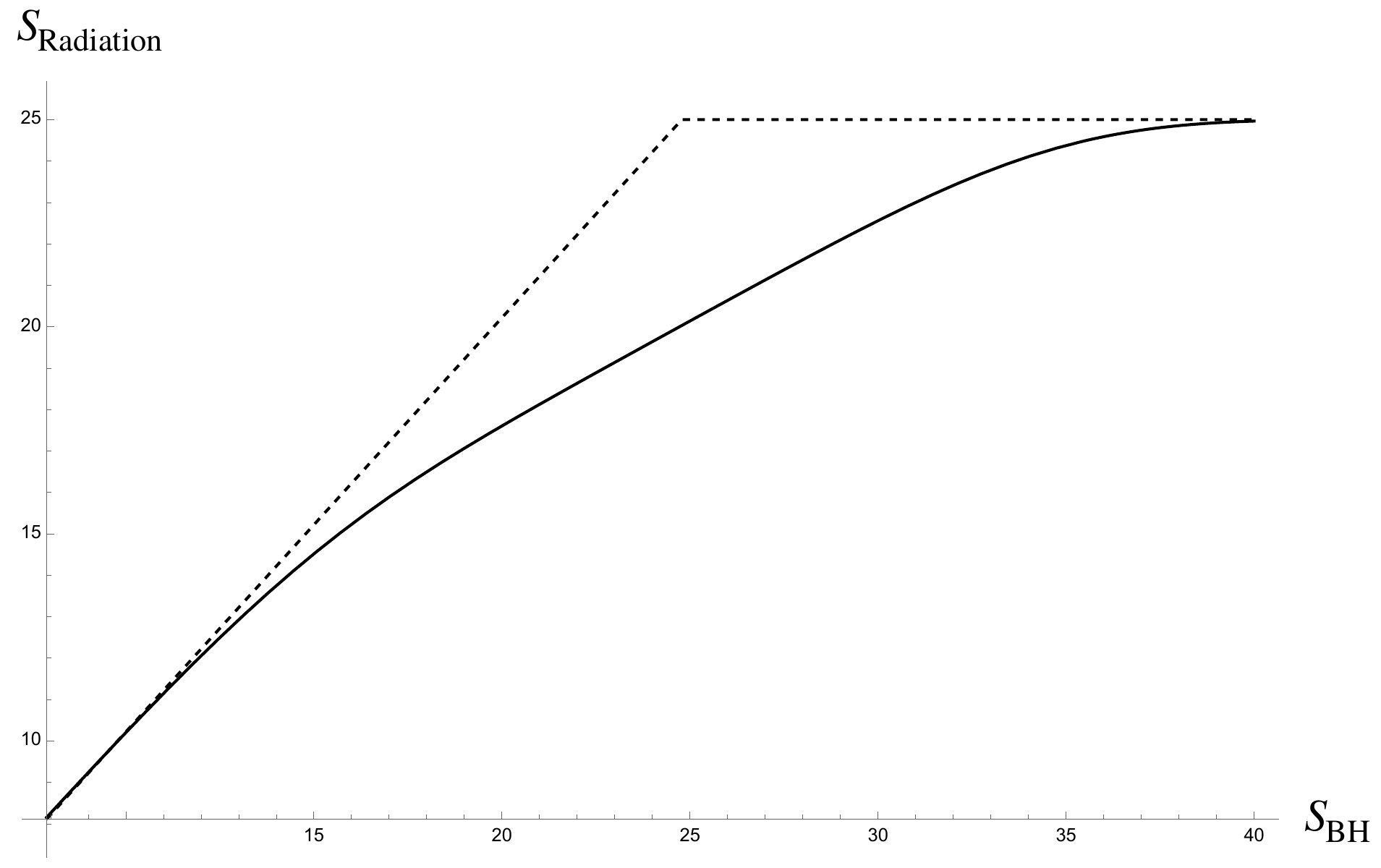}
    \caption{\textbf{A Page curve that largely deviates from the island formula.} The solid curve shows the values of the radiation entropy for various values of the BH entropy $S_\mathrm{BH}\approx S_0+4\pi^2/\beta$ with $\beta$ fixed to be $5$. The bulk entangled state is fixed to be a superposition of two branches with different Schmidt ranks, $2^{15}$ and $2^{35}$, and thus its von Neumann entanglement entropy is $25$. The transition doesn't occur at $\sbh=25$, in contrast to the prediction of the island formula~\eqref{eq:island_pssy} shown as the dashed curve. }
    \label{fig:page_curve}
\end{figure}

Given \eqref{eq:mainresult}, the radiation entropy simply equals to the (continuous) entropy of $\nu_{\tilde r}$ (cf.\eqref{eq:beyondisland}). 
Such a free convolution formula for the radiation entropy, unlike the island formula~\eqref{eq:island} or the QES formula, does not build upon the \emph{generalized entropy} of some extremal surface $\gamma$, 
$S_\mathrm{gen}(\gamma)=A[\gamma]/4G_N+S_\mathrm{bulk}$, such that its contribution in the partition function dominates over the rest. We learnt from our solution that generally there isn't a clean separation among the contributions so minimizing the generalized entropy over the candidate QES does not work. Instead, the partition functions associated with two competing QES, are \emph{freely convoluted} by the GPI, so \eqref{eq:mainresult} supersedes the island formula in the PSSY model. An example of such a Page curve\footnote{Unlike the usual Page curve, we vary the black hole hole entropy instead of the radiation dimension $k$ and the state of the bulk matter is fixed. Nonetheless, the same physics is captured because the Page curve in the PSSY model is only supposed to demonstrate the radiation entropy values at the snapshots of various configurations.} is plotted in Fig.~\ref{fig:page_curve} to demonstrate the large corrections. The departure from the island formula agrees well with the one-shot entropy analysis. (cf. Section~\ref{sec:oneshot}).\\

It seems that all we have claimed so far merely amounts to giving the replica trick \eqref{eq:replicatrick} a fancy name and a fancy symbol, but really the advantage of having \eqref{eq:mainresult} is that we can use the tools from free harmonic analysis to evaluate $\nu_{\tilde r}$. This is totally analogous to how we can use Fourier transforms to evaluate convolutions between probability distributions for classical (scalar) random variables.  

Generally, a free multiplicative convolution is difficult to evaluate even numerically. To do so, one needs to invoke a heavy set of machinery from the operator-valued free probability theory together with a couple of tricks. Fortunately, this particular free convolution at hand \eqref{eq:mainresult} is analytically tractable using the standard tools from free harmonic analysis. In Section~\ref{sec:solution} we show in detail how it can be done. Our result is summarized as an algorithm below.

For any $z$ in the upper complex plane $\mathbb{C}^+:=\{z\in\mathbb{C}|\mathrm{Im}(z)>0\}$, solve the following equation for ${y(z)}$ with the two input distributions $\nu_r, \nu_b$ (highlighted in red) supported on the positive real numbers $\mathbb{R}_+:=\{x\in\mathbb{R}|x\ge 0\}$,
\begin{equation}\label{eq:alg1}
    \int_{\mathbb{R}_+}\frac{\textcolor{red}{\nu_b}(x)\dd x}{z/x-k\int_{\mathbb{R}_+}\frac{ x'}{1-x'{y(z)}}\textcolor{red}{\nu_r}(x')\dd x'} = {y(z)}\ .
\end{equation}
This is a fixed-point equation for $y(z)$ that is numerically easy to solve. Then plug the solution $y(z)$ into the following equation
\begin{equation}\label{eq:alg2}
    G_{\nu_{\tilde r}}(z)=\frac1y\int_{\mathbb{R}_+}\frac{\textcolor{red}{\nu_r}(x)\dd x}{1-x{y(z)}}\ ,
\end{equation}
which gives the Cauchy transform of $\nu_{\tilde r}$, denoted as $G_{\nu_{\tilde r}}$, that is analytic on $\mathbb{C}^+$. The spectral distribution can be extracted using the Stieltjes inversion, 
\begin{equation}\label{eq:alg3}
    \nu_{\tilde r}(x)=-\frac{1}{\pi}\lim_{\epsilon\rightarrow 0}\mathrm{Im}\,G_{\nu_{\tilde r}}(x+i\epsilon)\,.
\end{equation}
The resulting distribution does not admit a closed-form expression for a generic $\nu_{\tilde r}$, and this algorithm gives the most explicit description that one can hope for.
In the context of random matrix theory, our solution also applies to resolving the spectrum of a class of random matrix ensemble, known as the separable sample covariance matrix. The same algorithm was obtained using the moment method in the random matrix literature~\cite{lixin2006spectral,hachem2006empirical,paul2009no,couillet2014analysis}. Here we offer a novel free probabilistic treatment that could potentially have independent interest.\\

The free convolution formula suggests that the quantum information encoded in the gravitational sector and the matter sector, or more generally the information encoded in two candidate QES,\footnote{It is merely an artifact of the PSSY model that among the two dominate saddle configurations, one only receives the contribution from the gravitational sector and the other only from the matter sector. We think generally the two sectors are not free from each other.} should be modelled as free random variables in a non-commutative probability space, which we choose to be a \emph{finite von Neumann algebra}. Voiculescu taught us that free random variables can be further modelled by large independent random matrices~\cite{voiculescu1991limit}, so we can deduce a random matrix/tensor model from the algebraic random variables for the PSSY model. We show that this random matrix model is nothing but a generalization of Page’s model, which is perhaps the very first random tensor network model used to study the information-theoretic aspects of gravity. This is discussed in Section~\ref{sec:page}. We provide more details on the interplay between random matrices and free probability in Appendix~\ref{app:random_matrices}.\\

It is legitimate to worry that the free convolution formula we had could be something very special to the PSSY model, and it is unclear if it can be generalized to realistic black holes. To address this concern, we should make an attempt to put the entropy formula in such a way that it doesn't explicitly entail the free convolution. We show that the convolution formula can be formulated as a generalized entropy, resembling the formulation due to Wall in his proof of the generalized second law~\cite{wall2012proof}. It equates, up to a state-independent constant, the radiation entropy to the relative entropy between the boundary state and the Hartle-Hawking state. In this sense, we difference in the radiation entropy precisely measures Hawking's surprisal in learning the actual state of the radiation while holding the hypothesis that it's a thermal state. This is discussed in Section~\ref{sec:holo}.\\

We also make contact with some relevant results in the literature, and perform some consistency checks on our proposal. In particular, we reproduce the original results of PSSY, and we show that the free convolution formula reduces to the island formula in regimes governed by the one-shot entropies, as expected for the general QES prescription~\cite{akers2021leading,wang2022refined}. This is discussed in Section~\ref{sec:checks}.\\

Lastly in Appendix~\ref{sec:additivity}, we discuss an observation that an old black hole encodes a channel that violates the additivity of minimum channel output entropy. The latter was a fundamental conjecture in quantum Shannon theory~\cite{shor2004equivalence} that is equivalent to the additivity of classical capacity of quantum channels, and it was proven false using a random channel construction by Hastings~\cite{hastings2009superadditivity}. Here we see that an old black hole naturally encodes such a random channel, matching a known counterexample to the conjecture~\cite{fukuda2022additivity}. Even though for an old black hole the subleading non-replica-symmetric saddles can safely be ignored as far as the entropies are concerned, the additivity violation is only visible when all the saddles are counted. Surprisingly, summing over all saddles can be important in regimes where we usually we don't expect it to be. \\

\textbf{Notations}\\

We use the lower case letters such as $r,b,c,u,\cdots$ to denote the non-commutative random variables, and they belong to a non-commutative probability spaces $(\W,\tau)$ which is a finite von Neumann algebra represented as bounded operators on an infinite-dimensional Hilbert space. It is equipped with a trace $\tau$ that gives the expectation values for the random variables. Their spectral distributions are denoted as $\nu_r,\nu_b,$ etc. The symbol $\mu$ is reserved for any specifically defined distributions. We use the upper case letters to denote the finite dimensional (random) matrices as $R,B,C,U,\cdots$. We denote the normalized trace as $\ttr:=\frac1N\tr$. For density operators under this notation convention, we shall not assign additional symbols to the density matrices, but instead directly use the system symbol to denote the its density matrix (such as $R:=\rho_{\mathsf R}$, etc). To avoid confusions, we use the Serif fonts, such as $\mathsf{B, R}$, to denote the systems. The entropies of density operators are then simply denoted as $S(R),  S_\al(R)$, etc. We shall use the tilde symbols $\tilde R$ and $\tilde B$ to denote density matrices of the radiation and the black hole in the fundamental description (at the boundary), and we use symbols $R$ and $B$ in the effective description (in the bulk). \\

\textbf{Related works}\\

The large correction to the QES formula was first pointed out by AP~\cite{akers2021leading}. An example of a bulk state in an incoherent superposition was given in support of the argument, but the general behaviour is left as indefinite. Later, the large correction is also shown for a similar non-flat bulk state in a coherent superposition of two branches in the PSSY model.  Both calculations use the method of Schwinger-Dyson equation with the resolvent, which was first used by PSSY to derive the Page curve in a canonical ensemble. However, this method does not seem to be applicable when we deal with general non-flat bulk states. This work should be viewed as a generalization of the resolvent method by leveraging the power of free probability. The basic idea that underlies this work already appears in the dissertation of the author~\cite{wang2022thesis}. In this paper, a further developed study is presented. Free probability has also been used in computing various distinguishability measures for random states that resemble the black hole microstates~\cite{kudler2021distinguishing}.

A similar result is obtained by Cheng et al for random tensor networks (RTN). They consider RTN with bonds of non-flat entanglement spectrum modelling area fluctuations~\cite{cheng2022random}. In a situation where there are two minimal cuts homologous to a boundary subregion, the spectral distribution of the boundary reduced state asymptotically equals to the free multiplicative convolution of the limiting spectral distributions of the bonds at the two cuts, which are further convoluted with a Marchenko-Pastur distribution. We will elaborate on how their result compares to ours in Section~\ref{sec:cheng}. In this work, we go one step further by showing how this particular convolution can actually be evaluated in Section~\ref{sec:solution}.

\section{A free probability primer}\label{sec:primer}

Free probability theory is a probability theory of non-commutative random variables equipped with a special notion of independence called \emph{freeness}. Before going into freeness, we first need to introduce some basics about non-commutative probability theory. It is usually formulated abstractly in terms of the algebra of random variables and their expectations. The abstraction allows the framework to accommodate classical probability theory, quantum theory and free probability theory. Here we give a minimal introduction of the relevant notions to set some backgrounds. We shall approach the subject following the treatments in~\cite{tao2012topics,mingo2017free,speicher2019lecture}. Experts should feel free to skip over this section.

\subsection{Non-commutative probability spaces}
The most general non-commutative probability space is defined on a unital (i.e. it contains the multiplicative identity $1$) $*$-algebra $\mathcal{A}$, which is algebra $\mathcal{A}$ over $\mathbb{C}$ endowed with an involution $*$ such that for all $a,b\in\mathcal{A}$ and $z\in\mathbb{C}$, 
\begin{equation}
    (za)^*=\bar z a^*,\ (a+b)^*=a^*+b^*,\ (ab)^*=b^*a^*,\ (a^*)^*=a,\ 1^*=1\ ,
\end{equation}
and elements satisfying $a=a^*$ are called \emph{self-adjoint}.

Examples of $*$-algebra that are particularly relevant to us are complex $N\times N$ matrices with the involution being the conjugate transpose; and bounded complex-valued random variables over some probability space $(\Omega,\mathcal{F},P)$, denoted as $L^{\infty}(\Omega,\mathcal{F},P)$, with the involution being the complex conjugate.

For a $*$-algebra to be a probability space, we need to add an expectation that is mathematically a unital $*$-linear (i.e. $\varphi(a^*)=\overline{\varphi(a)}$) function $\varphi:\mathcal{A}\to\mathbb{C}$.  
\begin{defn}
A non-commutative probability space\footnote{The term ``non-commutative'' should throughout be interpreted as ``potentially'' non-commutative, because it can also model commutative random variables. } $(\mathcal{A},\varphi)$ is defined as a unital $*$-algebra $\A$ equipped with a unital $*$-linear functional $\varphi$ that is positive and faithful.\footnote{A glossary of jargons: unital: $\varphi(1)=1$, positive: $\varphi(aa^*)\ge 0$ and faithful: $\varphi(aa^*)=0\iff a=0$. } 
\end{defn}
We call such a linear functional $\varphi$ a \emph{state} on $\A$. In probabilistic terms, $\A$ is an algebra of non-commutative random variables, and the state gives their expectation values. Technically, what we defined is called a \emph{$*$-probability space}.

A familiar example of a non-commutative probability space is quantum theory $(\A,\varphi)$, where $\A\subset\mathcal{B}(\h)$ is the operator algebra of observables as a subalegebra of bounded operators on a Hilbert space $\h$, and a vector in Hilbert space $\ket{\varphi}\in\h$ defines the state $\varphi(a):=\bra{\varphi}a\ket{\varphi}, a\in\A$. The state gives the expectation value for the observable $a$. In particular, we can consider a positive-operator-valueds-measure (POVM) set $\{m_i\}_{i\in I}$ and then $\psi(m_i)$ gives the Born rule probability of observing the outcome $i$.

This definition of probability space is flexible enough to allows us to talk about both commutative and non-commutative random variables under the same footing. For example, we can take $L^\infty(\Omega,\mathcal{F},P)$ defined above, which we henceforth abbreviate $L^{\infty}(\Omega)$, and we take the linear function to be the expectation $\E[\cdot]$ induced by the measure $P$. Another example is the algebra of $N\times N$ (deterministic) complex matrices, $(M_N(\mathbb{C}),\ttr)$, where $M_N(\mathbb{C})$ is the algebra of complex matrices of size $N$ and $\ttr$ is the normalized trace, $\ttr(X):=\frac1N\tr X$ for any $X\in M_N(\mathbb{C})$. 

The more interesting example is to put these together to accommodate random matrices,
\begin{equation}
   (L^{\infty}(\Omega)\otimes M_N(\mathbb{C}),\E\otimes\ttr)\ .
\end{equation} 
It is the algebra of $N\times N$ random matrices with individual matrix entry being a random variable on $(\Omega,\mathcal{F},P)$, and it is equipped with the expectation,
\begin{equation}\label{eq:traceN}
    \E\otimes\ttr(a):=\E[\ttr\,a]\,\quad a\in L^{\infty}(\Omega)\otimes M_N(\mathbb{C})\ .
\end{equation}

In the above example for random matrices, the state  $\varphi=\E\otimes\ttr$ has the important property of being \emph{tracial}. It means that $\varphi(ab)=\varphi(ba)$. Since our application of free probability theory concerns only the tracial case, we shall now restrict to such states. It is often desirable to have a \emph{unique} tracial state as providing the notion of expectation. Furthermore, later we would like to be able to discuss functions of random variables, so we better work with an algebra that is completed to pave the way for functional calculus. To this end, we can complete the $*$-algebra to a von Neumann algebra, which is also referred as a $W^*$-algebra.\footnote{One can also build a $C^*$-probability space on a $C^*$-algebra, see Chapter 3 in~Nica-Speicher\cite{nica2006lectures}.}

A von Neumann algebra $\W$ is defined as a unital $*$-algebra of bounded operators on a Hilbert space $\mathcal{B}(\h)$ that is closed in the weak operator topology.\footnote{The weak operator topology is the operator topology induced by the seminorms: $a\mapsto (v,aw)$ for any $v,\omega\in\h$. Namely, a sequence/net of bounded operators $a_i\in\W$ converges to $a\in\W$ in weak operator topology if $(v,a_iw)$ converges to $(v,aw)$ for any $v,\omega\in\h$.} We denote the commutant of $\W$ in $\mathcal{B}(\h)$ as $\W':=\{b\in\mathcal{B}(\h)|ab=ba, \forall a\in\W \}$. The bicommutant theorem says that a $*$-algebra $\W$ is a von Neumann algebra if and only if $\W=(\W')'$. The center of $\W$ is defined as $Z(\W):=\W\cap\W'$ and we say $\W$ is a factor if the center is trivial, i.e.  $Z(\W)=\mathbb{C}1$. 

\begin{defn}
A (tracial)\footnote{More generally, we can allow non-tracial state $\tau$ for a $W^*$-probability space. Since we shall only consider tracial situations, we often omit ``tracial'' in describing a $W^*$-probability space.} $W^*$-probability space $(\W,\tau)$ is defined as a von Neumann algebra $\W$ equipped with the a unital tracial normal\footnote{A state $\tau$ on $\W$ is normal if we have $\tau(\sup_\la a_\la)=\sup_\la\tau(a_\la)$ for any increasing nets $(a_\la)_{\la\in\Lambda}$ of self-adjoint operators $a_\la\in\W$.} state $\tau$.
\end{defn}
Such a unital tracial normal state is \emph{unique} when $\W$ is a \emph{finite} factor, which means it is either Type I$_n$ or Type II$_1$.\footnote{(Hyperfinite) factors can be classified based on the properties of their projections. Here we give a brief summary. A Type I factor has a minimal projection, and is isomorphic to the algebra bounded operators on some Hilbert space $\mathcal{B}(\h)$. A Type~II factor has no minimal projection but has a finite projection, and it is further a Type~II$_1$ factor if all the projections are finite. Lastly, a Type III factor has no finite projections. cf. e.g. Jones~\cite{jones2003neumann} for detailed explanations.} A Type I$_n$ factor is isomorphic to the familiar matrix algebra $M_n(\mathbb{C})$. A Type II$_1$ factor only has representations as a subalgebra of bounded operators on an infinite dimensional Hilbert space $\h$. Moreover, there is no irreducible representations, that is to say that every vector in $\h$ is entangled between the $\W$ and its commutant $\W'\in\mathcal{B}(\h)$. Nonetheless, trace does exist for every element in a Type II$_1$ factor, so we can still make sense of density operators and entropies, which we shall introduce later.  A Type II$_1$ algebra is also the setting where free probability was first invented by Voiculescu to study the free group factors isomorphism problem~\cite{voiculescu1985symmetries,voiculescu2005free}.

Unless otherwise specified, we will by default work with a $W^*$-probability space (built on a finite von Neumann algebra) as the stage for free probability theory. For example, note that the random matrix algebra $L^{\infty}(\Omega)\otimes M_N(\mathbb{C})$ is a direct integral of Type I$_N$ factors with a center $L^{\infty}(\Omega)$, equipped with a trace $\E\otimes\ttr$, so it is indeed a $W^*$-probability space. 
 
A non-commutative probability space $(\W,\tau)$ can be abstractly defined without the need to concretely set up an underlying sample space and event space as we did for the random matrices. This allows to make universal statements at the abstract level, a desirable property especially convenient for studying random matrices that often manifests universal features asymptotically. Freeness, as we shall introduce later, is a good example of such an abstraction describing the universal behaviour of independently distributed random matrices.

\subsection{Non-commutative distributions, density operators and entropies}\label{sec:entropies}

For a classical random variable with compact support, knowing all its moments is equivalent to knowing its distribution. For instance, given the moments $\E(X^n)$ of a real-valued random variable $X$, we can find a probability distribution $\mu_X$ such that
\begin{equation}
    \E(X^n) = \int x^n\mu_X(x)\ \dd x\ .
\end{equation}
Similarly, for a collection of scalar random variables $X_1,\ldots,X_k$, we have a joint probability distribution $\mu_{X_1,\ldots,X_k}$ that produces all the joint moments,
\begin{equation}
    \E (X_{i_1}\cdots X_{i_n}) = \int x_{i_1}\ldots x_{i_n}\ \mu_{X_1,\ldots,X_k}(x_{i_1},\ldots, x_{i_n})\ \dd x_{i_1}\ldots \dd x_{i_n},\ \forall n\in\mathbb{N}\ \mathrm{and}\  i_1,\ldots i_n\in [k]\ .
\end{equation}

Note that we can make sense of the probability distribution defined above in two different ways. Given an underlying Kolmogorov triple $(\Omega,\mathcal{F},P)$, $\mu_{X_1,\ldots,X_k}$ is nothing but the push forward of $P$ under the measurable map $\omega\in\Omega\mapsto(X_1(\omega),\ldots,X_k(\omega))\in\mathbb{R}^k$. We call this an \emph{analytic distribution} and this is what the distributions mean in the integral formulas above.

More abstractly, if we choose to work without referring to an underlying sample space but only with the expectations of our random variables, then the notion of probability distribution should generally be understood algebraically as the following linear map 
\begin{equation}
   \mu_{X_1,\ldots,X_k}: p\in \mathbb{C}[x_1,\ldots, x_k]\mapsto \E[p(X_1,\ldots, X_k)]\in \mathbb{R}
\end{equation}
where $\mathbb{C}[x_1,\ldots, x_k]$ denotes the ring of commutative polynomials in $k$ indeterminates $x_1,\ldots, x_k$.

We shall call this the \emph{algebraic distribution} and this is the appropriate notion that can be generalized to the non-commutative setting, in which the Kolmogorov triple is usually abstracted away. 

Given a collection of non-commutative self-adjoint random variables $a_1,\ldots,a_k\in (\W,\tau)$, their \emph{non-commutative distribution} is thus defined as the linear map
\begin{equation}
   \mu_{a_1,\ldots,a_k}: p\in \mathbb{C}\langle x_1,\ldots, x_k\rangle\mapsto \tau(p(a_1,\ldots, a_k))\in \mathbb{C}
\end{equation}
where $\mathbb{C}\langle x_1,\ldots, x_k\rangle$ denotes the ring of non-commutative polynomials in $k$ indeterminates $x_1,\ldots, x_k$. By linearity, we can equivalently define the non-commutative distribution as the collection of all their \emph{joint moments} $\tau(a_{i_1}\cdots a_{i_n}), \forall n\in\mathbb{N}, i_1,\ldots i_n\in [k]$. 

Generally, we do not have the analytic counterpart to a non-commutative distribution. Nonetheless, this is possible for the case of a single bounded\footnote{\label{ft:radius}The spectral radius $\rho(a)$ of a self-adjoint random variable in $(\W,\tau)$ is defined as $\rho(a):=\lim_{n\to\infty}|\tau(|a|^{2n})|^{1/2n}$, and the Cauchy-Schwarz inequality implies that the moments are (exponentially) bounded by $\rho(a)$, $\tau(a^n)\le\rho(a)^n$. We say a self-adjoint random variable $a$ is bounded if its spectral radius is finite.} random variable.\footnote{\label{ft:normal} Self-adjointness is not necessary here for the analytical distribution to exist. We choose to focus on self-adjoint variables here because these are the most relevant for us. Generally, for any normal operator (i.e. it commutes with its involution) in a *-probability space, an analytic distribution can also be constructed from the \emph{$*$-moments}, $\tau(a^ka^{*l}), \forall k,l\in\mathbb{N}$. See e.g. Chapter 1 in~\cite{nica2006lectures}.} We can find an analytic distribution from its moments as in the classical case. Mathematically this is known as a \emph{Hausdorff moment problem}, which asks for a unknown compactly supported distribution that has the given moments. The solution is unique if it is solvable.\footnote{For a collection of moments $\{m_n\}_{n=0}^\infty$ to define a unique probability distribution, it needs to satisfy Carleman's condition, $\sum_{n=1}^\infty m_{2n}^{-1/2n}=\infty$. This condition is always satisfied for a compactly supported probability distribution. } 

Consider a bounded single self-adjoint random variable $a\in(\W,\tau)$. Given its moments $\tau(a^n)$, one can find a unique measure $\nu_a$ with compact support on $\mathbb{R}$ such that
\begin{equation}\label{eq:analytic_distribution}
    \tau(p(a))=\int p(x) \nu_a(x) \dd x \ 
\end{equation}
for any polynomial $p\in\mathbb{C}\langle x\rangle$. The support is explicitly given by the spectral radius of $a$ as $[-\rho(a),\rho(a)]$ where $\rho(a)$ is the spectral radius of $a$ (cf. Footnote~\ref{ft:radius}). 

In fact, this claim holds generally for a $*$-probability space.\footnote{See Theorem 2.5.8 in~\cite{tao2012topics} for the more general statement.} However, the advantage of completing a $*$-algebra to a von Neumann algebra is that we can use these polynomials as building blocks to extend \eqref{eq:analytic_distribution} to any continuous functions on the support of $\nu_a$. We can define
\begin{equation}\label{eq:fundational_calculus}
    \tau(f(a)):= \int f(x) \nu_a(x) \dd x
\end{equation}
The operator $f(a)$ on the LHS is defined via a unital $*$-algebra homomorphism $\Phi_a$ mapping from $f\in L^\infty(\mathbb{R},\nu_a)$ to $f(a)\in\W$. $\Phi_a$ is known as the \emph{functional calculus} for $a$. Since polynomials are dense in the continuous functions over supported over $[-\rho(a),\rho(a)]$, \eqref{eq:fundational_calculus} uniquely determines $\Phi_a$. (cf. Theorem 3.1 in Nica-Speicher~\cite{nica2006lectures}.) We shall refer to this analytic distribution $\nu_a$ as the \emph{spectral distribution} of $a$. Such an extension to continuous functions is necessary for us to study spectral functions like the von Neumann entropy. \\

We now discuss how to define entropies for states acting on a $W^*$-probability space. This was first studied by Segal~\cite{segal1960note}. (See also a recent article on this subject~\cite{longo2022note}.) In a $W^*$-probability space, the \emph{density operator} $a\in (\W,\tau)$\footnote{When the density operator $a$ is unbounded, it does not belong to the algebra $\W$ itself but is \emph{affiliated} to the algebra, which means that bounded functions of $a$ belong to $\W$. For simplicity, we assume that our density operators have bounded spectra. } of a normal faithful state $\varphi$ on $\W$ is defined via the expectation $\tau$ via $\varphi(x)=\tau(xa)$ and has normalization $\tau(a)=1$. Such assignment of the density operators is unambiguous as the expectation $\tau$ is unique.

The von Neumann entropy of the state $\varphi$ is then defined as usual
\begin{equation}\label{eq:vnent}
   S(\varphi) = S(a) :=  -\tau(a\log a) := -\int x\log x\ \nu_a(x)\dd x. 
\end{equation}
where we've used \eqref{eq:fundational_calculus} for $f(x)=-x\log x$. 

Similarly, the $\al$-R\'enyi entropies are defined as
\begin{equation}\label{eq:renyient}
   S_\al(\varphi) = S_\al(a) :=  \frac{1}{1-\al}\log\tau(a^\al):=\frac{1}{1-\al}\log\int x^\al\ \nu_a(x)\dd x,\ \forall \al\in\mathbb{R}_+\ . 
\end{equation}

The von Neumann entropy so defined take values in $[-\infty,0]$, where the maximal is obtained at the tracial state $\tau$ itself, $S(\tau)=S(1)=0$. It differs from the entropy of density matrices in the usual quantum information convention. Consider now the usual convention for a finite dimensional matrix algebra $(M_N(\mathbb{C}),\tr)$, and the density matrix $A\in(M_N(\mathbb{C}),\tr)$ of a tracial state $\varphi$ is normalized by $\tr A=1$, against the density matrix $A_N\in(M_N(\mathbb{C}),\ttr)$ of the same state $\varphi$ normalized by $\ttr A_N=1$. It follows that $A_N=NA$ and 
\begin{equation}\label{eq:conti_entropy}
S(A) := -\tr A\log A = \log N- \tau A_N\log A_N = S(A_N)+\log N\approx S(a)+\log N \ .
\end{equation}
Therefore, as we go to infinite dimensions $N\to\infty$, if we have $A_N\to a\in (\W,\tau)$, then $S(A)$ typically diverges whereas $S(A_N)\to S(a)$ remains finite. 

We are able to compute the entropy $S(a)$ directly from the freely convoluted spectral distribution. In order to obtain the standard entropy value $S(A)$ for some $N\times N$ density matrix $A$, we simply add back $\log N$. This is a good approximation if $N$ is large enough such that $S(A_N)$ is close to its asymptotic value $S(a)$. \\

Finally, we'd like to briefly sketch how to define the relative entropy between two faithful normal states $\varphi$ and $\psi$ on a von Neumann algebra $\W$. Readers can refer to the review article~\cite{witten2018aps} for details.

Consider two faithful normal states $\varphi$ and $\psi$ on a von Neumann algebra $\W$ that is not necessarily finite, they can be represented as cyclic and separating vectors $\ket{\varphi}, \ket{\psi}\in \h$ in some Hilbert space $\h$,\footnote{Cyclic means $\{a\ket{\varphi}|a\in\W\}$ is dense in $\h$ and separating means $a\ket{\varphi}=0$ if and only if $a=0$. Such a (standard) representation can be constructed using the GNS construction~\cite{gelfand1943imbedding,segal1947irreducible}.} such that for any $x\in\W$,
\begin{equation}
    \varphi(x) = \bra{\varphi} x\ket{\varphi},\quad\psi(x) = \bra{\psi} x\ket{\psi}\ .
\end{equation}
Now we need to introduce the Tomita operator $S_{\varphi|\psi}$, defined via
\begin{equation}
    S_{\varphi|\psi}\,a\ket{\varphi} = a^*\ket{\psi}\ ,\forall a\in\W
\end{equation}
The Tomita operator admits the following polar decomposition,
\begin{equation}
    S_{\varphi|\psi} = J_{\varphi|\phi}\Delta_{\varphi|\psi}^\frac12
\end{equation}
where $J_{\varphi|\psi}$ is an anti-unitary operator that sends $\W$ to $\W'$, known as the relative modular conjugation, and $\Delta_{\varphi|\psi} = S_{\varphi|\psi}^*S_{\varphi|\psi}$ is the relative modular operator. Now we can define the relative entropy
\begin{equation}\label{eq:araki}
    D(\varphi||\psi):=-\bra{\varphi}\log\Delta_{\varphi|\psi} \ket{\varphi}\ .
\end{equation}

For a finite von Neumann algebra $(\W,\tau)$, the relative entropy can also be expressed in terms of the corresponding density operators using the more familiar Umegaki's formula,
\begin{equation}\label{eq:umegaki}
    D(\varphi||\psi)= D(a||b):=\tau(a\log a) - \tau(a\log b). 
\end{equation}
where $a$ and $b$ are the density operators of $\varphi$ and $\psi$ respectively.

\subsection{Freeness}\label{sec:freeness}
We have discussed how to make sense of the spectral distribution of a single bounded self-adjoint random variable. Can one go beyond the case of a single random variable? This calls for an appropriate notion of independence among non-commutative random variables. As we shall see, the very notion of \emph{freeness} is an appropriate one that is alternative to the more familiar tensor independence, and free probability theory is the study of \emph{free} non-commutative random variables. 
\begin{defn}[Freeness]
A collection of non-commutative random variables $a_1,\ldots,a_d$ in a $W^*$-probability space $(\W,\tau)$ is free if
\begin{equation}\label{eq:free}
    \tau([p_1(a_{i_1})-\tau(p_1(a_{i_1}))]\cdots[p_k(a_{i_k})-\tau(p_k(a_{i_k}))]) = 0
\end{equation}
for any polynomials $p_1,\ldots p_k$ and indices $i_1,\ldots i_k\in[d]$ with no adjacent ones equal. 
\end{defn}

Let's unpack this definition of freeness. In words, it says ``the alternating product of centered random variables is centered''. As advocated by Speicher, we can think of freeness as a  convenient rule to allow the joint moments to be expressed in the individual moments of the individual variables. For example, the definition \eqref{eq:free} implies that for free $a$ and $b$,
\begin{multline}
\tau\left(ab\right) = \tau\left(a\right)\tau\left(b\right),\quad\tau\left(aba\right) = \tau\left(a^2\right)\tau\left(b\right),\quad\tau\left(aabb\right)=\tau\left(a^2\right)\tau\left(b^2\right)\\
    \tau\left(abab\right)=\tau\left(a^2\right)\tau\left(b\right)^2+\tau\left(b^2\right)\tau\left(a\right)^2- \tau\left(b\right)^2\tau\left(a\right)^2.
\end{multline}
This is not immediately obvious that \eqref{eq:free} implies one can break down the joint moment of any word into the individual moments of its free letters, but we shall later show that this is indeed true. Since the moments define the non-commutative distributions, freeness should be understood as a notion of independence in the sense that the joint distribution of several free random variables is completely determined by the distribution of individual ones.

Note however that freeness is distinct from the notion of (tensor/Cartesian) independence, that concerns scalar random variables\footnote{Scalars are operators that are proportional to the identity.} or commutative objects more generally. Even for a non-commutative probability theory, such as quantum theory $(\A,\varphi)$, independence concerns commuting self-adjoint operators representing observables that are simultaneously measurable. We say two commuting observables $a$ and $b$ are independent in $(\A,\varphi)$ if $\varphi(a^{m_1}b^{n_1}a^{m_2}b^{n_2}\cdots)=\varphi(a^{m_1+m_2+\cdots})\varphi(b^{n_1+n_2+\cdots})$.\footnote{Rather than talking about independence between two observables for a given state, one can also discuss notions of independence between two observable subalgebras associated with different subsystems for any states. They are commonly formulated as the ability to independently prepare states for each subsystem. There are various ramifications of such notions and commutativity between the subalgebras are not generally demanded. See~\cite{redei2010quantum} and references therein.} This doesn't happen say when $\varphi$ is a Bell state shared by Alice and Bob and $a, b$ are spacelike separated local observables of Alice and Bob. Their observations (measurement outcomes) are therefore correlated.

On the other hand, freeness~\eqref{eq:free} gives a distinct rule for how the joint moments should factorize for non-commutative random variables. We can use the following example to illustrate their difference. Consider two independent commuting random variables in a $W^*$-probability space $(\W,\tau)$ with zero means but positive second moments. Then $\tau(abab)=\tau(a^2)\tau(b^2)>0$, so they cannot be free because freeness demands $\tau(abab)=\tau(a^2)\tau(b)^2+\tau(a)^2\tau(b^2)-\tau(a)^2\tau(b)^2=0$.  In fact, one can show that for a pair of commutative random variables to be free, one of them has to be a scalar, and a scalar is free from anything.\footnote{See for example Proposition 1.10 and 1.11 in~\cite{speicher2019lecture}.} Hence, freeness and independence are almost orthogonal notions and neither is stronger than the other. 

Despite their sheer difference, freeness is a sensible counterpart of independence that is tailor-made for non-commuting random variables. This is because the key theorems and properties about independent random variables naturally have their free analogs for free random variables. Examples are various limit theorems, such as the free central limit theorem~\cite{voiculescu1986addition} and the law of large numbers~\cite{lindsay1997some,haagerup2013law}, the free convolutions of distributions~\cite{voiculescu1986addition,voiculescu1987multiplication} and classification of infinitely divisible
and stable laws~\cite{bercovici1992levy,bercovici1999stable}. An example that is particularly relevant for us is the \emph{free Poisson limit theorem}. It is the non-commutative analog of the Poisson limit theorem that asserts the sum of Bernoulli random variables (coin flips) asymptotes to a Poisson random variable. We will discuss it later in Section~\ref{sec:circulars}. \\

Conversely, given any probability measure $\nu$ with compact support on $\mathbb{R}$, it's always possible to construct a $W^*$-probability space $(\W,\tau)$, such that there exists some self-adjoint $a\in\W$ and $\nu_a=\nu$. (cf. e.g. Facts 4.2 in~\cite{speicher2019lecture}). 

Furthermore, with a collection of non-commutative probability spaces $(\W_i,\tau_i)_{i\in [n]}$, one can always amalgamate them into a larger probability space $(\W,\tau)$, such that $\W_i\subset\W$, $\tau(a_i)=\tau_i(a_i), \ \forall a_i\in\W_i$ and any collection of random variables draw individually from these subalgebras, $\{a_i\}, a_i\in\W_i$, are free. (cf. e.g. Lemma 2.5.20 in Tao~\cite{tao2012topics}). This is achieved by the \emph{free product},\footnote{Here we stick to the standard terminology. Note that the term free product is used instead for the free multiplicative convolution in Cheng et al~\cite{cheng2022random}, which should not be confused with the algebraic operation we are referring to here.} denoted as $\W=*_{i\in [n]}\W_i, \tau=*_{i\in [n]}\tau_i$. We shall spare any details on free product here and interested readers shall consult chapters $5$-$7$ in Nica-Speicher~\cite{nica2006lectures}.  It allows us to introduce free random variables of specified distributions in a joint probability space, just like we are used to do in classical probability theory. The only difference is that we now replace the Cartesian/tensor product by free products. \\

Free probability was born in the field of operator algebra~\cite{voiculescu1985symmetries}. Nonetheless, it was later realized by Voiculescu that it has an intimate connection with random matrix theory.  In Voiculescu's seminal work in 1991~\cite{voiculescu1991limit}, he proved the first result connecting between the asymptotic random matrices and free probability theory. (cf. Theorem~\ref{thm:Voiculescu} for the precise statement in Appendix~\ref{app:random_matrices})

Roughly speaking, Voiculescu's theorem implies that random matrices with classically independent matrix elements tend to be free in the asymptotic limit. In fact, the asymptotic freeness would not hold without the non-commutativity as we argue above. Heuristically, we can say that any finite-dimensional random matrices are still too commutative to be free. When studying an abstract non-commutative random variable, it is often useful to use a sequence of random matrices as a model for concreteness. We will later also make contact with random matrices  for a better understanding of the free probabilistic calculations.  We postpone more details regarding this connection to Appendix~\ref{app:random_matrices}.\\

At the level of moments, it is somewhat cumbersome to describe freeness using \eqref{eq:free}. We now introduce an alternative characterization in terms of the \emph{free cumulants}. We first define the notion of non-crossing (NC) partitions.\footnote{The non-crossing partitions are in one-to-one correspondence with the non-crossing permutations: For each NC partition $\pi=\{V_i\}_i$, we let the NC permutation be defined with cycles $(V_i)_i$ such that each cycle $(V_i)$ consists of the elements in the block $V_i$ arranged in an increasing order. }
\begin{defn}\label{def:nc}
A partition $\pi$ of the set $[n]$ is defined as $\pi=\{V_i\}_i$ such that the blocks $V_i$ satisfy $V_i\ne\emptyset\,\,\forall i$, $V_i\cap X_j=0\,\, \forall i\ne j$ and $\cup_i V_i=[n]$. A partition $\pi$ is non-crossing if it does not have $a_1<b_1<a_2<b_2$ for some $a_1,a_2\in V_i$ and $b_1,b_2\in V_j$ with $i\ne j$. 
\end{defn}
We denote the set of NC partitions of $[n]$ as $\mathrm{NC}_n$. We use $|\pi|$ to denote the cardinality of $\pi\in \mathrm{NC}_n$.
Given a finite set $[n]$ of size $n$, consider now doubling it with additional elements $\bar 1,\ldots,\bar n$ and interlace them with $1,\ldots,n$ in an alternating way, $1,\bar 1,\ldots,n,\bar n$. For any $\pi\in \mathrm{NC}_n$, its \emph{Kreweras complement}~\cite{kreweras1972partitions}, $\bar\pi\in NC(\bar 1,\ldots,\bar n)\cong \mathrm{NC}_n$, is defined as the biggest\footnote{There is a natural partial order defined for NC partitions, called the reverse refinement order. We say $\pi_1\le\pi_2$ if each block of $\pi_1$ is contained in the blocks of $\pi_2$. } element in $NC(\bar 1,\ldots,\bar n)$ such that $\pi\cup\bar\pi\in NC(1,\bar 1,\ldots,n,\bar n)$.\footnote{The corresponding permutation of $\bar\pi$ is equivalently given by $\pi^{-1} \eta$ where $\eta$ is the cyclic permutation~\cite{biane1997some}. }  See Fig.~\ref{fig:kreweras_b} for an illustration. Note that generally for any two partitions $\pi_1,\pi_2\in\mathcal{P}_n$, we have $|\pi_1|+|\pi_2|\le |\pi_1 \pi_2|+n$. We have the equality saturated $|\pi|+|\bar\pi|=n+1$ for and only for a NC partition and its Kreweras complement.\footnote{One can define a metric on the permutations known as the \emph{Cayley distance}, $d(\pi_1,\pi_2):=n-|\pi_1^{-1} \pi_2|$, which also counts the minimal number of transpositions needed to transform $\pi_1$ to $\pi_2$. Then equation $|\pi_1|+|\pi_2|\le |\pi_1 \pi_2|+n$ follows from the triangle inequality of the Cayley distance.}\\

We are now ready to define free cumulants.
\begin{defn}
For a unital linear function $\tau:\W\mapsto\mathbb{C}$, the free cumulants are multilinear maps $\kappa_n: \W^n\mapsto \mathbb{C}$ defined recursively via 
\begin{equation}\label{eq:m-c0}
    \tau(a_1\cdots a_n) = \sum_{\pi\in \mathrm{NC}_n}\prod_{V\in\pi} \kappa_{|V|}(\{a_i\}_{i\in V})
\end{equation}
where $V$ is a block of a NC partition $\pi$ and $|V|$ denotes its cardinality.
\end{defn}
We shall often drop free and just call them cumulants. We refer \eqref{eq:m-c0} as the \emph{moment-cumulant formula}. There is only one term in the sum involving the highest cumulant $\kappa_n$, thus the formula can inductively be inverted for the cumulants in terms of the moments.\footnote{More precisely, the set of non-crossing partitions forms a lattice with respect to refinement order and the cumulants are given by the Mobius inversion with respect to this order.}
Therefore, \eqref{eq:m-c0} is indeed the defining formula for free cumulants. 

We are often interested in the relation between the moments and cumulants of a single random variable $a$, we can define the abbreviations for the $n$-th moment and $n$-th cumulant of $a$ as
\begin{equation}
    m_n(a) := \tau(a^n) \, \quad \kappa_n(a) := \kappa_n(a,\ldots,a) \ .
\end{equation}
We then have a special case of the free moment-cumulant formula \eqref{eq:m-c0},
\begin{equation}\label{eq:m-c}
    m_n(a) = \sum_{\pi\in \mathrm{NC}_n}\prod_{V\in\pi} \kappa_{|V|}(a) \ .
\end{equation}

We shall remark that for a single self-adjoint operator $a$, the moments $m_n(a)$ are identical to the classical moments of a random variable $X$ with the probability distribution given by the spectral distribution of $a$. Because of the distinction between freeness and independence, however, the corresponding classical cumulants, denoted as $u_n(X)$, are defined differently from $\kappa_n(a)$. The same formula~\eqref{eq:m-c} is used up to changing the NC partitions to all partitions of $[n]$, denoted as $\mathcal{P}_n$, yielding the classical moment-cumulant formula,
 \begin{equation}\label{eq:classicalm-c}
     m_n(X) = \sum_{\pi\in \mathcal{P}_n}\prod_{V\in\pi} u_{|V|}(X) .
 \end{equation}
 
Hence, one can see that $\kappa_n$ and $u_n$ with $n\ge 4$ are different from each other.  The classical cumulants $u_n(X)$ are also known as the Ursell functions in statistic physics or the connected correlation functions in QFT. In a Feynman diagram expansion of a partition function in QFT, the $n^\mathrm{th}$ cumulant $u_n(\phi)$ defined for a field $\phi$ gives the $n$-connected Green functions that connect exactly $n$ sources. A cumulant $u_n(\phi)$ should thus be understood as capturing genuine $n$-partite interactions, as opposed to independence or correlations that can be decomposed into interactions among fewer parties. Analogously, we will see that the free cumulants are precisely computed by the replica wormholes in the planar gravitational saddles of the replica trick GPI.\\

Here comes the punchline: \emph{Freeness is equivalent to vanishing mixed cumulants}. 
\begin{thm}[Speicher~\cite{speicher1994multiplicative}]\label{thm:cumufree}
The fact that $a$ and $b$ in $(\W,\tau)$ are free is equivalent to vanishing $\kappa_n(x_1,\ldots,x_n)$ whenever $n\ge 2$, $x_i=a$ or $b\;\;\forall i$ and $x_i=a, x_j=b$ for some $i,j$. 
\end{thm}

It's now clear that any joint moments of free random variables can be written in terms of the individual cumulants, because the mixed cumulants vanish. Since each individual cumulants is a polynomial of the individual moments via inverting the moment-cumulant formula, we have established the desired property that the joint moments are polynomials of the individual moments for free random variables. This drastic simplification is what freeness buys us. 

As discussed previously, from the individual moments of a collection of free self-adjoint (or normal cf. Footnote~\ref{ft:normal}) random variables, we can extract the spectral distribution of each random variable. Their freeness further allows us to extract the analytic distributions for some combinations of free random variables via tools that we will now come to.


\subsection{Free harmonic analysis}\label{sec:free}

We now introduce some tools from free harmonic analysis that will prove useful.  

\subsubsection{The Cauchy transform}
We start with the \emph{Cauchy-Stieltjes transform},\footnote{The Stieltjes transform usually refers to the Cauchy transform with a minus sign.} or simply the Cauchy transform, of a distribution $\mu$ on $\mathbb{R}$,
\begin{equation}\label{eq:cauchy}
    G_{\mu}(z):=\int_I \frac{\dd \mu(x)}{z-\lambda}, \quad z\in \mathbb{C}\setminus I\,,
\end{equation}where $I$ denotes the support of $\mu$ on the real line. 

The Cauchy transform is defined on both the upper and lower complex half-plane, denoted as $\mathbb{C}^+:=\{z:\mathrm{Im}(z)>0\}$ and $\mathbb{C}^-:=\{z:\mathrm{Im}(z)<0\}$ respectively.  However, note that $G_{\mu}(z)^*=G_{\mu}(z^*)$ so we only need to focus on the domain $\mathbb{C}^+$. In particular, it is a holomorphic function from $\mathbb{C}^+$ to $\mathbb{C}^-$. We can extract the spectral distribution $\mu(x)$ via the inverse transform, known as the \emph{Stieltjes inversion},
\begin{equation}\label{eq:reversecauchy}
    \mu(x)=-\frac{1}{\pi}\lim_{\epsilon\rightarrow 0}\mathrm{Im}\,G_\mu(x+i\epsilon)\,.
\end{equation}

Consider now a self-adjoint operator $a\in(\W,\tau)$. The Cauchy transform of its spectral distribution can be regarded as a moment generating function for $a$,
\begin{equation}\label{eq:cauchy2}
   G_{\nu_a}(z)=\sum^\infty_{n=0}\frac{m_n(a)}{z^{n+1}} \ .
\end{equation}
In the random matrix theory, the Cauchy transform for the spectral distribution of a random matrix $A\in (L^\infty(\Omega)\otimes M_N(\mathbb{C}),\E\otimes\ttr)$ is also known as the (trace of) resolvent of $A$.

\subsubsection{Free additive convolution and $R$-transform.}\label{sec:+convo}

Freeness allows us to break down the moments of any polynomial of several free random variables in terms of the cumulants/moments of each individual random variable. It's natural to ask about the spectral distribution $\nu_{a+b}$ of the sum of two free self-adjoint random variables $a$ and $b$. This distribution is known as the \emph{free additive convolution} of $\nu_a$ and $\nu_b$~\cite{voiculescu1986addition}, 
\begin{equation}\label{eq:convo+}
    \nu_{a+b}=\nu_a\boxplus\nu_b
\end{equation}
where the notation is chosen to indicate that this spectral distribution can be obtained from their spectral distributions $\nu_a$ and $\nu_b$ thanks to the freeness. The free additive convolution~$\boxplus$ is associative and commutative operation. Note this operation is defined here for any two compactly supported distributions on $\mathbb{R}$,\footnote{It can also be defined for distributions with non-compact support on $\mathbb{R}$~\cite{bercovici1993free}. } Therefore, it only depends on the spectral distributions $\nu_a$ and $\nu_b$ rather than the specific variables $a$ and $b$, because different random variables can share the same spectrum.

In order to extract the convoluted distribution, we look at its cumulants, $\kappa_n(a+b)$. Recall that free cumulants are multilinear maps and the mixed free cumulants for free random variables vanish, from which it immediately follows that, for any $n$,
\begin{equation}\label{eq:cumulantsum}
    \kappa_n(a+b) = \kappa_n(a) + \kappa_n(b) \ .
\end{equation}
Therefore, we obtain a simple relation between $a+b$ and the individual $a$ and $b$ in terms of the free cumulants. Now we need to make an analytic connection back to the moments via the \emph{$R$-transform}. For a deterministic spectral distribution $\nu_a$, its $R$-transform is defined as
\begin{equation}\label{eq:rtransform0}
    R_{\nu_a}(z) = \sum_{n=1}^\infty \kappa_n(a) z^{n-1} \ ,
\end{equation}
The $R$-transform can be viewed as a free cumulant generating function. Note that unlike the Cauchy transform that is analytic on the entire $\mathbb{C}^+$, the $R$-transform is usually only analytical for small $z$ and the domain depends on the measure $\nu_a$. See Chapter 3 in Mingo-Speicher~\cite{mingo2017free} for a detailed account of its analytic domain. 

Then \eqref{eq:cumulantsum} implies the convolution $\boxplus$ becomes additive under the $R$-transform,
\begin{equation}\label{eq:rtransconvo}
    R_{\nu_a\boxplus\nu_b}(z) = R_{\nu_a}(z)+R_{\nu_b}(z) \ .
\end{equation}

Recall that the Cauchy transform~\eqref{eq:cauchy} is like a moment generating function. Speicher showed that it is related to $R$-transform via~\cite{speicher1994multiplicative},
\begin{equation}\label{eq:rtransform}
    R_{\nu_a}(G_{\nu_a}(z))+G_{\nu_a}(z)^{-1}=z \ .
\end{equation}
Treating $G_{\nu_a}$ as a variable $z$, we can write it as
\begin{equation}
    R_{\nu_a}(z)+z^{-1}=G^{-1}_{\nu_a}(z)
\end{equation}
where $G^{-1}_{\nu_a}(z)$ is the inverse function of the Cauchy transform. It can be further re-written as
\begin{equation}
    G_{\nu_a}(R_{\nu_a}(z)+1/z)=z \ .
\end{equation}
Hence, the Cauchy transform is the functional inverse of $R_{\nu_a}(z)+1/z$.\footnote{This is sometimes used as an alternative definition for the $R$-transform.}  

The equation~\eqref{eq:rtransform} connects the $R$-transform and the Cauchy transform and thus the spectral distribution. Let's summarize the algorithm for free additive convolution: 
\begin{alg}\label{alg+}
Free additive convolution $\nu_a\boxplus\nu_b$ via the $R$-transform: \\
1. compute the Cauchy transform from the spectrum measures $\nu_a$ and $\nu_b$ using \eqref{eq:cauchy};\\
2. compute the $R$-transforms using \eqref{eq:rtransform} and apply \eqref{eq:rtransconvo} to obtain $R_{\nu_a\boxplus\nu_b}$;\\
3. find the Cauchy transform and then extract $\nu_a\boxplus\nu_b$ via the  Stieltjes inversion \eqref{eq:reversecauchy}.
\end{alg}

\subsubsection{Free multiplicative convolution and $S$-transform.}\label{sec:xconvo}

Similarly, one can also find the spectral distribution of the product  of two free positive self-adjoint random variables $a$ and $b$, $ab$ or $ba$. In particular, let's consider the positive operators $\sqrt{b}a\sqrt{b}$ or $\sqrt{a}b\sqrt{a}$ that share the same spectrum. It is given by the \emph{free multiplicative convolution}~\cite{voiculescu1987multiplication},
\begin{equation}
     \nu_{\sqrt{a}b\sqrt{a}}=\nu_{\sqrt{b}a\sqrt{b}} = \nu_a\boxtimes\nu_b
\end{equation}
where the notation is chosen to indicate that this spectral distribution can be obtained from their spectral distributions $\nu_a$ and $\nu_b$ thanks to the freeness. Just like free additive convolution, the free multiplicative convolution $\boxtimes$ is an associative and commutative operation. Note this operation is defined for any two compactly supported distributions on $\mathbb{R}_+$.\footnote{It can also be defined for distributions with non-compact support on $\mathbb{R}_+$~\cite{bercovici1993free} or distributions supported on the unit circle in $\mathbb{C}^+$~\cite{bercovici1992levy}. One can also relax the condition that one distribution can be supported on $\mathbb{R}$ provided the other distribution is supported $\mathbb{R}_+$~\cite{rao2007multiplication}.} Therefore, it only depends on the spectral distributions $\nu_a$ and $\nu_b$ rather than the specific variables $a$ and $b$. The convoluted distribution has bounded support if $\nu_a$ and $\nu_b$ have bounded support.

Let's see how this works with the help of the \emph{$S$-transform}. Analogous to the $R$-transform, which can help us calculate the free additive convolution, the free convolution can be computed by the $S$-transform, where the convolution turns into a multiplication.  We first define another moment function $\omega$ just as an intermediate tool,
\begin{equation}\label{eq:momentf}
    \omega_{\mu}(z):=z G_{\mu}(z)-1=\sum_{n=1}^\infty\frac{m_n(\mu)}{z^n}\,.
\end{equation}
The $S$-transform $S_{\mu}(z)$ is then defined as
\begin{equation}\label{eq:stransform}
     S_{\mu}(z) := \frac{1+z}{z\omega_{\mu}^{-1}(z)}
\end{equation}
where $\omega_{\mu}^{-1}(z)$ is the inverse function of \eqref{eq:momentf}. Generally, the $S$-transform is only analytic in the neighbourhood of $0$.  See for instance \cite{bercovici1993free} for more details its analytic domain. We shall also note that the $S$-transform and $R$-transform are related via the composition inverse of the functions 
\begin{equation}\label{eq:composition_inverse}
    x=zR_{\mu}(z) ,\quad  z=xS_{\mu}(x) \ .
\end{equation}
Using this, one can transfer between the $R$-transform and the $S$-transform, and it shall prove very useful later.

The $S$-transform allows us to compute the free multiplicative convolution with
\begin{equation}\label{eq:xconvolution}
    S_{\nu_a\boxtimes\nu_b}(\omega) =S_{\nu_a}(\omega)S_{\nu_b}(\omega)\,.
\end{equation}

We then invert \eqref{eq:stransform} to obtain $\omega_{\nu_a\boxtimes\nu_b}(z)$, thus $G_{\nu_a\boxtimes\nu_b}(z)=(\omega_{\nu_a\boxtimes\nu_b}(z)+1)/z$ and finally the spectral distribution $\nu_a\boxtimes\nu_b$ via the Stieltjes inversion~\eqref{eq:reversecauchy}. Let's summarize the algorithm for the free multiplicative convolution,
\begin{alg}\label{algx}
Free multiplicative convolution $\nu_a\boxtimes\nu_b$ via the $S$-transform: \\
1. compute the moment functions  \eqref{eq:momentf} $\omega_{\nu_a}$ and $\omega_{\nu_b}$ from the Cauchy transforms~\eqref{eq:cauchy};\\
2. compute the $S$-transforms using \eqref{eq:stransform} and apply \eqref{eq:xconvolution} to obtain $S_{\nu_a\boxtimes\nu_b}$;\\
3. find the Cauchy transform and then extract $\nu_a\boxtimes\nu_b$ via the Stieltjes inversion \eqref{eq:reversecauchy}.
\end{alg}

Above, we took Voiculescu's approach to introduce the free multiplicative convolution in terms of the $S$-transforms~\cite{voiculescu1987multiplication}. The following result due to Nica-Speicher provides a complementary combinatorial perspective on the free multiplicative convolution~\cite{nica2006lectures}. Let $(\W,\tau)$ be a non-commutative probability space and $a, b\in (\W,\tau)$ such that they are free. We have the moments of the product $ab$,
\begin{equation}\label{eq:freemulti1}
    m_n\left(ab\right)=\sum_{\pi\in \mathrm{NC}_n}\left(\prod_{V\in\pi}\kappa_{|V|}(a)\cdot \prod_{\bar V\in\bar\pi} m_{|\bar V|}\left(b\right)\right)\ ,
\end{equation}
and free cumulants
\begin{equation}\label{eq:freemulti2}
    \kappa_n(ab)=\sum_{\pi\in \mathrm{NC}_n}\left(\prod_{V\in\pi}\kappa_{|V|}(a)\cdot \prod_{\bar V\in\bar\pi} \kappa_{|\bar V|}(b)\right)\ .
\end{equation}
Notice that these formulas look exactly like what the replica trick GPI gives us~\eqref{eq:replicatrick}. This is the key observation of this work and we shall match the details later in Section~\ref{sec:convolution}.\\

In practice, the free multiplicative convolution and the free additive convolution via the $S, R$-transforms can be hard to evaluate either analytically or even numerically. The difficulty lies in dealing with the functional inverses when converting the $S,R$-transforms back to the Cauchy transforms.\footnote{There are some efforts from the machine learning community in developing algorithms to directly handle inverting the $S$-transforms~\cite{pennington2018emergence,reda2021free}.} In fact, one can circumvent from dealing with the $S, R$-transforms directly and instead resort to the subordinate functions~\cite{belinschi2007new}. They have better analytic properties and boil down to fixed-point equations that are numerically tractable. (cf. Chapter 5 in~\cite{speicher2019lecture} for an introduction). 

There is a general procedure developed to evaluate any polynomials of a collection of free random variables. (cf. Chapter 10 in Mingo-Speicher~\cite{mingo2017free} and \cite{mai2017analytic} for details.) The procedure builds on the operator-valued free probability theory~\cite{speicher1998combinatorial}, that generalizes the scalar-valued case that we've been using. It then uses a linearization trick to formulate any polynomial as a linear combination of free variables with matrix coefficients, turning the problem into an addition of operator-valued free random variables~\cite{haagerup2005new,helton2018applications}. Then, instead of the $R$-transforms, one can implement the subordination formalism to compute the sum~\cite{belinschi2017analytic}. Later it turns out that we do not have to implement this general machinery for our problem. 

\subsubsection{Free compression}
Consider a projection $p\in (\W,\tau)$ of trace $\tau(p)<1$ acting as $pap$ on $a\in (\W,\tau)$. It has the spectral distribution $\tau(p)\delta_1 + (1-\tau(p))\delta_0$. These projected random variables form a subalgebra $(p\W p,\tau_p)\subset (\W,\tau)$ where $\tau_p:=\tau/\tau(p)$ is the renormalized trace and $p$ is the unit element of $p\W p$. This projected subalgebra forms a non-commutative probability space and is referred as a \emph{compression} of $(\W,\tau)$. The compression inherits the structural properties of~$(\W,\tau)$. For instance, if $(\W,\tau)$ is a $W^*$-probability space, so is $(p\W p,\tau_p)$. In particular, if $(\W,\tau)$ is a Type II$_1$ factor, so is $(p\W p,\tau_p)$.

Consider now that $p$ is free from some $a\in (\W,\tau)$. We call $pap$ a \emph{free compression} of $a$ by $p$. We are interested in applying the free multiplicative convolution to compute the spectral distribution, $\nu_{pap/\tau(p)}$, of $pap/\tau(p)\in(\W,\tau)$, where it's customary to normalize the compressed variable by $\tau(p)$. 
 One can show that
\begin{equation}\label{eq:freecompression1}
  \nu^p_{pap/\tau(p)} = \nu_a\boxtimes (\tau(p)\delta_{1/\tau(p)} + (1-\tau(p))\delta_0) = \tau(p)\nu_a^{\boxplus 1/\tau(p)}+ (1-\tau(p))\delta_0
\end{equation}
where the  $\boxplus$ power denotes a $1/\tau(p)$-fold free additive convolution, provided $1/\tau(p)$ is an integer. If we restrict to the subalgebra $(p\W p,\tau_p)$ that the random variable is compressed to, we switch to the renormalized trace $\tau_p$ and the corresponding spectral distribution, denoted as $\nu^p_{pap/\tau(p)}$, is simply
\begin{equation}\label{eq:freecompression2}
    \nu^p_{pap/\tau(p)} = \nu_a^{\boxplus 1/\tau(p)}.
\end{equation}
Surprisingly, we see that a free compression $p$ essentially implements a $1/\tau(p)$-fold free additive convolution up to a rescaling or $\tau(p)$. In particular, we have the following relation for the cumulants defined w.r.t. $\tau_p$,
\begin{equation}
    \tau(p)\kappa_n(pap/\tau(p)) = \kappa_n(a),\quad \forall n\in\mathbb{N}\ ,
\end{equation}
and thus the $R$-transforms satisfy
\begin{equation}
     \tau(p)R_{\nu^p_{pap/\tau(p)}} = R_{\nu_a}\ .
\end{equation}

It's also useful to write \eqref{eq:freecompression2} as
\begin{equation}\label{eq:freecompression3}
    \nu^p_{pap} = \nu_a^{\boxplus 1/\tau(p)}\circ D_{\tau(p)}.
\end{equation}
where $D_{\tau(p)}$ denotes a dilation (rescaling) of the distribution. This operation is defined to rescale any distribution $\nu(x)$ as
\begin{equation}\label{eq:rescaling}
    \nu\circ D_\la(x) := \la\nu(\la x)\ .
\end{equation}
Equation \eqref{eq:freecompression3} follows from the general fact that if one rescales a self-adjoint operator by $a$, then its spectral distribution is dilated by $1/\la$,
\begin{equation}
    \nu_{\la a} = \nu_a\circ D_{1/\la}\ .
\end{equation}

In fact, we can use \eqref{eq:freecompression2} to extend the definition of the free additive convolution power to any real number $t\ge 1$ with $\tau(p)=1/t$.\footnote{This is possible, for instance, in a Type $\mathrm{II}_1$ factor which has no minimal projection. }  Moreover, one can use the free compression to generate, from a given compactly supported $\nu$, a $\boxplus$-semigroup\footnote{See~\cite{belinschi2005partially} for a free multiplicative counterpart.} of compactly supported probability measures~\cite{nica1996multiplication}. It is denoted as, $(\nu_t)_{t\ge 1}$, and it satisfies
\begin{equation}\label{eq:semigroup}
    \nu_{t+s} = \nu_t\boxplus\nu_s\ , \quad \nu_1=\nu\ .
\end{equation}

Here is how the $\boxplus$-semigroup can be constructed. Starting from a given $\nu$, we assign a self-adjoint random variable $a$ with this spectral distribution. We also introduce a projection of trace $1/t$ and put $a$ and $p$ in a common probability space $(\W,\tau)$ under the free product such that $p$ and $a$ are free. Then we consider $pap\in (p\W p,\tau_p)$ and define $\nu_t:=\nu^p_{tpap}=\nu^{\boxplus t}$. The semigroup property~\eqref{eq:semigroup} is manifestly satisfied. We will later see an important family of distributions for which the semigroup extends also to $0<t<1$.

\subsection{Circular, semi-circular, quarter-circular and free Poisson distributions}\label{sec:circulars}

In this subsection, we introduce some canonical examples of probability distributions in free probability theory that are relevant for our discussions later. These are closely related to some canonical random matrix ensembles and we shall mention some in passing. The details on random matrices are provided in Appendix \ref{app:random_matrices}.

One of the most important and random matrix ensemble is the Gaussian Unitary Ensemble (GUE) of Hermitian matrices. An $N\times N$ GUE matrix $S_N$ have i.i.d. complex Gaussian matrix elements with the real and complex parts independently distributed with zero mean and variance $1/2N$, and the diagonal are i.i.d. Gaussian-distributed with zero mean and variance $1/N$. The spectral distributions of a sequence of GUE $(S_N)_N$ converges almost surely to the semi-circular distribution,
\begin{equation}\label{eq:semicircle}
    \mu_{\mathrm{s}}(x):=(2\pi)^{-1}\sqrt{(4-x^2)}\
\end{equation}
with support $x\in[-2,+2]$. We call any element $s\in (\W,\tau)$ a \emph{semi-circular element} if its spectral distribution $\mu_s$ is given by the semi-circular distribution. It moments are given by the Catalan numbers,
\begin{equation}
   m_n(s)=\lim_{N\to\infty} \E\ \ttr(S^n_N) = \frac{1}{n+1}\binom{2n}{n}\ \ \text{for even}\ n,  \ \ \text{and zero otherwise}\ .
\end{equation}
Its free cumulants are given by
\begin{equation}
    \kappa_n(s) = \delta_{n,2}\ .
\end{equation}

The semi-circular distribution is thus most easily characterized by that its only non-zero cumulant is $\kappa_2$. This fact directly leads to the \emph{free central limit theorem}. The classical theorem says that for a collection of $n$ independent random variables $x_1,\ldots,x_n$ with zero mean and unit variance. Their average $(\sum_{i=1}^n x_i)/\sqrt{n}$ converges (in distribution) to a standard normal random variable. The free probability counterpart is the free central limit theorem, which asserts that for a collection of free self-adjoint random variables $\{a_i\}_{i=1}^n$ in $(\W,\tau)$ with zero mean and unit variance ($\kappa_2(a_i)=1$), their average $(\sum_{i=1}^n a_i)/\sqrt{n}$ converges (in distribution) to a \emph{semicircular element} $s\in (\W,\tau)$.\footnote{We can sketch here the proof for this theorem. The spectral distribution of the addition of a collection of free random variables is described by the free additive convolution, under which the free cumulants add. Therefore, all the free cumulants have linear growth in $n$ whereas the denominator $\sqrt{n}$ suppresses $\sum_i^n\kappa_m(a_i)$ by $n^{m/2}$, so only the first ($m=1$) and second cumulant ($m=2$) survive in the limit $n\to\infty$. Since the first cumulant equals to the first moment which vanishes for all $a_i$, we have only a non-zero $\kappa_2=1$. Hence, the limiting distribution is semicircular and the free central limit theorem follows.}\\

A \emph{circular element} $c\in (\W,\tau)$ is defined to be
\begin{equation}
    c:=\frac{s_1+i s_2}{\sqrt{2}}
\end{equation}
for two free semi-circulars $s_1, s_2$. Its spectral distribution is the uniform distribution over the unit disk on the complex plane, called the \emph{circular distribution},
\begin{equation}\label{eq:circle}
    \mu_{\mathrm{c}}(z):=\pi^{-1}\id_{|z|\le 1}\ .
\end{equation}

They can be modelled by large random matrix ensemble $C_N$ of form 
\begin{equation}
    C_N=S_N^{(1)} +i S_N^{(2)}
\end{equation}
for two $N\times N$ independent GUE matrices $S_N^{(1)}$ and $S_N^{(2)}$.  $C_N$ is called the (square) \emph{Ginibre ensemble} and they have i.i.d. matrix elements that follow the complex Gaussian distribution of zero mean and variance $1/N$. Its limiting spectral distribution is given by the circular distribution. The Ginibre ensemble has an important property that its joint distribution of the matrix elements are invariant under any unitary action from either left or right.  We say it is both \emph{left and right unitarily invariant.}\footnote{See, for instance, Lemma 1 in~\cite{mezzadri2007generate}.} \\

Now consider the modulus of a circular element, known as a \emph{quarter-circular element}
\begin{equation}
    q:=|c|:=\sqrt{cc^*}\ .
\end{equation}
Its spectral distribution follows the \emph{quarter-circular distribution},
\begin{equation}\label{eq:quartercircle}
    \mu_{\mathrm{q}}(x):=\pi^{-1}\sqrt{(4-x^2)}
\end{equation}
with support $x\in[0,2]$. The spectral distribution is also shared by the modulus of a semi-circular element $|s|$. Its moments are given by $m_n(q) = \frac{2^n\Gamma(1/2+n/2)}{\sqrt{\pi}\Gamma(2+n/2)}$. It's more interesting to look at the moments of its square, which are given by the Catalan numbers,
\begin{equation}
   m_n(q^2)= \frac{1}{n+1}\binom{2n}{n}\ ,\quad\forall n
\end{equation}
and the cumulants are given by
\begin{equation}
    \kappa_n(q^2) = 1\ ,\quad\forall n\ .
\end{equation} 

A useful lemma due to Voiculescu claims that a circular element admits the polar decomposition~\cite{voiculescu1991limit},
\beq\label{eq:polardecomp}
c=uq
\eeq
in terms of a Haar unitary $u$ and a quarter-circular element $q$ that are free from each other. An element $u\in(\W,\tau)$ is called a Haar unitary if $u^*u=uu^*=1$ and $\tau(u^n)=\delta_{0,n}, \forall n\in\mathbb{N}$. Its random matrix counterpart is of course the Haar unitary ensemble, also known as the circular unitary ensemble. \\



The quarter-circular distribution \eqref{eq:quartercircle} is a special instance of the \emph{free Poisson distribution}. Let us motivate it with the \emph{free Poisson limit theorem}. It is the non-commutative analog to the Poisson limit theorem that asserts the sum of Bernoulli random variables (coin flips) asymptotes to a Poisson random variable. The free Poisson distribution can be obtained from multiple free additive convolutions of the Bernoulli distribution, which has two outcomes $0$ and $\alpha$. Then $\alpha$ is the jump size and $\lambda$ is the rate of obtaining the outcome $\alpha$. We have\footnote{The classical Poisson distribution is obtained by taking the classical $*$-convolution power. }
\begin{equation}\label{eq:poissonlimit}
    \mu_{\lambda,\alpha} =\lim_{N\to\infty} \left[(1-\lambda/N)\delta_0+(\lambda/N)\delta_\alpha\right]^{\boxplus N}\ ,
\end{equation}
which is known as the \emph{free Poisson distribution} and it admits a close-form expression,
\begin{equation}\label{eq:poisson}
    \mu_{\lambda,\alpha}:=
    \begin{cases}
      (1-\lambda)\delta_0 + \tilde\mu , &\mathrm{if}\quad 0\le\lambda\le 1 \\ 
      \tilde\mu , &\mathrm{if}\quad \lambda>1
    \end{cases}
\end{equation}
where $\tilde\mu(x):=\frac{1}{2\pi\alpha x}\sqrt{(x_+ - x)(x - x_-) }\mathbf{1}_{\left[x_-, x_+\right]} \ , \quad x_\pm:=\alpha(1\pm\sqrt{\lambda})^2$. \\

Its moments are given by
\begin{equation}\label{eq:poisson_mom}
    m_n(\mu_{\lambda,\alpha}) = \alpha^n\sum_{k=1}^n \lambda^{n-k} N(n,k)
\end{equation}
where $N(n,k):=\binom{n}{k}\binom{n}{k-1}/n$ are the Narayana numbers. When $\lambda=\alpha=1$, the moments reduce to the Catalan numbers of the quarter-circular distribution. The free Poisson distribution is most concisely characterized by its cumulants,
\begin{equation}\label{eq:poisson_cumu}
    \kappa_n(\mu_{\lambda,\alpha}) = \lambda\alpha^n\ .
\end{equation}
The free Poisson distribution is also known as the Marchenko-Pastur (MP) distribution under a change of variables. The MP distribution describes the limiting spectrum of a Wishart ensemble which has the form $X_NX_N^*$ where $X_N$ is a (rectangular) $N\times M$ Ginibre ensemble with i.i.d. matrix entries of zero mean and variance $\alpha/M$ and $M/N\to\la$. \\

Consider a generalization of the free Poisson limit theorem such that the jump size is not a deterministic value $\alpha$, but rather follows some probability distribution $\nu$,
\begin{equation}\label{eq:compoundpoissonlimit}
    \mu_{\lambda,\nu} =\lim_{N\to\infty} \left[(1-\lambda/N)\delta_0+(\lambda/N)\nu \right]^{\boxplus N}\ .
\end{equation}
This is known as the \emph{free compound Poisson distribution}. They are most easily characterized by the cumulants. 
\begin{equation}\label{eq:compound_cumulants}
    \kappa_n(\mu_{\lambda,\nu}) = \lambda m_n(\nu)\ .
\end{equation}
A special case of a compound Poisson distribution is when $\lambda=1$. Then the cumulants of $\mu_{\lambda,\nu}$ are exactly given by the moments of $\nu$. In this case,  suppose $\nu$ is the spectral distribution of some random variable $a$, then $\mu_{\lambda,\nu}$ is the spectral distribution of $cac^*$. Namely, the conjugation by circular elements swaps moments to cumulants. To see how, consider the general moment formula of the free multiplicative convolution between $a$ and $cc^*$, which is the square of a quarter-circular element $q^2$. Consider the general moment formula of the free multiplicative convolution \eqref{eq:freemulti1},
\begin{equation}
\begin{aligned}
   m_n(cac^*)&=\sum_{\pi\in \nc_n}\left(\prod_{V\in\pi}\kappa_{|V|}(q^2)\cdot \prod_{\bar V\in\bar\pi} m_{|\bar V|}(a)\right)=\sum_{\pi\in \nc_n}\prod_{\bar V\in\bar\pi} m_{|\bar V|}(a)=\sum_{\pi\in \nc_n}\prod_{V\in\pi} m_{|V|}(a)\ ,
\end{aligned}
\end{equation}
which implies \eqref{eq:compound_cumulants} via the definition of free cumulants~\eqref{eq:m-c}.\\

The natural random matrix sequence with the limiting spectral distribution of a free compound Poisson distribution is the sample covariance matrix ensemble,\footnote{\label{ft:sample_cov} In statistics, $A_N$ represents the covariance matrix of a centered random vector $\mathbf{y}$ with $N$ data entries, $A_N=\E \mathbf{yy}^*$, and we can write the random vector as $\mathbf{y}=\sqrt{A_N}\mathbf{x}$ where $x$ is some vector with centered i.i.d. entries with unit variance. A sample covariance matrix $M^{-1}\sum_i^{M} \mathbf{y}_i\mathbf{y}_i^*$ is an estimator for $A_N$ with $M$ i.i.d. draws of $\mathbf{y}$. Alternatively, we can write the sample covariance matrix as \eqref{eq:sample_cov}.}
\begin{equation}\label{eq:sample_cov}
  \sqrt{A_N}X_NX_N^*\sqrt{A_N}\ ,
\end{equation}
where $X_N$ is an $N\times M$ Ginibre ensemble with i.i.d. matrix entries of zero mean and variance $1/M$. $A_N$ is a sequence of independent positive semidefinite matrices $A_N$ with the the limiting spectral distribution $\nu$ . \\

A compound Poisson distribution is a canonical example of \emph{free infinitely divisible distributions}, which are probability distributions that can be decomposed as\footnote{A similar notion of infinite divisibility also exists for the free multiplicative convolution.}
\begin{equation}
    \mu = \mu_{1/k}^{\boxplus k}
\end{equation}
for any $k\in \mathbb{N}$. We can then make use of $\mu_{1/k}$ to define any fractional convolution power $\mu^{\boxplus l/k}$ as $\mu_{1/k}^{\boxplus l}$. By continuity,  we obtain a one-parameter family of spectral distributions, $(\mu_{t})_{t> 0}$, that actually forms a $\boxplus$-semigroup under the associative action of the free multiplicative convolution $\boxplus$, $\mu_{s+t}=\mu_{s}\boxplus\mu_{t}$. 

Conversely, given a such a semigroup $(\mu_{t})_{t> 0}$, any element $\mu_t$ is infinitely divisible. Hence, free infinite divisibility is equivalent to the existence of a $\boxplus$-semigroup for $t\ge 0$. Therefore, we see that while every compactly supported probability measure belongs to a $\boxplus$-semigroup with $t\ge 1$, the ones with the extended parameter range $t>0$ are very special. See Chapter~3 in Hiai-Petz~\cite{hiai2006semicircle} for a good discussion of free infinite divisibility.

It is then evident that the $R$-transforms of $\mu_t$ are simply multiples of the $R$-transform of $\mu_1=\mu$,
\begin{equation}
    R_{\mu_t}=tR_{\mu}\ , \ t\ge 0 \ .
\end{equation}
It turns out that one can be more explicit about the $R$-transform of an infinitely divisible $\mu$. It admits the following form, known as the free L\'evy-Khintchine formula,\footnote{This is the free analog of the classical L\'evy-Khintchine formula that characterizes the characteristic functions of infinitely divisible distributions.}
\begin{equation}\label{eq:Levy-Khintchine}
    R_{\mu}(z) = \kappa_1(\mu) + \int_{\mathbb{R}} \frac{z}{1-xz} \eta(x)\dd x,\quad\forall z\in \mathbb{C}\setminus\mathbb{R}\cup\{0\}\ ,
\end{equation}
for some positive finite measure $\eta$ on $\mathbb{R}$. Rather than being analytic only for small $|z|$, such $R_{\mu}(z)$ is analytic on both $\mathbb{C}^\pm$, and it maps $\mathbb{C}^\pm\to\mathbb{C}^\pm$.

In particular, any compound Poisson distribution $\mu_{\lambda,\nu}$ can be decomposed as 
\begin{equation}
    \mu_{\lambda,\nu} = \mu_{\lambda/k,\nu}^{\boxplus k}
\end{equation}
and the semigroup of distributions is defined as $\mu_t:=\mu_{t\lambda,\nu}$ . Using the cumulants \eqref{eq:compound_cumulants}. Its $R$-transform reads
\begin{equation}\label{eq:poisson_rtransform}
    R_{\mu_{\la,\nu}}(z) = \lambda x\sum_{n=0}^\infty \int x^nz^n\nu(x)\dd x = \int\frac{\lambda x}{1-xz}\nu(x)\dd x\ .
\end{equation}
which admits the form of the L\'evy-Khintchine formula \eqref{eq:Levy-Khintchine} with $\kappa_1=\lambda m_1(\nu)$ and $\eta(x) =x^2\lambda\nu(x)$.

In sum, thanks to the free infinite divisibility, the  $R$-transforms of the free compound Poisson distributions are relatively simple to work with. Unlike the free Poisson distribution, a free compound Poisson distribution doesn't generally have a closed-form expression, but a closed-form expression for the $R$-transform is already good enough for many purposes.

\section{Results}\label{sec:results}
We now come to the main observation that the replica trick GPI should be understood as a free multiplicative convolution of two probability distributions, one of which is determined by the reservoir Hamiltonian or any quantum processing we apply to the Hawking radiation collected, and the other is determined by the JT gravity+EOW brane theory. We need to sort out some technical issues in order make them match. Then we use the tools from free harmonic analysis to evaluate the convolution. 

We shall also infer from the convolution formula a random matrix model that allows us to compute any spectral functions of the radiation density matrix. Surprisingly, it fits perfectly with Page's original proposal addressing how an evaporating black hole should be treated information-theoretically. To support the validity of the free convolution formula, we perform some consistency checks with some known results in the literature. Finally, we show how to re-formulate the entropy formula such that it is free of free probability, but resembles Wall's formula for the generalized entropy.
 
\subsection{Replica trick as a free multiplicative convolution}\label{sec:convolution}
We would like to match \eqref{eq:replicatrick} with \eqref{eq:freemulti1}. The main idea is that the collection of gravitational partition functions $\{Z_n\}$, obtained from the  connected replica wormholes, should be viewed as free cumulants rather than moments.

These $Z_n$ partition functions are defined by the boundary condition that consists of alternative JT asymptotic boundaries and the EOW branes. They are computed by PSSY based on an earlier calculation by Yang~\cite{yang2019quantum}. The results read
\begin{multline}\label{eq:gravZn}
    Z_n=\int_0^\infty\dd E\,\rho(E)y(E)^n\ ,\\
    y(E):=e^{-\beta E}2^{1-2\mu}|\Gamma(\mu-1/2+i\sqrt{2E})|^2, \rho(E):=e^{S_0}\sinh(2\pi\sqrt{2E})/(2\pi^2)
\end{multline}
where $y(E)$ is a Boltzmann-type factor that is influenced by the EOW brane, and $\rho(E)$ is the density of states. After normalization with $Z_1^n$. We can rewrite it as the $n^\mathrm{th}$ moment of some distribution, 
\begin{equation}
    \frac{Z_n}{Z_1^n} = \int_0^{y(0)} \frac{\rho(\tilde E(y))}{-y'(\tilde E(y))} (y/Z_1)^n \dd y  =\int_0^{x(0)} \frac{\rho(E(x))}{-y'(E(x))} x^n \dd x =: \int_0^{x(0)} \nu_b(x) x^n \dd x 
\end{equation}
where $\tilde E(y)$ is the inverse function of $y(E)$, $x:=y/Z_1$, $E(x)=\tilde E(xZ_1)$ in the second equality and lastly we've defined a distribution function
\begin{equation}\label{eq:nub}
    \nu_b(x):=\frac{\rho(E(x))}{-y'(E(x))}\ .
\end{equation}
Note that $y'<0$ as $y(E)$ is a decreasing function in $E$. Note that $\rho(E)$ is not integrable so is $\nu_b$ whose zeroth moment diverges at $x\to 0$, so $\nu_b$ is \emph{not} a probability distribution, but only a positive measure over $[0,x(0)]$. 

The Bekenstein-Hawking entropy of the JT black hole + EOW brane system is given by
\begin{equation}\label{eq:bhentropy}
  \sbh:=-\partial_N(Z_n/Z_1^n)|_{n=1}=-\int_{0}^{x(0)} x\log x\, \nu_b(x)\,\dd x\ .
\end{equation}
In other words, the BH entropy is the entropy of the measure~$\nu_b$. In particular, $\mu\gg 1/\beta$, we have
\begin{equation}
    \sbh = S_0+4\pi^2/\beta+\O(1)\ .
\end{equation}


It's more handy to work with a probability distribution, so we let $\nu_{b_N}$ be an \emph{regulated} probability distribution  support on $[x(E_N),x(0)]$,
\begin{multline}\label{eq:truncated_distribution}
    \nu_{b_N}(x)  := \frac{\rho_N(E(x))}{-y'(E(x))N}\circ D_{1/N}=\frac{\rho_N(E(x/N))}{-y'(E(x/N))N^2} , 0<\quad x(E_N)\le x\le x(0), \\
   \mathrm{where}\quad \rho_N(E) := \frac{2^{S_0}}{2\pi^2}\sinh\left(2\pi\sqrt{\frac{E(E_N-E)}{E_N}}\right), \quad 0\le E\le E_N
\end{multline}
and where $D_{1/N}$ is the dilation operation defined in~\eqref{eq:rescaling},  $E_N$ is determined via the constraint $\int^{E_N}_0\dd E\rho_N(E)=N$. As $N\to\infty$,\footnote{This is also known as the double scaling limit in JT gravity~\cite{saad2019jt}.} we have the \emph{weak convergence} of the distributions,\footnote{We say a sequence of distributions $\mu_N$ on $\mathbb{R}$ converges weakly (or in weak topology) to some limiting distribution $\mu$, if for any bounded continuous function $f$, we have $\lim_{N\to\infty}\int f(x)\mu_N(x)\dd x=\int f(x)\mu(x)\dd x$.}
\begin{equation}\label{eq:limits_nu}
    \rho_N\to\rho,\quad N\nu_{b_N}\circ D_{N}\to \nu_b\ .
\end{equation}

The partition function $Z_n$ is evaluated in the large $N$ limit, which poses some difficulties for us to treat it probablistically. We thus work with the regulated probability distribution $\nu_{b_N}$ and take the large $N$ limit in the end. We think of this procedure as \emph{undoing} the large $N$ limit. Though we try to be concrete in \eqref{eq:truncated_distribution}, we shall see that the details of the regularization is not relevant for us, and any other regularization scheme can do equally well. The only requirement we demand is that the moments should converge to the moments of $\nu_b$ large $N$ limit. That is, for
\begin{equation}\label{eq:moments_of_b}
   m_n(\nu_{b_N}) := \int_{x(E_N)}^{x(0)} \nu_{b_N}(x) x^n \dd x\ ,
\end{equation}
we have
\begin{equation}
  (Z_n/Z_1^n =)\ m_n(\nu_b) = \lim_{N\to\infty}m_n(N\nu_{b_N}\circ D_{N}) = \lim_{N\to\infty}N^{1-n}m_n(\nu_{b_N}) \ .
\end{equation}
which follows from \eqref{eq:limits_nu} for our regulated $\nu_{b_N}$.

Recall now the replica trick GPI~\eqref{eq:replicatrick},
\begin{equation}\label{eq:pssypartitionfunction}
\begin{aligned}
    \tilde Z_n&=\sum_{\pi\in \mathrm{NC}_n}\left(\prod_{V\in\pi}Z_{|V|}\cdot\prod_{\bar V\in\bar\pi}k^{|\bar V|}\tr R^{|\bar V|}\right)\\
    &=k\sum_{\pi\in \mathrm{NC}_n}\left(\prod_{V\in\pi}k^{(|V|-1)}m_{|V|}(\nu_b)Z_1^{|V|}\cdot\prod_{\bar V\in\bar\pi}\frac1k\tr\left(kR\right)^{|\bar V|}\right)\ .
\end{aligned}
\end{equation}

The moments $\frac1k\tr\left(kR\right)^n$ of the given bulk radiation density matrix $R$ define a probability distribution $\nu_r$. Denote the spectral values of $R$ as $\{\la_i\}_{i=1}^k$, we have
\begin{equation}
    \nu_r(x):= \frac{1}{k}\sum_{i=1}^k \delta(x-k\la_i)\ .
\end{equation}

Let's use the moments $\tilde Z_n/kZ_1^n$ to define a probability distribution $\nu_{\tilde r}$, which we will soon confirm that it's indeed a valid probability distribution. The normalization $kZ_1^n$ is to impose the normalization that $m_0(\nu_{\tilde r})=m_1(\nu_{\tilde r})=1$ as we want for the radiation density operator $\tilde r$, which we shall discuss later. We therefore have the moments of $\nu_{\tilde r}$,
\begin{equation}\label{eq:convolution_formula1}
    m_n(\nu_{\tilde r}):=\frac{\tilde Z_n}{kZ_1^n}=k^n\sum_{\pi\in \nc_n}\left(\prod_{V\in\pi}\frac1k m_{|V|}(\nu_b)\cdot \prod_{\bar V\in\bar\pi} m_{|\bar V|}(\nu_r)\right)\ .
\end{equation}
We can also define the moments of the corresponding distribution $\nu_{\tilde r_N}$ at finite $N$,
\begin{equation}\label{eq:convolution_formula1_l}
    m_n(\nu_{\tilde r_N}):=\frac{k^n}{N^n}\sum_{\pi\in \nc_n}\left(\prod_{V\in\pi}\frac{N}{k}m_{|V|}(\nu_{b_N})\cdot \prod_{\bar V\in\bar\pi} m_{|\bar V|}(\nu_r)\right)\ ,
\end{equation}
and they satisfy
\begin{equation}\label{eq:r_moment_convergence}
    m_n(\nu_{\tilde r}) = \lim_{N\to\infty}m_n(\nu_{\tilde r_N})\ .
\end{equation}

Comparing the free multiplicative convolution formula~\eqref{eq:freemulti1} with \eqref{eq:convolution_formula1_l}, we see that $\nu_{\tilde r_N}$ is indeed a valid probability distribution 
\begin{equation}\label{eq:distribution_canonical1}
   \nu_{\tilde r_N} = \left(\nu_r \boxtimes\mu_{\frac{N}{k},\nu_{b_N}}\right)\circ D_{\frac{N}{k}} =\nu_r \boxtimes\left(\mu_{\frac{N}{k},\nu_{b_N}}\circ D_{\frac{N}{k}}\right) \ 
\end{equation}
where the compound Poisson distribution $\mu_{\frac{N}{k},\nu_{b_N}}$ follows from the free cumulants $(N/k)m_n(\nu_{b_N})$ in \eqref{eq:convolution_formula1_l}, and the factor $(k/N)^n$ in front gives rise to a dilation $D_{N/k}$. In the second equality we put the dilation inside the second factor of the free convolution.\footnote{This is legit because $(\nu_a\boxtimes\nu_b)\circ D_\la$ is the spectral distribution of $\la\sqrt{a}b\sqrt{a}$ for some free random variables $a,b$ and a scalar $\la$. Clearly, we can associate the scalar multiple to $b$ (or $a$), yielding $\nu_a\boxtimes(\nu_b\circ D_\la) = (\nu_a\boxtimes\nu_b)\circ D_\la$.}

Note that the factors $\prod_{V\in\pi}(N/k)m_{|V|}(\nu_{b_N})$ in \eqref{eq:convolution_formula1_l} are the moments of the distribution $(N/k)\nu_{b_N}$ which integrates to $N/k$. Since $(N/k)\nu_{b_N}$ has bounded support, it is the unique distribution with moments $(N/k)m_n(\nu_{b_N})$. It is therefore not a probability distribution either. Nevertheless, the main point here is that they constitute a legit set of cumulants of a free compound Poisson distribution. 

In the large $N$ limit, $N\to\infty$, the cumulants $(k/N)^{n-1}m_n(\nu_{b_N})$ converge
\begin{equation}
    (k/N)^{n-1}m_n(\nu_{b_N})\to k^{n-1}m_n(\nu_b)\ ,
\end{equation}
so the limiting distribution of \eqref{eq:distribution_canonical1} is well-defined. We have
\begin{equation}\label{eq:distribution_canonical0}
    \nu_{\tilde r}:=\lim_{N\to\infty}\nu_{\tilde r_N} = \nu_r \boxtimes\left( \lim_{N\to\infty}\mu_{\frac{N}{k},\nu_{b_N}}\circ D_{\frac{N}{k}}\right)\ .
\end{equation}
We can further denote the limiting distribution in the bracket as 
\begin{equation}
    \mu_{\nu_b}:= \lim_{N\to\infty}\mu_{\frac{N}{k},\nu_{b_N}}\circ D_{\frac{N}{k}}
\end{equation}
and we obtain the result claimed in our introduction~\eqref{eq:mainresult}, that the physical radiation spectrum is given by the convolution of the spectral data of the gravitational sector and the matter sector respectively.
\beq
\nu_{\tilde r}=\nu_r\boxtimes \mu_{\nu_b}\ .
\eeq
We should emphasize that the distribution $\mu_{\nu_b}$ and thus the end result $\nu_{\tilde r}$ are independent of the way we regularize $\nu_b$ in $\nu_{b_N}$. We simply made an explicit choice in \eqref{eq:truncated_distribution} but the details of such regularization is irrelevant. 

Now we have a justification for the replica trick as a well-posed moment problem. The $\nu_{\tilde r}$ defined in \eqref{eq:convolution_formula1} is the legit probability distribution consistent with the moments given by the normalized partition functions $\tilde Z_n/kZ_1^n$ in \eqref{eq:replicatrick}. Furthermore, $\mu_{\nu_b}$ has compact support because both $\nu_r$ and $\mu_{\nu_b}$ are compactly supported. Hence, the replica trick is mathematically justified as a soluble Hausdorff moment problem which admits the \emph{unique} solution. \\

Let us also quickly mention the results for a microcanonical ensemble with a fixed energy $E\pm\Delta E$, for which the partition function becomes much simpler,
\begin{equation}\label{eq:pf}
    Z_n[E,\Delta E] =2^{(1-n)S}Z^n_1[E,\Delta E]\ ,
\end{equation}
where $S:=\log(\Delta E\rho(E)e^{S_0})$ is the entropy of the microcanonical ensemble. It corresponds to the partition functions \eqref{eq:convolution_formula1_l} in canonical ensemble $\nu_{\tilde r_N}$ for $N=2^S$ and $\nu_{b_N}$ being flat. We denote the radiation spectral distribution as $\nu_{\tilde r}$ and its moments read
\begin{equation}
    m_n(\nu_{\tilde r}) =  (2^S/k)\sum_{\pi\in \nc_n}\prod_{V\in\pi}k2^{-S}m_{|V|}(r)\ .
\end{equation}
The corresponding spectral distribution is
\begin{equation}\label{eq:micro_spec_dist}
    \nu_{\tilde r} = (2^S/k)\mu_{k/2^S,\nu_r}+(1-2^S/k)\delta_0\ ,
\end{equation}
which is a special case of the general result \eqref{eq:distribution_canonical1} we had.\\

Given the spectral distribution $\nu_{\tilde r}$ of the radiation density operator $\tilde r$ in the fundamental (boundary) description, its von Neumann entropy $S(\tilde r)$ is given by \eqref{eq:vnent},
\begin{equation}\label{eq:beyondisland}
\begin{aligned}
    S(\tilde r) =-\tau(\tilde r\log \tilde r) = -\int_{\mathbb{R}_+}\dd x\ \nu_r\boxtimes\mu_{\nu_b}(x)\ x\log x.
\end{aligned}
\end{equation}
and the R\'enyi entropies are given by \eqref{eq:renyient},
\begin{equation}\label{eq:beyondislandrenyi}
    S_\al(\tilde r) :=\frac{1}{1-\al}\log \tau(\tilde r^\al) = \frac{1}{1-\al}\log\int_{\mathbb{R}_+}\dd x\ \nu_r\boxtimes\mu_{\nu_b}(x)\ x^\al\ .
\end{equation}

We can relate them to the (standard) von Neumann entropy of the $k\times k$ radiation density matrix $\tilde R$ using \eqref{eq:conti_entropy},
\begin{equation}
    S(\tilde R) = S(\tilde r) + \log k\ .
\end{equation}
\begin{equation}
   S_\al(\tilde R) = S_\al(\tilde r) + \frac{\al}{\al-1}\log k\ .
\end{equation}

The convolution formula for entropy~\eqref{eq:beyondisland} supersedes the island formula and provides a more accurate account of the radiation entropy in the PSSY model. It applies to any entanglement spectrum of the bulk state. See Fig.~\ref{fig:page_curve} for an example. The Page curve can be obtained once we are given a sequence of input spectral distributions of the bulk state at different times. 

Using the free multiplicative convolution, we can now resolve the spectrum even when the QES is not defined. Note that the generalized entropy is no longer relevant in our convolution formula~\eqref{eq:beyondisland}.  Islands and QES appear when the convolution factorizes. This happens when a particular summand dominates in the convolution expansion~\eqref{eq:convolution_formula1}, yielding a \emph{factorized} expression for the moments, and then the generalized entropy would show up in the entropy calculations. Such special cases hide away the more fundamental fact that the partition functions are described by a free multiplicative convolution between the quantum information encoded in the two competing QES. We will discuss when does the convolution factorize in Section~\ref{sec:oneshot}.

\subsection{Evaluating the free multiplicative convolution formula}\label{sec:solution}

We now want to evaluate free multiplicative convolution~\eqref{eq:distribution_canonical0} via the $S$-transforms following Algorithm~\ref{algx}. As mentioned, this algorithm is tricky to implement for two arbitrary probability distributions. Fortunately for us, we can reformulate the calculation in terms of the $R$-transforms, which are better understood and easier to handle for the free compound Poisson distributions that we are dealing with. Therefore, translating everything to the $R$-transforms is the key to our derivation below.

We are eventually interested in the spectral distribution of \eqref{eq:distribution_canonical0}, but let's first examine the case at finite $N$ \eqref{eq:distribution_canonical1}.
\begin{equation}
    \nu_r \boxtimes\left(\mu_{\frac{N}{k},\nu_{b_N}}\circ D_{\frac{N}{k}}\right) =: \nu_r \boxtimes\mu_{\nu_b;N}
\end{equation}
where we denote the distribution in the bracket as $\mu_{\nu_b;N}$.
We proceed by computing the $S$-transforms $S_{\nu_r}$ and $S_{\mu_{\nu_b;N}}$. 
Compare the formal series of the moment function \eqref{eq:momentf} and the $R$-transform \eqref{eq:rtransform0}, we have
\begin{equation}
    \omega_{\nu_r}(1/\mathfrak{z}) =   \mathfrak{z} R_{\mu_{1,\nu_r}}(\mathfrak{z})
\end{equation}
where we use the fact that the free compound Poisson distribution has the moments of $\nu_r$ as cumulants $\kappa_{n}(\mu_{1,\nu_r}) = m_n({\nu_r})$. Let $x:=\omega_{\nu_r}(1/\mathfrak{z})$, we have
\begin{equation}
    S_{\nu_r}(x)=\mathfrak{z}(x)(1+x)/x
\end{equation}
where $\mathfrak{z}(x)$ is defined via
\begin{equation}
     \mathfrak{z}(x)R_{\mu_{1,\nu_r}}(\mathfrak{z}(x))=x\ .
\end{equation}\\

Now consider the $S$-transform of $S_{\mu_{\nu_b;N}}$. Using \eqref{eq:composition_inverse}, we have 
\begin{equation}
  S_{\mu_{\nu_b;N}}(x)=z(x)/x
\end{equation}
where $z(x)$ is defined via
\begin{equation}
    z(x)R_{\mu_{\nu_b;N}}(z(x))=x\ .
\end{equation}
Then it follows that
\begin{equation}\label{eq:y-x-z}
    S_{\nu_r\boxtimes\mu_{\nu_b;N}}(x) = S_{\nu_r}(x)S_{\mu_{\nu_b;N}}(x)\implies \frac{1+x}{xy(x)} = \frac{z(x)}{x}\frac{1+x}{x}\mathfrak{z}(x) \implies y(x)\mathfrak{z}(x)z(x)=x\ .
\end{equation}
We would like to solve for the inverse function $x(y)$, so to obtain the Cauchy transform via \eqref{eq:momentf} defined for any $y\in\mathbb{C}^+$.
\begin{equation}
G_{\mu_{\nu_r\boxtimes\mu_{\nu_b;N}}}(y)=y^{-1}+x(y)/y  =  y^{-1}+z(y)\tilde{z}(y)\ ,   
\end{equation}
where we've used \eqref{eq:y-x-z}, and the functions $z(y),\tilde{z}(y)$ are given by
\begin{equation}
        \begin{cases}
          \mathfrak{z} R_{\mu_{1,\nu_r}}(\mathfrak{z})=&x=yz\mathfrak{z}\\
          zR_{\mu_{\nu_b;N}}(z)=&x=yz\mathfrak{z}
        \end{cases}
        \implies
        \begin{cases}
          R_{\mu_{1,\nu_r}}(\mathfrak{z}(y))&=yz(y)\\
          R_{\mu_{\nu_b;N}}(z(y))&=y\mathfrak{z}(y)
        \end{cases}\ .
\end{equation}
Now we have a system of equations only in terms of the $R$-transforms. We can solve for $\mathfrak{z}(y)$ by plugging the first equation to the second and eliminate $z(y)$,
\begin{equation}\label{eq:step_eliminate}
    R_{\mu_{\nu_b;N}}[R_{\mu_{1,\nu_r}}(\mathfrak{z}(y))/y] = y\mathfrak{z}(y)\ .
\end{equation}

Since both $\mu_{1,\nu_r}$ and $\mu_{\nu_b;N}$ are free compound Poisson distributions and their $R$-transforms admit the simple form of the L\'evy-Khintchine formula \eqref{eq:poisson_rtransform},
\begin{multline}
    R_{\mu_{1,\nu_r}}(z) = \int_{\mathbb{R}_+}\frac{x}{1-xz}\nu_r(x)\dd x,\\ R_{\mu_{\nu_b;N}}(z) =\int_{\mathbb{R}_+}\frac{(N/k) x}{1-xz}\nu_{b_N}\circ D_{\frac{N}{k}}(x)\dd x=\int_{\mathbb{R}_+}\frac{(N/k)^2 x}{1-xz}\nu_{b_N}(Nx/k)\dd x  \ . 
\end{multline}
Plugging these formulas to \eqref{eq:step_eliminate}, we have for any $y\in\mathbb{C}^+$,
\begin{equation}
    \int_{\mathbb{R}_+}\frac{(N/k)^2x\,\nu_{b_N}(Nx/k)}{y-x\int_{\mathbb{R}_+}\frac{ x'}{1-x'\mathfrak{z}(y)}\nu_r(x')\dd x'}\dd x = \mathfrak{z}(y)\ .
\end{equation}
Now taking $N\to\infty$,  we have the following equation that just depends on the two input distributions, $\nu_r$ and $\nu_b$, corresponding to the moments feed into the replica trick~\eqref{eq:replicatrick},
\begin{equation}\label{eq:fixed-point-eq2}
    \int_{\mathbb{R}_+}\frac{x\,\nu_b(x/k)/k^2}{y-x\int_{\mathbb{R}_+}\frac{ x'}{1-x'\mathfrak{z}(y)}\nu_r(x')\dd x'}\dd x\stackrel{x/k\to x}{=} \int_{\mathbb{R}_+}\frac{\,\nu_b(x)\dd x}{y/x-k\int_{\mathbb{R}_+}\frac{ x'}{1-x'\mathfrak{z}(y)}\nu_r(x')\dd x'} = \mathfrak{z}(y)\ .
\end{equation}
Note that this is a fixed-point equation for $\mathfrak{z}(y)$ that is numerically easy to solve via an iteration algorithm, and the inputs to the algorithm are the distributions $\nu_r$ and $\nu_b$ that are ready to use.

Using $\mathfrak{z}(y)$, the Cauchy transform is given by
\begin{equation}
    G_{\nu_r\boxtimes\mu_{\nu_b}}(y)=y^{-1}+\mathfrak{z}(y)R_{\mu_{1,\nu_r}}(\mathfrak{z}(y))/y=\frac1y\int_{\mathbb{R}_+}\frac{\nu_r(x)\dd x}{1-x\mathfrak{z}(y)}\ . 
\end{equation}
The spectral distribution $\nu_{\tilde r}=\nu_r\boxtimes\mu_{\nu_b}$ can then be extracted using the Stieltjes inversion formula~\eqref{eq:reversecauchy}. We thus obtain the algorithm \eqref{eq:alg1}-\eqref{eq:alg3} claimed in the introduction.\footnote{As commented earlier in Section~\ref{sec:xconvo}, the free multiplicative convolution is usually tricky to solve even just numerically. Nonetheless, our problem is still tractable without resorting to the machinery of the operator-valued free probability theory and the subordination formalism.}\\

We should remark that such random matrices are known as the \emph{separable sample covariance matrices}, which have applications in multivariate statistical inference~\cite{lixin2006spectral,hachem2006empirical,paul2009no,couillet2014analysis} and wireless communication~\cite{tulino2004random}. They read,
\begin{equation}\label{eq:sep_sample_cov}
  \sqrt{A_N}X_N B_N X_N^*\sqrt{A_N}\ ,
\end{equation}
where $X_N$ is a $N\times M$ Ginibre matrix with variance $1/M$, and the $N\times N$ matrix $A_N$ and the $M\times M$ matrix $B_N$ are deterministic and positive semidefinite. They are the sample covariance matrix of $M$ samples of $N$ data points packaged in the matrix $\sqrt{A_N}X_N\sqrt{B_N}$.\footnote{As compared to the sample covariance matrix \eqref{eq:sample_cov} mentioned earlier (cf. Footnote~\ref{ft:sample_cov}), now we allow correlations $B_N$ among the $M$ samples. However, these correlation is independent from the correlation $A_N$ among the $N$ data points. More explicitly, consider the covariance matrix $\E \mathbf{YY}^*$ of the all the sample data in stacked in one $NM\times 1$ vector $\sqrt{A_N}X_N\sqrt{B_N}\to \mathbf{Y}:=\mathrm{vec}(\sqrt{A_N}X_N\sqrt{B_N})$. Then it follows that the covariance matrix is separable, $\E \mathbf{YY}^* = A_N\otimes B_N$, and hence the name. } We will find in the next subsection that the random matrix ensemble fundamentally describing the radiation state $\tilde R$~\eqref{eq:matrixmodel1}, as inferred from the convolution formula, indeed has the form of a separable sample covariance matrix.

The limiting spectral distribution of \eqref{eq:sep_sample_cov} is given by the same algorithm we presented. This result was previously obtained in random matrix literature under various weaker assumptions on the random matrix ensemble $X_N$~\cite{lixin2006spectral,hachem2006empirical,paul2009no,couillet2014analysis}.  Here we've presented a novel and much simpler way to resolve the spectrum using free probabilistic methods. 

\subsection{A random matrix perspective}\label{sec:page}
\subsubsection{A model in free random variables}\label{sec:freevariables}
We now look for a density operator $\tilde r$ in some $W^*$-probability space that has the spectral distribution of~\eqref{eq:distribution_canonical0}. We start by looking for the density operators $\tilde r_N$ with the the spectral distribution at finite $N$ \eqref{eq:distribution_canonical1}, and then obtain $\tilde r$ as the limit.

Consider a collection of $W^*$-probability spaces $(\W_N,\tau_N)_N$ containing the following random variables $r_N,p_N,b_N,c\in(\W_N,\tau_N)$, where $p_N$ is a projector free from the collection of the positive self-adjoint operator $b_N$ that has the spectral distribution of $\nu_{b_N}$ as in~\eqref{eq:truncated_distribution}.\footnote{Note that we cannot assign $m_n(\nu_b)$ as moments to any random variable because $\nu_b$ is not a probability distribution.}; and $c$ is a circular element free from all $r_N, p_N,b_N$.\footnote{As discussed earlier in the end of Section~\ref{sec:freeness}, the desired $W^*$-probability space can always be constructed to accommodate these free random variables. We can actually be more concrete by realizing $(\W_N,\tau_N)$ as the free product of two Type $I$ matrix algebras representing the gravitational sector and the matter sector respectively, $(\W_N,\tau_N)=M_N(\mathbb{C}) * M_k(\mathbb{C})$. The former hosts the observables of the JT + EOW brane system at finite $N$ and the latter hosts the observables of the $k$-dimensional radiation reservoir.} The positive self-adjoint operator $r_N$ live in the compression $(p_N\W_N p_N,\tau_{p_N})$, where $\tau_{p_N}:=\tau/\tau(p_N)=(N/k)\tau$, and it is also free from $p_N$. It has the spectral distribution of $\nu_{r_N}$, which is defined as
\begin{equation}\label{eq:distribution_r}
  \nu_{r_N}:=(k/N)\nu_r\circ D_{k/N} + (1-k/N)\delta_0\ .
\end{equation}
It's evident that $\tau(b_N) =\tau(r_N)= 1$, so they are density operators.

We seek for a positive self-adjoint operator that shares the finite-$N$-spectral distribution~\eqref{eq:distribution_canonical1}. We postulate that, up to a unitary,\footnote{We can only determine the density operator up to a unitary transformation, because the replica trick only provides the information about the spectrum.} this operator is given by
\begin{equation}\label{eq:operatororder1}
  \tilde r_N := \sqrt{r_N}cb_Nc^*\sqrt{r_N}\in(\W_N,\tau_N)\ .
\end{equation}
Let's check if \eqref{eq:operatororder1} gives us what we want. It is positive and has unit trace,
\begin{equation}
    \tau(\tilde r_N) = \tau(\sqrt{r_N}cb_Nc^*\sqrt{r_N}) = \tau(r_N)\tau(cc^*)\tau(b_N)=1
\end{equation}
where we used freeness to factorize the trace, and $\tau(cc^*)=1$. So it's a legit density operator. We have $cb_Nc^*\in(\W_N,\tau_N)$ whose spectral distribution is governed by a free compound Poisson distribution $\mu_{1,\nu_{b_N}}$.  It is then conjugated by $\sqrt{r_N}$ yielding the spectral distribution,
\begin{equation}\label{eq:embedded_distribution}
    \nu_{\tilde r_N} = \nu_{r_N}\boxtimes \mu_{1,\nu_{b_N}} =(k/N) \nu_r\boxtimes \mu_{N/k,\nu_{b_N}}\circ D_{k/N} + (1-k/N)\delta_0
\end{equation}
whose non-zero part gives \eqref{eq:distribution_canonical1}. 

To extract the solely the spectral distribution \eqref{eq:distribution_canonical1}, consider now the $p_N$ that projects onto the support of $r_N$ and $\tilde r_N$. Note that both $r_N$ and $\tilde r_N$ live in the subalgebra $(p_N\W_N p_N,\tau_{p_N})$. We have $\tau_{p_N}(r_N^n) = m_n(\nu_r)$. The projected $p\tilde r_N p$ reads
\begin{equation}\label{eq:operatororder1_projected}
    p_N\tilde r_N p_N = \sqrt{r_N}p_Ncb_Nc^*p_N\sqrt{r_N}\in(p_N\W_N p_N,\tau_{p_N})
\end{equation}
Since the projection $p_N$ acts as a free compression on $cb_Nc^*$, we have $p_Ncb_Nc^*p_N\in(p_N\W_N p_N,\tau_{p_N})$ whose spectral distribution is $\mu_{N/k,\nu_{b_N}}\circ D_{k/N}$. This follows from~\eqref{eq:freecompression3}. Then 
\begin{equation}
    \nu^p_{p_N\tilde r_N p_N} = \nu_r\boxtimes \mu_{N/k,\nu_{b_N}}\circ D_{k/N} = \nu_{\tilde r_N}
\end{equation}
which matches \eqref{eq:distribution_canonical1} as we want. 

We now take the large $N$ limit. Let $\tilde r\in(\W,\tau)$ be a random variable that has the spectrum $\nu_{\tilde r}$~\eqref{eq:distribution_canonical0}. Consider the limit of the sequence of the freely compressed variables $(p_N\tilde r_N p_N)_N\in (p_N\W_N p_N,\tau_{p_N})_N\subset (\W_N,\tau_N)$.\footnote{This is only possible for a $W^*$-probability space if $\W$ is of Type II$_1$ because it has no minimal projection.} Then it follows from~\eqref{eq:distribution_canonical0} that $p_N\tilde r_Np_N$ converges to $\tilde r$ in distribution,\footnote{Convergence in distribution means all the moments converge. See Definition~\ref{def:convergence} for the precise definition.}
\begin{equation}\label{eq:convergence_of_r}
    p_N\tilde r_N p_N\stackrel{N\to\infty}{\longrightarrow} \tilde r\ .
\end{equation}
Let's check the trace of $p_N\tilde r_N p_N$,
\begin{multline}
    \tau_{p_N}(p_N\tilde r_Np_N) = \tau_{p_N}(\sqrt{r}p_Ncb_Nc^*p_N\sqrt{r})=\tau_{p_N}(r)\tau_p(p_Ncb_Nc^*p_N)\\=\tau(p_Ncb_Nc^*p_N)/\tau(p_N)=\tau(cc^*)\tau(b_N) = 1
\end{multline}
where we switched to the algebra $(\W_N,\tau_N)$ in the third equality and used freeness several times. It follows from the convergence that $\tilde r$ also has unit trace, so this is the density operator we want. \\


Let us now examine the complementary black hole system. Since the global state on the black hole-radiation is chosen to be pure, the black hole density operator $\tilde b_N$ shares the same spectrum as the radiation system. We can flip the the operator ordering in \eqref{eq:operatororder1} and posit that $\tilde b_N$ is given by
\begin{equation}\label{eq:operatororder2}
    \tilde b_N: = \sqrt{b_N}c^*r_Nc\sqrt{b_N}\in (\W_N,\tau_N)\ ,
\end{equation}
which shares the spectral distribution~\eqref{eq:embedded_distribution} as $\tilde r_N$. We should think of $\tilde b_N$ as the density operator for the black hole at finite $N$. However, the large $N$ limit makes it unclear if one can assign a density operator $\tilde b$ to the black hole that shares this spectrum. This is unlike how we get $\tilde r$ which is defined with the help of a sequence of free compressions. In fact, we claim no sensible density operator exists for the black hole at the large $N$ limit.

Our claim is supported by the fact that in the large $N$ limit, the distribution~\eqref{eq:embedded_distribution} has unbounded higher moments $n>1$, indicating that there is not a meaningful density operator that the sequence of $(\tilde b_N)_N$ converges to.  Similarly, we have $\nu_{b_N}\to \delta_0$ at the large $N$ limit, so there isn't a density operator that $(b_N)_N$ converges to either. 

If it is the case that $b_N$ and  $\tilde b_N$ do belong to the algebra of boundary observables at finite $N$, these limiting traces being either infinity or zero indicates that the large $N$ algebra of boundary observables is no longer finite, such as a Type III von Neumann algebra or a Type II$_\infty$ von Neumann algebra, so it ceases to qualify as a $W^*$-probability space.

Nonetheless, physically we still expect that the entropy of black hole should be identical to the radiation entropy for a global pure state. Then how can we compute the black hole entropy directly from $\tilde b_N$? We shall provide an answer later in Section~\ref{sec:holo}. 

\subsubsection{A random matrix model}\label{sec:matrixmodel}

Physically, the radiation density operators $\tilde r,p_N\tilde r_Np_N$, etc should be modelled as density matrices, for the radiation reservoir is simply a $k$-dimensional system. They are, however, fundamentally described as \emph{random density matrices} following the intimate connection between free probability and random matrix theory. This supports the ensemble interpretation of the replica wormholes~\cite{penington2022replica,bousso2020gravity,bousso2020unitarity,marolf2020transcending,marolf2021observations}. 

On the other hand, the black hole system can be approximately described in a boundary random matrix theory~\cite{saad2019jt}. Therefore, we can also meaningfully assign (random) density operators to model the random variables $b_N,\tilde b_N$ at least for finite $N$. In the large $N$ limit, the operator algebra describing the boundary theory is presumably no longer a finite von Neumann algebra, so we lose the trace to define density operators. Therefore, let us stay at finite $N$ and discuss the random matrix model that describes the outcomes of the replica trick calculation.

We shall assume that $N$ is large enough $N\gg k$ to approximate any large $N$ limit result. For instance, the value of $\sbh$ obtained from $B_N$ is arbitrarily close to~\eqref{eq:bhentropy}. At finite $N$, we can also embed the given $k$-dimensional radiation density matrix $R$ in $M_N(\mathbb{C})$, so we can discuss the black hole and radiation density matrices on equal footing. 

The following random matrix ensemble models \eqref{eq:operatororder1},\footnote{\label{ft:noW*} Note that $L^{\infty-}(\Omega):=\bigcap_{p=1}^\infty L^p(\Omega)$ denotes the algebra of classical random variables that have all its moments bounded. It is larger than $L^{\infty}(\Omega)$ and in fact not a von Neumann algebra. We need to work with it instead of $L^{\infty}(\Omega)$ because the Ginibre matrix has Gaussian entries which have unbounded support but bounded moments. Therefore $(L^{\infty-}(\Omega)\otimes M_N(\mathbb{C}),\E\otimes\ttr)$ is only a $*$-probability space but not a $W^*$-probability space. In the end of the day, since we only demand the convergence in distribution, it's legit that the limit is described by a finite von Neumann algebra albeit the random matrices are not. (cf. Appendix~\ref{app:random_matrices}).}
\begin{equation}\label{eq:matrixmodel1}
   \tilde R_N:=\sqrt{R_N}C_N B_N C^*_N\sqrt{R_N} \in (L^{\infty-}(\Omega)\otimes M_N(\mathbb{C}),\E\otimes\ttr)
\end{equation}
where $\tilde R_N$ is the $N\times N$ boundary radiation density matrix; $R_N\in M_N(\mathbb{C})$ is the embedded version of $R\in M_k(\mathbb{C})$ in an arbitrary $k$-dimensional subspace of $\h$ that represents $M_N(\mathbb{C})$; $B_N$ is a $N\times N$ diagonal random matrix (in an arbitrary basis) with the spectral distribution $\nu_{b_N}$\footnote{More precisely, we say demand the empirical spectral distribution (cf. Appendix~\ref{app:random_matrices}) of $B_N$ to be $\nu_{b_N}$. Here the basis can be chosen arbitrarily because the Ginibre ensemble is both left- and right-unitarily invariant.};
and $C_N$ is a $N\times N$ Ginibre matrix with variance $1$. As we've alluded to, the ensemble $R_N$ or $P_N\tilde R_NP_N$ has the form of a separable sample covariance matrix~\eqref{eq:sep_sample_cov}. 

Similarly, the random matrix model of the black hole state at finite $N$ \eqref{eq:operatororder2} reads\footnote{\label{ft:correlation} 
Here we treat the Ginibre matrix $C_N$ to be independent from the ones used for \eqref{eq:operatororder1}. We do so because the replica trick GPI does not tell us how $\tilde R$ and $\tilde R_N$ are correlated, so we will not care about their potential correlations. }
\begin{equation}\label{eq:matrixmodel2}
   \tilde B_N:= \sqrt{B_N}C_NR_N C^*_N\sqrt{B_N}\in (L^{\infty-}(\Omega)\otimes M_N(\mathbb{C}),\E\otimes\ttr)\ .
\end{equation}

The radiation density matrix at the large $N$ limit is modelled by a $k\times k$ density matrix under the rank-$k$ projection $P_N$ onto the support of $R_N$. It is given by
\begin{equation}
    \tilde R :=  \lim_{N\to\infty} P_N\tilde R_NP_N  \in (L^{\infty-}(\Omega)\otimes M_k(\mathbb{C}),\E\otimes\ttr).
\end{equation}

 We claim that $\tilde R_N, \tilde B_N, \tilde R$ model in finite dimensions the density operators $\tilde r_N, \tilde b_N, \tilde r$ as defined in \eqref{eq:operatororder1}, \eqref{eq:operatororder2}, \eqref{eq:convergence_of_r}, in the sense that the spectral distributions of the former is well approximated by the spectral distributions of the latter at large $k$ and $N$. We picked the normalization that $\E\tilde B_N, \tilde R_N$ and $\tilde R$ are all normalized with the standard trace ($\tr$) to a good approximation at large $k$ and $N$, so they can be viewed as density matrices.\footnote{Note that $\ttr B_N=m_1(\nu_{b_N})=1$, so unlike $R_N$, $B_N$ is not a density matrix in the usual normalization convention.} We shall provide the justification in Appendix~\ref{app:random_matrices}.

The entropies of $\tilde R, \tilde R_N, \tilde B_N$ are then also random variables on the underlying probability space $(\Omega,\mathcal{F},P)$. For a separable sample covariance matrix, it is known that the almost surely spectrum of the random matrix converges (weakly) to the freely convoluted distribution $\nu_{\tilde r}$~\cite{lixin2006spectral,paul2009no,couillet2014analysis}. Namely, a typical draw of say $\tilde R$ can have its spectral distribution arbitrarily close to $\nu_{\tilde r}$ with high probability at large enough $k$ and $N$. Therefore, we shall use $S(\tilde R)$ to denote both  its expectation value and its \emph{typical} entropy value. 

When $R$ is maximally mixed, $R_N=P_N/k$, so we have
\begin{equation}
    \tilde R:=\lim_{N\to \infty} X_N B_N X_N^* 
\end{equation}
where $X_N=P_NC_N/k$ is a $k\times N$ Ginibre ensemble with variance $1/k$. It matches with the random matrix model proposed by PSSY in their Appendix D (eqn. D.10). In this view, the PSSY model with non-flat bulk entanglement spectrum simply amounts to conjugating the random matrix with the additional matrix $R$. 

Let us put the random matrix ensemble in a more familiar form in which a matrix model is usually written. According to PSSY, we can rewrite the large $N$ limit of $B_N$ in terms of the random JT Hamiltonian $H_N$ found by \cite{saad2019jt} in the double scaling limit as follows,
\begin{equation}
    Z_n/Z_1^n=m_n(\nu_b)=\lim_{N\to\infty}N^{1-n}\E\ttr B_N^n = \lim_{N\to\infty} \int \dd H_N\, (\tr\, x(H_N)^n) e^{-N\tr V_N(H_N)}
\end{equation}
where we used \eqref{eq:moments_of_b} and $x(\ \cdot\ )$ is the spectral function $y(E)$ in \eqref{eq:gravZn} rescaled by $1/Z_1$, and the integral is over $N\times N$ Hermitian matrices $H_N$ with the measure controlled by the a sequence of potentials $(V_N(H_N))_N$ that are determined by $(\nu_{b_N})_N$.\footnote{This sequence diverges in accordance with the fact that $(\nu_{b_N})_N$ approaches zero.} We can then compute any spectral function $\E f(H)$ of the JT Hamiltonian via 
\begin{equation}
    \E f(H) = \lim_{N\to\infty} \int \dd H_N\, f(H) e^{-N\tr V_N(H_N)}\ .
\end{equation}

On the other hand, we can write the expectation taking over the Ginibre ensemble as 
\begin{equation}
    \E_{C_N}(\ \cdot\ ) = \pi^{-N^2}\int\dd C_N (\ \cdot\ ) e^{-\tr C_NC_N^*}\ .
\end{equation}
For a rectangular $k\times N$ Ginibre ensemble $X_N$, we have
\begin{equation}
    \E_{X_N}(\ \cdot\ ) = \pi^{-Nk}\int\dd X_N (\ \cdot\ ) e^{-\tr X_NX_N^*}\ .
\end{equation}

Say we are interested any spectral function $f(\tilde R)$, such as the von Neumann entropy. Putting things together, it can be evaluated using the following matrix integral,
\begin{equation}
    \E f(\tilde R) = \lim_{N\to\infty}\pi^{-Nk}\int\dd X_N\dd H_N f(RX_N x(H_N)X_N^*R)\,e^{-N\tr V_N(H_N) - \tr X_NX^*_N}
\end{equation}
where $R$ is the given deterministic $k\times k$ bulk radiation density matrix.

\subsubsection{A generalized Page's model}
The above random matrix models can also be packaged together in a random tensor network. We now take a slight detour to introduce some basics of RTN. Generally, a RTN is used as a model for capturing the information-theoretic aspects of the AdS/CFT duality, and it is very successful in reproducing features like the Ryu-Takayanagi formula and the subregion-subregion duality~\cite{hayden2016holographic,harlow2018tasi,hayden2019learning,akers2021leading,akers2021quantum,cheng2022random}. 

A RTN is defined for a given a connected undirected graph $(V,E)$. For each edge-vertex pair $(e,x)$, we assign a Hilbert space $\h_{e,x}$. Attached to each edge $e=(xy)\in E$ is a bipartite system $\h_e:=\h_{e,x}\otimes \h_{e,y}$ with dimension $D_e$, which we refer as the bond dimension. An entangled pure state $\ket{\phi_e}\in \h_e$ is assigned to each edge $e\in E$. Usually, they are chosen as Bell states. The vertices are divided into bulk vertices $V_b$ and boundary vertices $V_\partial$.  At each bulk vertex, we introduce a projection to a random Gaussian vector\footnote{A random Gaussian vector has i.i.d. complex Gaussian entries with zero mean and unit variance. It is unnormalized, and it is equivalent to a Haar random vector up to the normalization. Nonetheless, the norm $||\psi_x||$ is one on expectation and it is asymptotically normalized at large dimensions in the sense that the fluctuations of the norm tends to zero.} $\ket{\psi_x}\in\bigotimes_{e\sim x}\h_{e,x}$ where $e\sim x$ denote any edge $e$ connected to $x$. In this way, the RTN defines a state in the boundary Hilbert space,
\begin{equation}
    \ket{\rho_\partial}:=\bigotimes_{x\in V_b}\bigotimes_{e\in E} (I_{V_\partial}\otimes\bra{\psi_x}) \ket{\phi_e}\in\bigotimes_{x\in V_\partial} \h_x\ .
\end{equation}
Because of the random Gaussian projection, $\ket{\rho_\partial}$ is only normalized in expectation with small fluctuations at large bond dimensions.

\begin{figure}
    \centering
    \includegraphics[width=.95\linewidth]{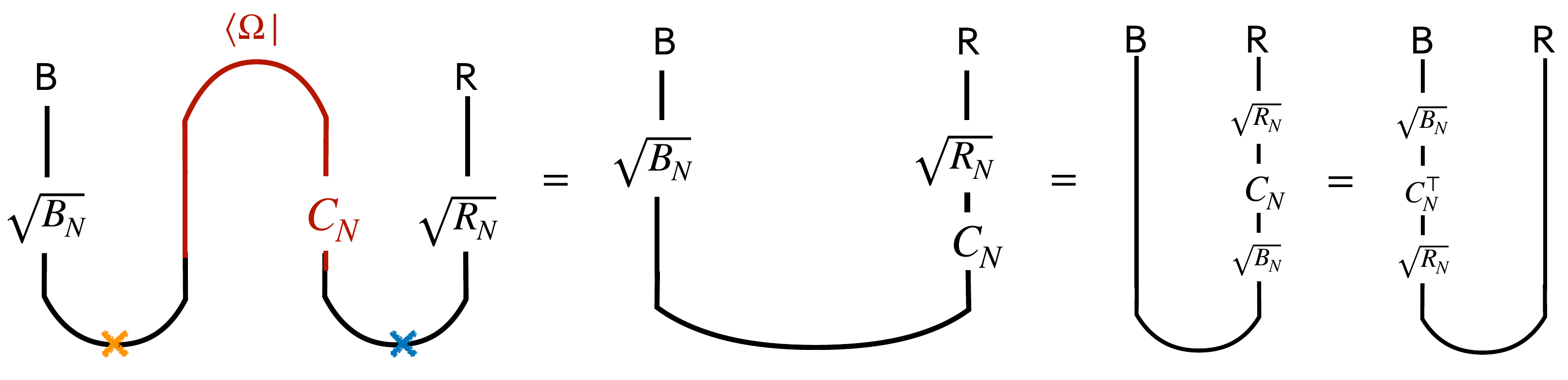}
    \caption{\textbf{The PSSY model as a generalized Page's model.} The left tensor diagram shows the RTN for the PSSY model with non-flat entanglement spectrum, where the red part indicates the projection to a Gaussian random vector. The middle diagram shows that it generalizes Page's model by allowing the two edges to have non-flat spectrum. One can shift the operators all to the radiation system or the black hole system, so when tracing out either $\mathsf B$ or $\mathsf R$, we obtain the consistent random matrix models deduced from our free convolution formula. The two phases described by island formula corresponds to make the blue cut or the orange cut which entail an entropy contribution of $\sbh$ or $S(R)$ respectively.}
    \label{fig:rtn}
\end{figure}

The RTN of the PSSY model consists of two edges, one bulk vertex and two boundary vertices (cf. Fig.~\ref{fig:rtn}). The two edges states are given by
\begin{equation}
    \ket{\sqrt{R_N}}_{\mathsf{RR'}} := (\sqrt{R_N})_{\mathsf{R}}\otimes I_{\mathsf{R'}}\ket{\Omega}_{\mathsf{RR'}}\ ,\quad \ket{\sqrt{B_N}}_{\mathsf{BB'}} := (\sqrt{B_N})_{\mathsf{B}}\otimes I_{B'}\ket{\Omega}_{\mathsf{BB'}}
\end{equation}
where the Bell state $\ket{\Omega}$ has dimension $N$ in both states but the Schmidt rank of them are $|\mathsf R|=k$ and $|\mathsf B|=N$ respectively.  
The density matrices $B_N$ and $R_N$ are diagonal in the basis used to define the Bell state $\Omega$. They represent the purifications of the effective radiation density matrix $R_N$, and of the effective black hole density matrix $B_N$. If a minimal (QES) surface cuts through one of the two bonds, it entails an entropy contribution of $S(R_N)=S(R)$ or $\sbh$ respectively. 

The random projection acts on the purifying systems $\mathsf{R'B'}$,
\begin{equation}
    \ket{\psi}_{\mathsf{R'B'}} = (C_N)_{\mathsf{R'}}\otimes I_{\mathsf{B'}} \ket{\Omega}_{\mathsf{R'B'}} 
\end{equation}
Then the RTN defines the state
\begin{equation}
    (I_{\mathsf{BR}}\otimes\bra{\psi}_{\mathsf{R'B'}})\ket{\sqrt{R_N}}_{\mathsf{RR'}}\ket{\sqrt{B_N}}_{\mathsf{BB'}}
\end{equation}
which equals to either of the following two expressions,
\begin{equation}
    (\sqrt{R_N}C_N\sqrt{B_N})_{\mathsf{R}}\otimes I_{\mathsf{B}} \ket{\Omega}_{\mathsf{RB}} ,\quad I_{\mathsf R}\otimes (\sqrt{B_N}C_N^\top\sqrt{R_N})_{\mathsf{B}}\ket{\Omega}_{\mathsf{RB}}
\end{equation}
where $C_N$ (or $C_N^\top$) is a $N\times N$ Ginibre ensemble. The equalities follow from manipulating the tensor diagrams in Fig.~\ref{fig:rtn}.\footnote{When moving the operators from $\mathsf B$ to $\mathsf R$ or vice verse, they generally pick up a transpose defined w.r.t. to basis chosen in defining the Bell state. The diagonal matrices $B_N$ and $R_N$ are not affected, so it only affects the Ginibre matrix. Nonetheless, since the entropies of a Ginibre matrix is i.i.d., $C_N^\top$ is the same random ensemble as any $N\times N$ Ginibre ensemble. } Let's denote its marginal states as $\tilde B_N$ and $\tilde R_N$ respectively as usual for the fundamental description of the respective density matrices. It is evident that they match with \eqref{eq:matrixmodel1} and \eqref{eq:matrixmodel2} respectively.\footnote{Note that the RTN also determines how the reduced random density states are correlated, but we will ignore this aspect of the RTN model as the gravity calculation does not inform us about it. cf. Footnote~\ref{ft:correlation}.}\\

This RTN is manifestly a generalization of the model proposed by Page to deduce the Page curve~\cite{page1993average,page1993information}. Page considered a microcanonical scenario, where the $B_N$ state is flat and $\log N$ measures the entropy of the remaining black hole. Also, $R_N$ is chosen to be flat. Then one simply has a random bipartite state on $\mathsf{BR}$ and Page considered the entropy $S(\tilde R_N)$ as one vary the relative sizes between $N$ and $k$. Basically, the RTN model we described generalizes Page's model by allowing the two bond states to be non-flat.

In this view, we should think of the two competing saddles as the competition between two ``areas'', with one given by the standard BH area term and the other given by the bulk entanglement between the brane and the radiation. The island formula describes the situation in which the radiation entropy can be described by a clean cut of one of the two bonds. This happens when the contribution coming from contracting the $\mathsf B$ system is separated from $\mathsf R$, in the sense that the resulting spectrum of $\tilde R$ depends only on $B_N$ or $R_N$ but not both. We will later see how these separations are mathematically characterized by the one-shot entropies. 

Generally, due to the non-flatness of the area spectrum, there isn't such a clean cut between the two sources of correlations. Nonetheless, the random Ginibre ensemble help set the two sources free, so we can exploit this freeness to calculate the spectrum of \eqref{eq:matrixmodel1}. This is the general takeaway message from our work. See more discussions in Section~\ref{sec:discussion}. To what extent freeness is relevant and how useful it is in AdS/CFT shall be explored in future work.

\subsection{Hawking's surprisal}\label{sec:holo}

Hopefully we have convinced you that the free convolution formula for the radiation entropy~\eqref{eq:beyondisland} doesn't entail the notion of generalized entropy. Nonetheless, it can actually be interpreted as a generalized entropy if formulated in the right way. Consider the following formulation of the generalized entropy in Wall's proof of the generalized second law~\cite{wall2012proof}. Wall equates the generalized entropy associated with any cut $\gamma$ on the event horizon of a (classically) stationary black hole as follows,\footnote{The formula is valid for a classically stationary black hole for which the contribution of the classical area increase is turned off, which is the interesting case concerning the generalized second law. }
\begin{equation}\label{eq:wall}
    S_\mathrm{gen}(\gamma) = A(+\infty)/4G_N-S(\psi|_\mathcal{A_\gamma}||\psi_\mathrm{HH}|_\mathcal{A_\gamma})\ ,
\end{equation}
where $\psi$ is the quantum state of the matter field in consideration, $\psi_{\mathrm{HH}}$ is the Hartle-Hawking state describing a black hole in thermal equilibrium, the relative entropy is evaluated for the subalgebra $\mathcal{A_\gamma}$ associated with the spacetime subregion external to $\gamma$, and $A(+\infty)$ is the horizon area at infinity. It is independent of the state $\psi$ so it drops out when we study the evolution of $S_\mathrm{gen}$ along the horizon. The relative entropy term is the important one. Its change packages together the variations of the entropy of quantum fields and their black-reactions to the horizon area, and the generalized second law simply follows from its monotonicity under restriction to subalgebras associated with nested spacetime subregions~\cite{lieb1973proof,araki1976relative,petz1986quasi}.\\

The idea is that although the free convolution formula for entropy~\eqref{eq:beyondisland} does not have the form of an area piece plus a bulk entropy piece, it could nevertheless match with Wall's formula of the generalized entropy~\eqref{eq:wall}. This was also the idea recently used to show that in de Sitter spacetime, the entropy of any semiclassical state\footnote{One can meaningfully assign an entropy to states because the operator algebra of observables in de Sitter is shown to be finite, more specifically, a Type II$_1$ factor. } is matching with the generalized entropy of the cosmological horizon~\cite{chandrasekaran2022algebra}.

We are interested in the relative entropy between the black hole state $\tilde b_N$ and the black hole thermal state $b_N$. Since  We can interpret this quantity as \emph{Hawking's surprisal} in learning the actual black hole state $\tilde b_N$, given his original belief that the black hole is basically a canonical thermal state $b_N$.  First of all, recall the definition~\eqref{eq:operatororder2} of $\tilde b_N$,\footnote{Unlike entropies that are spectral functions, the relative entropy not only depends on spectra of $\tilde b_N$ and $b_N$, but also the relative phases between them. Hence, we have made a choice in defining $\tilde b_N$ to be \eqref{eq:operatororder2} that fixes the relative phase between $\tilde b_N$ and $b_N$. }
\begin{equation}
    \tilde b_N: = \sqrt{b_N}c^*r_Nc\sqrt{b_N}\in (\W,\tau)\ .
\end{equation} 
Technically, since we do not have appropriately defined random variables for the black hole and radiation in the asymptotic limit, we consider the relative entropy $S(\tilde b_N||b_N)$ at finite $N$, and then take the large $N$ limit. At finite $N$, Umegaki's formula~\eqref{eq:umegaki} of the relative entropy applies as $\tilde b_N$ and $b_N$ are density operators in a $W^*$-probability space $(\W,\tau)$, 
\begin{equation}
    S(\tilde b_N||b_N) = \tau(\tilde b_N\log \tilde b_N) - \tau(\tilde b_N\log b_N)\ .
\end{equation}

The first term gives 
\begin{equation}
\tau(\tilde b_N\log \tilde b_N)=-S(\tilde b_N) = -S(\tilde r_N)
\end{equation}
where we used the fact that the $\tilde b_N$ shares the spectral distribution with $\tilde r_N$.

The second term gives 
\begin{multline}
    \tau(\tilde b_N\log b_N)=\lim_{n\to 0}\tau(\tilde b_N b^n_N)/n= \lim_{n\to 0}\tau(c^*r_Ncb_N^{n+1})/n\\ =\tau(r_N)\tau(cc^*)\lim_{n\to 0}\tau(b_N^{n+1})/n =S(b_N)
\end{multline}
where we used the replica trick for the logarithm, substituted in \eqref{eq:operatororder2}, and used freeness to compute the traces; in the last step we used $\tau(r_N)=1, \tau(cc^*)=1$. 

Together we have,
\begin{equation}
    S(\tilde b_N||b_N) = S(b_N)-S(\tilde r_N) = (S(b_N)+\log N) - (S(\tilde r_N)+\log N)\ .
\end{equation}
where we added a pair of $\pm\log N$ to each term to prepare for taking the large $N$ limit.

Now we take the large $N$ limit. The second term in bracket gives,
\begin{equation}
    \lim_{N\to\infty} S(\tilde r_N)+\log N = S(\tilde r)+\log N=S(\tilde R)\ ,
\end{equation}
where the last step follows from \eqref{eq:conti_entropy}.

The first term in bracket gives,
\begin{equation}
 \begin{aligned}
     \lim_{N\to\infty}S(b_N)+\log N=& -\lim_{N\to\infty}\int\dd x\ \nu_{b_N}(x) x\log x+ \log N \\
     =& -\lim_{N\to\infty}\int N^{-1}\dd x\ \left(N^2\nu_{b_N}(x)\right) (x/N)\log(x/N)  \\
     \stackrel{y:=x/N}{=}& -\lim_{N\to\infty}\int \dd y\ \left(N^2\nu_{b_N}(yN)\right) y\log y  \\
     \stackrel{\eqref{eq:limits_nu}}{=}& -\int \dd y\ \nu_b(y) y\log y =\sbh\ ,
 \end{aligned}
 \end{equation}
 which is nothing but the BH entropy~\eqref{eq:bhentropy}. \\

Altogether we have
\begin{equation}\label{eq:b2gen}
     \lim_{N\to\infty}-S(\tilde b_N||b_N) = S(\tilde R)-\sbh\ .
\end{equation}
The minus relative entropy on the LHS could be interpreted as the black hole/radiation entropy measured with reference to the black hole background, and the RHS says that it simply equals to the radiation entropy $S(\tilde R)$ with the BH entropy subtracted.

Put \eqref{eq:b2gen} differently, we have
\begin{equation}\label{eq:b2gen2}
    S(\tilde R) = \sbh -  \lim_{N\to\infty}S(\tilde b_N||b_N)\ ,
\end{equation}
which is now a formula for the radiation entropy that doesn't explicitly involve free probability. Unlike the free convolution formula~\eqref{eq:beyondisland}, \eqref{eq:b2gen2} manifestly obeys the Bekenstein bound~\cite{Bekenstein1973,bekenstein1981universal}, i.e. $S(\tilde R)\le \sbh$, also known as the central dogma of black hole physics~\cite{almheiri2020entropy}. This follows from the positivity of the relative entropy.

This formula addresses the question we left in section~\ref{sec:page}. How one can make sense of the black hole entropy at the large $N$ limit in terms of $\tilde b_N$ which doesn't have a sensible large $N$ limit? The answer is that we should simply compute Hawking's surprisal. Although both $\tilde b_N$ and $b_N$ at large $N$ are no longer legit random variables in a finite von Neumann algebra, the limiting relative entropy, $\lim_{N\to\infty}S(\tilde b_N||b_N)$, survives and remains finite. \\

The limiting relative entropy already resembles the term in Wall's formula, where the reference state is also the thermal state of the black hole. In order to make the match more manifest, we can try to be more explicit about the limiting relative entropy with a formal construction. Let the GNS Hilbert space constructed from the tracial state $\tau$ be $\tilde\h_\mathrm{TFD}$. Any faith normal states admit a representation as a cyclic and separating state vector $\ket{\id}$. The state vector for the state $x\mapsto\tau(b_N x)$ is
\begin{equation}
    \ket{\psi_N}:=\ket{\sqrt{b_N}}:=\sqrt{b_N}\ket{\id}\ ,
\end{equation}
which is a finite $N$ version of the thermofield double state (i.e. the canonical purification of the thermal state). We can obtain other states from the action of the operators in $\W$ or its commutant $\W'\subset\mathcal{B}(\tilde\h_\mathrm{TFD})$) on $\ket{\sqrt{b_N}}$,
\begin{equation}
    a\ket{\psi_N} = \ket{a\sqrt{b_N}},\quad a\in\W\ ;\quad a'\ket{\psi_N} = \ket{\sqrt{b_N}a'},\quad a'\in\W'\ .
\end{equation}

Then various inner products are given by
\begin{equation}\label{eq:}
    \bra{\psi_N}b^*a\ket{\psi_N} = \psi_N(b^*a) = \tau(b^*ab_N) ,\quad a,b\in\W
\end{equation}
\begin{equation}
    \bra{\psi_N}a'^*a\ket{\psi_N} = \tau(a'^*\sqrt{b_N}a\sqrt{b_N}),\quad a\in\W, a'\in\W'
\end{equation}

Another the state vector we care about in $\tilde\h_\mathrm{TFD}$ is the one for $\tilde b_N$,
\begin{equation}
    \ket{\tilde \psi_N}:=\sqrt{r_N}c\ket{\psi_N}=\ket{\sqrt{b_N}c\sqrt{r_N}},\quad r_N,c\in\W'\ .
\end{equation}
Then we can write the relative entropy equivalently using Araki's formula~\eqref{eq:araki},
\begin{equation}
   S(\tilde b_N||b_N) = -\bra{\tilde \psi_N}\log\Delta_{\tilde\psi_N|\psi_N}\ket{\tilde \psi_N}\ .
\end{equation}
where $\Delta_{\tilde\psi_N|\psi_N}$ is the relative modular operator.

Now we follow a standard procedure to define the Hilbert space $\h_\mathrm{TFD}$ at the large $N$ limit. A pedagogical description is given in Section 3 of~\cite{witten2021does}. The basic idea is to apply the GNS construction for the state $\psi_\mathrm{HH}$ obtained from the limits of the finite $N$ counterparts $\psi_N$. In particular the inner product between $a\ket{\psi_\mathrm{HH}}$ and $b\ket{\psi_\mathrm{HH}}$ are obtained from the limit of the corresponding finite $N$ inner products $\bra{\psi_N}b^*a\ket{\psi_N}=\psi_N(b^*a)$. $\psi_\mathrm{HH}$ is called the thermofield double state, also known as the Hartle-Hawking (HH) state for a thermal black hole. Then we complete the algebra $\W$ to a von Neumann algebra $\W$ acting on $\h_\mathrm{TFD}$ in the weak topology. Unlike $\W$, the so obtained large $N$ algebra $\A$ is generically infinite. 

We can also repeat the procedure to define the large $N$ limit of $\ket{\tilde \psi_N}$. The constructed GNS Hilbert space would be the same as $\h_\mathrm{TFD}$, as long as for any $a\in\W$ there exists some $b\in\W$ such that the inner product $\braket{a\psi_N}{b\tilde \psi_N}$ has a finite large $N$ limit. We shall assume that this is the case.  

Since all the inner products in $\h_\mathrm{TFD}$ are defined via the limits, we have
\begin{equation}
   \lim_{N\to\infty}-\bra{\tilde \psi_N}\log\Delta_{\tilde\psi_N|\psi_N}\ket{\tilde \psi_N} = -\bra{\tilde \psi} \log\Delta_{\tilde\psi|\psi_\mathrm{HH}} \ket{\tilde \psi}\ (= S(\tilde \psi||\psi_\mathrm{HH}))\ ,
\end{equation}
so formally, under the assumption that we could put both $\tilde\psi$ and $\psi_\mathrm{HH}$ in the Hilbert space $\h_\mathrm{TFD}$, we could rewrite \eqref{eq:b2gen2} as
\begin{equation}\label{eq:b2genN}
   S(\tilde R) =\sbh- S(\tilde \psi||\psi_\mathrm{HH})\ .
\end{equation}

In the PSSY model, the JT black hole is stationary so $\sbh$ can be matched with the constant term $A(+\infty)/4G_N$ in \eqref{eq:wall}. Therefore, our formula \eqref{eq:b2genN}, or the less explicit \eqref{eq:b2gen2}, matches with Wall's formula~\eqref{eq:wall} and thus can be viewed as a generalised entropy. Just like \eqref{eq:wall}, the more relevant is the relative entropy term if we only care about the evolution of the radiation entropy, as in the Page curve for example. Here the relative entropy term contains implicitly the free convolution formula in its first argument.

For more realistic evaporating black holes, matter fields back-react on the geometry creating area fluctuations. This is not captured by the PSSY model, and would make the GPI analysis much harder. Therefore, being able to evaluate the exact radiation entropy of a realistic evaporating black hole using free multiplicative convolution is perhaps too good to hope for. The upshot of the relative entropy formula \eqref{eq:b2genN} is that it might be applicable in more general settings.

\subsection{Sanity checks}\label{sec:checks}
We now perform some sanity checks to our result to show its consistency with some other relevant results in the literature.

\subsubsection{The island formula and the one-shot entropies}\label{sec:oneshot}
When is the island formula~\eqref{eq:island_pssy} valid? The general answer was proposed by AP, who claim that the QES prescription in AdS/CFT is only valid in regimes governed by the \emph{one-shot entropies}~\cite{akers2021leading}. These entropies characterize information processing tasks in the one-shot regime where the only one copy of the source is involved. This is in contrast to how the Shannon/von Neuamnn entropy is relevant in the asymptotic regime where many identical and independent copies of the source are assumed. 

In the context of the island formula in the PSSY model, the relevance of these one-shot entropies was first discussed in~\cite{wang2022refined}. The main point is that the island formula only holds when a black hole is still young such that the \emph{smooth min-entropy} of its emitted radiation (in the effectively description) is below $\sbh-\O(\sqrt{\beta})$, or when a black hole gets old enough such that the \emph{smooth max-entropy} of emitted radiation (in the effectively description) is above $\sbh+\O(\sqrt{\beta})$. 

We have shown that the free convolution formula offers an accurate value of $S(\tilde R)$ in any regime. Therefore, the convolution formula should reduce to the island formula under the above one-shot entropic conditions. We have observed this phenomenon in Fig.~\ref{fig:page_curve} where the transitions occur at $\sbh=15$, corresponding to the min-entropy, and $\sbh=35$, corresponding to the max-entropy. These transition are further smoothed out by the finite temperature corrections. Our goal in this subsection is to show that the convolution factorizes precisely under the above mentioned entropic conditions. Instead of directly comparing the size of each summand to single out the dominant term, we resort to the generalized Page's model introduced previously that realizes this spectral distribution.\footnote{This strategy is the same as how AP argued the relevance of the one-shot entropies in the corrected QES prescription~\cite{akers2021leading} using RTN and the decoupling theorems, whereas a direct comparison among the saddle contributions is carried out in a replica trick GPI derivation in~\cite{wang2022refined}.} \\

We first introduce the one-shot entropies, which for us specifically refer to the smooth min-entropy and the smooth max-entropy.\footnote{The max-entropy is usually defined with the R\'enyi-1/2 entropy rather than the R\'enyi-0 (Hartley) entropy that we used, as the former is preferred for the duality relations that are particularly useful when dealing with the conditional versions. We stick to the latter for its better intuition as measuring the rank of the state. In any case, their difference is often small after smoothing. cf. Tomamichel~\cite{tomamichel2015quantum} for a comprehensive review.} 
\begin{equation}
    S_\ma^\eps(A)= \min_{A'\in\mathcal{B}_A^\eps} \log \mathrm{rank}(A'),\quad S_\mi^\eps(A)_\rho = \max_{A'\in\mathcal{B}_A^\eps} -\log \la_\ma(A')\ .
\end{equation}
where $\mathcal{B}_A^\eps$ denotes the $\eps$-ball around a density matrix $A$ in the purified distance.\footnote{The purified distance is a fidelity-based metric between density matrices defined as $\delta(A,B) := \sqrt{1-F(A,B)}$ where the fidelity is defined as $F(A,B):=(\tr\sqrt{\sqrt{A}B\sqrt{A}})^2$. For our purposes, the trace distance would work just fine as well.} The non-smoothed min/max-entropies are the two extreme points of R\'enyi $\alpha$-entropies for $\alpha=\infty$ and $\alpha=0$ respectively. For an $N\times N$ density matrix $A$, both $S_\ma^\eps(A)$ and $S_\mi^\eps(A)$ are upper bounded by $\log N$. Generally, one needs to deal with the conditional smooth entropies, but the setup in the PSSY model is relatively simply in that there are no quantum side information to condition on. So the above defined one-shot entropies will suffice for our purpose.

The operational relevance of these one-shot entropies was first pointed out in~\cite{renner2004smooth}.\footnote{See the more comprehensive discussion including the conditional one-shot entropies in~\cite{konig2009operational}.} The smooth min-entropy measures the maximum amount of randomness, which is $\eps$-close to uniform, that can be extracted from a single copy of an $N\times N$ density matrix $A$. More concretely, a random projection $\Pi_\mi$ of dimension smaller $2^{S_\mi^\eps(A)}$ can typically produce a uniform outcome, i.e. $N\Pi_\mi A \Pi_\mi\approx \Pi_\mi$,\footnote{\label{ft:decoupling} It also follows from the special case of a one-shot decoupling theorem (cf. Lemma~4.5 in \cite{berta2009single}) applied to a scenario without the reference system.} hence extracting  $S^\eps_\mi(A)$ bits of perfect randomness from a non-perfect source of $\log N$ bits.

On the other hand, we can faithfully compress a copy of the $N$-dimensional state $A$ into a ``smaller'' state with dimension $2^{S_\ma^\eps(A)}$ up to an error of $\O(\eps)$. More explicitly, we can project the state to the spectral support on the largest $2^{S^\eps_\ma}$ eigenvalues, such that the projected state $\Pi_\ma A\Pi_\ma$ (with appropriate renormalization) remains $\eps$-close to $A$. Note that such a projector $\Pi_\ma$ commutes with $A$. Since the projected operator has its trace $\eps$-close to $1$, we shall neglect the normalization issue here. \\

Consider now the random density matrix for the radiation \eqref{eq:matrixmodel1}. When $S^\epsilon_\mi(B_N)-1>S^\eps_\ma(R)$, for two independently chosen smoothing parameters $\epsilon$ and $\eps$, we can find a projector~$\Pi$ with rank $M$ such that $S^\epsilon_\mi(B_N)-1>M>S^\eps_\ma(R)$ and 
\begin{equation}
    \Pi R_N\Pi = \sqrt{R_N}\Pi \sqrt{R_N}\approx R_N
\end{equation}
which follows from $M>S^\eps_\ma(R)$. Note that $R\in(M_k(\mathbb{C}),\tr)$ and $R_N\in(M_N(\mathbb{C}),\tr)$ share the same values for any entropy. It follows that
\begin{multline}\label{eq:youngbh}
\tilde R_N =\sqrt{R_N}C_N B_N C^*_N\sqrt{R_N} \approx \Pi\sqrt{R_N}\Pi C_N B_N C^*_N\Pi\sqrt{R_N}\Pi \\ \approx \Pi\sqrt{R_N}\Pi U|C_N|B_N|C_N|U^*\Pi\sqrt{R_N}\Pi \approx \Pi\sqrt{R_N}\Pi\sqrt{R_N}\Pi\approx R_N
\end{multline}
where we replace each $R_N$ by $\Pi\sqrt{R_N}\Pi$ and use the polar decomposition $C_N=U|C_N|$ with a Haar random unitary $U$ (cf.\eqref{eq:polardecomp}). Then we have $\Pi U|C_N|B_N|C_N|U^*\Pi\approx\Pi$ under a random projection $\Pi U$ when $S^\epsilon_\mi(B_N)-1$. To see how the condition follows, recall that the min-entropy depends only on the largest eigenvalue. Since $|C_N|^2/N=C_NC^*_N/N$ asymptotes to a squared quarter-circular element has its spectral support over $[0,2]$, the min-entropy $S_\mi(|C_N|B_N|C_N|/N)$\footnote{The positive operator $|C_N|B_N|C_N|/N$ is normalized to a good approximation at large $k$, so we can treat it as a density matrix.} is only different from $S_\mi(B_N)$ by one bit, and so is the smooth min-entropy close by continuity. This explains the subtraction by one that we need to include in the condition above.\footnote{This order-one effect is somewhat universal. It already appears in Page's original calculation as the leading correction term~\cite{page1993average}.}  

Therefore, we have
\begin{equation}
    S(\tilde R)\approx S(\tilde R_N) \approx S(R_N)=S(R)\ ,
\end{equation}
so a young black hole satisfying the condition $S^\epsilon_\mi(B_N)-1>S^\eps_\ma(R)$ obeys the island formula~\eqref{eq:island_pssy}. Generally, the approximation error that \eqref{eq:youngbh} entails is set by the smoothing parameter~$\O(\eps)$.\\

When $S^\epsilon_\ma(B_N)<S^\eps_\mi(R)$, we simply invert the argument above and apply it to \eqref{eq:matrixmodel2}. There exists another projector $\Pi$ with rank $M$ such that $S^\epsilon_\ma(B_N)<M<S^\eps_\mi(R)-1$ and
\begin{equation}
    \Pi B_N\Pi = \sqrt{B_N}\Pi \sqrt{B_N}\approx B_N\ ,
\end{equation}
which follows from $M>S^\epsilon_\ma(B_N)$. It follows that
\begin{multline}\label{eq:oldbh}
\tilde B_N \approx \Pi\sqrt{B_N}\Pi C_N R_N C^*_N\Pi\sqrt{B_N}\Pi \\ \approx \Pi\sqrt{B_N}\Pi U|C_N|R_N|C_N|U^*\Pi\sqrt{B_N}\Pi \approx \Pi\sqrt{B_N}\Pi\sqrt{B_N}\Pi\approx B_N\ .
\end{multline}

Therefore, we have 
\begin{equation}
    S(\tilde R)\approx S(\tilde{R}_N) = S(\tilde B_N) \approx S(B_N)\ ,
\end{equation}
so an old black hole satisfying the condition $S^\epsilon_\ma(B_N)<M<S^\eps_\mi(R)-1$ obeys the island formula~\eqref{eq:island_pssy}. Again, the approximation error that \eqref{eq:oldbh} entails is set by the smoothing parameter~$\O(\eps,\epsilon)$.\\

This is the general result addressing when one can compute the entropy of a boundary reduced state by putting a minimal cut as in Fig.~\ref{fig:rtn} in the generalized Page's model, without invoking the free convolution formula. While we assume the density matrix $R$ to be arbitrary, we do know more about $B_N$ which is only parameterized by the temperature (assuming the brane tension is large $\mu\gg 1/\beta$).  In fact, $B_N$ essentially behaves like a projector of dimension $2^{\sbh}$ in the following sense. In the PSSY model, the relevant parameter is $\beta$. The smooth min/max-entropies of $B_N$ are roughly
\beq\label{eq:nearlyflat}
S_\mi^\epsilon(B_N)\approx\sbh-\O(1/\sqrt{\beta}), \quad S_\ma^\epsilon(B_N)\approx\sbh+\O(1/\sqrt{\beta})\ .
\eeq 
for $\epsilon\sim\O(\mathrm{poly}(\beta))$. This nearly flatness is in fact a generic feature of a holographic state, such as a thermal black hole state. Its smooth min/max-entropy is close to its von Neumann entropy, which is of order $1/G_N$, up to a subleading $1/\sqrt{G_N}$ correction~\cite{bao2019beyond}. In the PSSY model, the temperature plays the parametric role of $G_N$, and with dimensional analysis one can artificially introduce Newton's constant as $\beta\to\beta G_N$.

We thereby establish our claim made in the beginning of this subsection. The value of the smoothing parameters $\eps$ and $\epsilon$ can be chosen freely so that one obtain the most useful estimate of the entropy $S(\tilde R)$. For $B_N$, a good choice is something like $\epsilon\sim\O(\mathrm{poly}(\beta))$ that gives the estimate \eqref{eq:nearlyflat} up to an error of $\O(-\log\beta)$. Such finite temperature effect that smoothes out the Page transition is well understood~\cite{penington2022replica,murthy2019structure,marolf2020probing,dong2020enhanced}. The correction is of order $\O(1/\sqrt{\beta})$, which is a subleading effect as compared to the case of leading corrections $\sim\O(1/\beta)$ when the gap between the min-entropy and the max-entropy of the bulk radiation state is large as considered in~\cite{akers2021leading,wang2022refined}. 

While this finite temperature correction is usually treated in parallel to the corrections due to non-flat bulk spectrum, here we see that it's really the same phenomenon of enhanced replica symmetry breaking due to the non-flatness. We treated them on equal footing because a RTN  models them equally. It's just that the non-flatness of the thermal spectrum is mild so it only gives a subleading correction to the island formula, whereas the spectrum of the bulk state could set made arbitrary by quantum processing.

\subsubsection{The result of PSSY}\label{sec:knownresults}

As a second sanity check, we now reproduce the result concerning the radiation spectrum in the original work of PSSY. They considered a maximally entangled bulk state. $\nu_r=\delta_1$ and $m_n(\nu_r)=1, \forall n$, so the operator $r$ is the identity. Then the radiation spectrum is given by the gravitational sector only
\begin{equation}
    \nu_{\tilde r} = \mu_{\nu_b}= \lim_{N\to\infty}\mu_{\frac{N}{k},\nu_{b_N}}\circ D_{\frac{N}{k}},
\end{equation}
which is simply a compound Poisson distribution up to a rescaling. Its $R$-transform admits the general form of the L\'evy-Khintchine formula \eqref{eq:poisson_rtransform},
\begin{equation}
    R_{\mu_{\nu_b;N}}(z) = (N/k)\sum_{n=0}^\infty z^n\int_{\mathbb{R}_+} (N/k)x^{n+1}\nu_{b_N}(Nx/k)\dd x = \int_{\mathbb{R}_+}\frac{(N/k)^2x}{1-xz}\nu_{b_N}(Nx/k)\dd x\ .
\end{equation}
The large $N$ limit gives,
\begin{equation}
    R_{\mu_{\nu_b}}(z)=\int_{\mathbb{R}_+}\frac{x/k^2}{1-xz}\nu_{b}(x/k)\dd x \stackrel{x/k\to x}{=} \int_{\mathbb{R}_+}\frac{x}{1-kxz}\nu_{b}(x)\dd x\ .
\end{equation}
To extract the distribution, we resort to the Cauchy transform via \eqref{eq:rtransform},
\begin{equation}
    G_{\mu_{\nu_b}}(z)R_{\mu_{\nu_b}}(G_{\mu_{\nu_b}}(z)) = \int_{\mathbb{R}_+}\frac{ xG_{\mu_{\nu_b}}(z)}{1-kxG_{\mu_{\nu_b}}(z)}\nu_b(x)\dd x=zG_{\mu_{\nu_b}}(z)-1\ ,
\end{equation}
which, up to the normalization convention used, matches with PSSY's equation (2.34) that is derived from the Schwinger-Dyson equations for the resolvents.\\

To extract the spectral distribution, one needs to solve for the Cauchy transform. However, even in this case, a closed-form expression for the spectral distribution is not possible. PSSY offered an approximated solution and we now try to make a similar approximation by deriving it directly from our general convolution formula.

The approximation starts by working with the regulated probability distributions. Let us set $N=k$ in  \eqref{eq:convolution_formula1_l} as a first approximation,
\begin{equation}
\begin{aligned}
    m_n(\mu_{\tilde r_k})=&\sum_{\pi\in \nc_n}\left(\prod_{V\in\pi}m_{|V|}(\nu_{b_k})\cdot \prod_{\bar V\in\bar\pi} m_{|\bar V|}(\nu_r)\right)=\sum_{\pi\in \nc_n}\prod_{V\in\pi}m_{|V|}(\nu_{b_k})\\
     =&m_n(\nu_{b_k}) = \int_{x(E_k)}^{x(0)}\dd x\ x^n k^{n-1}\frac{\rho_k(E(x))}{-y'(E(x))} 
\end{aligned}
\end{equation}
where we used a simple approximation in the second line that $\pi=\eta$ dominates the sum. This is because $|\pi|\le|\eta|=n$ and $n\mapsto m_n(\nu_{b_k})$ is increasing for large $k$. In the last step, we used \eqref{eq:truncated_distribution} and $E_k$ is defined via
\begin{equation}
    \int^{E_k}_0\dd E\rho_k(E)=k\ .
\end{equation}

In order to make an explicit comparison with PSSY's result of the radiation spectrum, we need to switch to their normalization convention,
\begin{equation}
    \int D(x)\ \dd x =k, \quad\tr\tilde R^n=\int D(x)x^n\ \dd x\ .
\end{equation}
where $D(x)$ denote the spectral distribution of the radiation state in their convention.

On the other hand, we have
\begin{equation}
    k^{-1}\tr\left(k\tilde R\right)^n \approx m_n(\mu_{\tilde r_k}) \implies \tr\tilde R^n = \int_{x(E_k)}^{x(0)}\dd x\ x^n \frac{\rho_k(E(x))}{-y'(E(x))} \ ,
\end{equation}
so it follows that
\begin{equation}
    D(x) = \frac{\rho_k(E(x))}{-y'(E(x))}  = \int_0^{E_k} \dd E \rho_k(E) \delta(x-x(E))
\end{equation}
where we used the properties of delta functions to bring it to a form that matches with PSSY's equation (2.50) where $x(E)=y(E)/Z_1$ is denoted as $\omega(E)$ in PSSY.

We remark that we used a different regularization scheme from PSSY in making the approximation. Instead of using a smooth integrable distribution $\rho_k$ as in \eqref{eq:truncated_distribution}, they simply truncate $\rho_E$ at $E_k'$ when $\int_0^{E_k'}\rho(E)=k$ and replaces the contribution from $E\in(E_k',+\infty)$ by a delta function that gives a lowest possible spectral value $\lambda_0$. In our scheme,  $\lambda_0$ is replaced by $x(E_k)$. Nonetheless, the regularization schemes only differ at large $E$ for which $x(E)$ is in a small exponential tail. The qualitative result should not depend on the exact scheme used in making the approximation.

\subsubsection{The result of Cheng et al on random tensor networks}\label{sec:cheng}

We mentioned in our introduction that a similar result was obtained recently in the context of random tensor network by Cheng et al~\cite{cheng2022random}. They considered a generalized RTN model where the edges are no longer maximally entangled but rather having arbitrary entanglement spectrum, modeling the area fluctuations that one would have in AdS/CFT. In the case of there are two competing minimal cuts $\gamma_{1,2}$, the transition between them is no longer sharp. The details of the entropy of the reduced state on $\Gamma\subset V_\partial$ depend on the spectra $\mu_{1,2}$ of the bonds cut through by $\gamma_{1,2}$ .

This is essentially the same phenomenon that we are investigating for the PSSY model, as was made obvious in the previous subsection. There is, however, one conceptual difference between our starting points and it is a somewhat pedantic. The RTN is by construction finite dimensional and the above mentioned result is formulated for a sequence of RTNs with increasing bond dimensions and converging spectral distributions at the two cuts. The free probability description is thus invoked via Voiculescu’s theorem. We, however, do not have the convenience of a random matrix/tensor model as the fundamental description. Therefore, we take the first-principle rationale that GPI in the planar limit is the fundamental description, and then we do not need to worry about the convergence issues. Technically, this means we shall understand the GPI in terms of free random variables rather than random matrices.  

One can certainly question the validity of the planar limit, and indeed it is, after all, a compromise for the analytic tractability of the problem, because we do not have a controlled way to account for the small fluctuations due to the nonplanar diagrams. So we opt for the pragmatic approach to work in the planar limit, and we think it's conceptually cleaner, because we can translate the entire GPI calculation into the framework of non-commutative probability theory.  

Alternatively, if one takes the random matrix model, which can be deduced from the relevant free random variables, as the fundamental description for the PSSY model, one has essentially the same setup as the RTN model in Cheng et al. Then the free probabilistic calculation only serves as a leading order approximation. Heuristically, how well it approximates a finite dimensional RTN also gauges how well justified the planar approximation is.\\

Besides that, our result in Section~\ref{sec:convolution} and their results in Section 3.2 are almost identical. At the first sight, it might not be obvious that the results are actually identical. The main technical difference comes from that we are dealing with the thermal distribution at the large $N$ limit, which is itself not a proper probability distribution. Nonetheless, if we focus on our result at finite $N$~\eqref{eq:distribution_canonical1}, then a comparison can be made. Consider the operator we had~\eqref{eq:operatororder1}, whose non-zero part gives \eqref{eq:distribution_canonical1},
\begin{equation}\label{eq:comparison1}
    \sqrt{r_N}cb_Nc^*\sqrt{r_N}\ .
\end{equation}
We can reorder the product without changing its moments into the following product of two free factors
\begin{equation}
    r_N\cdot cb_Nc^* 
\end{equation}
where $r_N$ has the spectral distribution $\nu_{r_N}$~\eqref{eq:distribution_r}.

The second term shares the spectrum with $b_Nc^*c=b_Nq^2$, where we again have a product of two free factors. Recall that a squared quarter-circular element $q^2$ has the distribution~\eqref{eq:quartercircle}, which we denoted as $\mathrm{MP}(1)$ following Cheng et al. $b_N$ has the spectral distribution $\nu_{b_N}$~\eqref{eq:truncated_distribution}. Then the moments of \eqref{eq:comparison1} are shared with the moments of
\begin{equation}
    r_N\cdot b_N\cdot q^2\ ,
\end{equation}
and its moments define the distribution
\begin{equation}
    \nu_{r_N}\boxtimes\nu_{b_N}\boxtimes \mathrm{MP}(1)\ ,
\end{equation}
which matches exactly with their Theorem 3.4 in~\cite{cheng2022random}.

\section{Takeaway: Freeness is good}\label{sec:discussion}
We presented a concrete solution to the PSSY model with non-flat bulk entanglement spectrum using ideas from free probability theory. We showed that the free probabilistic framework is perhaps the most appropriate language to understand the model, bridging between the replica trick GPI calculation on one hand and the random matrix model on the other.

Aside from the solution to this particular problem, the main conceptual takeaway from our work is that freeness is a useful feature that can be exploited in studying how gravity processes quantum information, in parallel to the more familiar notion of independence. Though certainly more subtle and underlying than tensor independence, it is worth the attention. By making the PSSY model a bit more complicated, we are able to see freeness manifest in gravity that is otherwise hidden when we only consider a flat entanglement spectrum. As we mentioned, this has to do with a mathematical fact that the identity is trivially free from any other random variable, and this is the only situation where freeness and independence coincides. Therefore, a flat entanglement spectrum with the reduced density operator being an identity hides the fact that the degrees of freedom associated with two candidate surfaces in transition are not always independent but rather always free.

Generally, it is hard to characterize quantum correlation in gravity just like in any generic many-body quantum system. We usually rely on the independence between subsystems in order to make definitive statements. For instance, the reference that encodes any information that falls into a black hole is \emph{decoupled}, i.e. independent from, the radiation or the remaining black hole before or after the Page time respectively.  This allows us to tell where is the information encoded so we can in principle retrieve it from there.  

Over the past, we have made much progress in understanding how gravity stores and processes quantum information, starting from the seminal work of Page~\cite{page1993average,page1993information} and Hayden-Preskill~\cite{hayden2007black} towards the more recent quantum error correction with complementary recovery~\cite{almheiri2015bulk,dong2016reconstruction,harlow2017ryu}, (random) tensor networks~\cite{pastawski2015holographic, hayden2016holographic,bao2019beyond,hayden2019learning,dong2019flat,akers2019holographic} and entanglement wedge reconstruction as state-merging~\cite{akers2021leading}. From an information-theoretic perspective, an overarching technical core underlying these results is the decoupling theorems~\cite{berta2009single,berta2011quantum,dupuis2014one,dupuis2010decoupling}, which characterizes when do we have independence between certain subsystems. Then Uhlmann's theorem~\cite{uhlmann1976transition} would tell us how the quantum correlations is constrained so that we can characterize the whereabouts of the information. Another example is the quantum de Finetti theorem~\cite{caves2002unknown,konig2005finetti,christandl2007one,renner2007symmetry}, which uses the permutation symmetry of large collection of systems to deduce the independence among subsystems. This idea is recently used to explain the tension between the recent replica trick calculation of the Page curve and Hawking's original calculation~\cite{renner2021black}.

On the other hand, as we have emphasized in Section~\ref{sec:freeness}, freeness is a distinct notion of independence. In contrast to the standard tensor independence, that concerns commuting random variables, freeness is a notion of independence tailor-made for non-commuting random variables. Technically, it is defined via the free product instead of the tensor product.  As for the decoupling theorem, de Finetti theorem and Uhlmann's theorem, freeness comes with its own tricks that we can utilize. There is a whole package of free probabilistic tools that one can use to quantitatively characterize the information, as we have seen them in action. 

More importantly, freeness and independence are not mutually exclusive in applications as they work under different conditions and concern different objects. The work by Cheng et al is another good example of this. While freeness can be used to give precise results when the spectral distributions at the candidate minimal cuts converge, one can still make some useful claims about the boundary reduced state using the decoupling approach when the convergence is not assumed (cf. Section 4 in~\cite{cheng2022random}).   

From the RTN perspective, the intuition is that the random tensors not only facilitate the decoupling among different bulk subregions, but also set the information encoded in different QES free from each other. It would be interesting to explore the possibility of putting them to work in synergy in future works.

In quantum theory, which is usually treated as a non-commutative probability theory of observables, we are used to the notion of independence as captured by commuting observables. Now we have seen that it's also possible and fruitful to accommodate freeness in quantum theory, and especially quantum gravity.  We hold the vision that freeness can be very useful in quantum gravity. One hint comes from the relevance  of the free independence between two perturbation of a black hole separated longer than the scrambling time~\cite{chandrasekaran2022large}. We are also largely motivated by the fact that various recent GPI calculations call for random matrix models~\cite{saad2019jt,stanford2019jt,stanford2020more,johnson2021quantum}. In fact, people already came across free random variables in studying the large $N$ matrix models~\cite{gopakumar1995mastering,douglas1995free,douglas1995large}.
Free probability is very relevant here because of its intimate connection with random matrix theory, building on the remarkable insight of Voiculescu that independent random matrices tend to be free in the large $N$ limit. The theory of higher order freeness has also been developed to study the fluctuations of random matrices~\cite{mingo2006second,mingo2007second,collins2007second,borot2021analytic}.  We therefore believe that freeness should be ubiquitous in gravity, provided that we look in the right directions.

\acknowledgments
I am grateful to Roland Speicher for helping me with free probability theory, and Renato Renner for many stimulating discussions and valuable comments. I also thank Chris Akers, Patrick Hayden, Juan Maldacena, Tony Metger, Geoff Penington, Xiaoliang Qi, Pratik Rath, Stephen Shenker, Douglas Stanford, Edward Witten, and Zhenbin Yang for discussions.

\appendix

\section{Additivity violation of the minimum channel output entropy}\label{sec:additivity}
In this appendix, we discuss how an old black hole encodes a channel that violates the additivity of minimum channel output entropy.

In quantum information theory, understanding the quantum channel capacity has fundamental significance. It provides us fundamental insights regarding how much advantage that quantumness can offer over classical resources. One specific problem concerns the capacity of quantum channel for transmitting classical information. People want to know if entanglement can help given the access to multiple copies of the channel. It is conjectured that quantum entanglement does not enhance the channel capacity and the classical capacity of quantum channel is additive. The additivity is verified for many concrete channels but remains an open problem until a counterexample is found. 

The conjecture is not proved false directly. The counter example concerns an equivalent problem of the additivity of the minimum channel output entropy, which we shall now briefly introduce. The minimum channel output entropy\footnote{Here $S_\mi$ should not be confused with the min-entropy. We apologize for the clash of notations.} for a channel $\mathcal{N}$ is defined as\footnote{This is a convex optimization problem so the inputs are restricted to pure states w.l.o.g.}
\begin{equation}
    S_\mi(\mathcal{N}) := \min_\psi S(\mathcal{N}(\psi))\ .
\end{equation}

Shor showed that asking about channel additivity is equivalent to, among several other equivalents statements, asking if $S_\mi$ is additive for any two channels $\mathcal{N},\mathcal{M}$~\cite{shor2004equivalence}.
\begin{equation}
    S_\mi(\mathcal{N}\otimes\mathcal{M}) \stackrel{?}{=} S_\mi(\mathcal{N}) +  S_\mi(\mathcal{M})
\end{equation}
This is proved false by Hastings~\cite{hastings2009superadditivity} who showed that
\begin{equation}\label{eq:additivity_violation}
    S_\mi(\mathcal{N}\otimes\mathcal{\bar N})< S_\mi(\mathcal{N}) +  S_\mi(\mathcal{\bar N})
\end{equation}
for some $N$-dimensional channel $\mathcal{N}$ and its conjugate channel $\mathcal{\bar N}$.\footnote{The conjugate channel of a channel $\mathcal{N}(\, \cdot\, ):=\sum_i K_i(\, \cdot\, ) K_i^*$ is defined as $\mathcal{\bar N}(\, \cdot\, ):=\sum_i \bar K_i(\, \cdot\, ) \bar K_i^*$.}

We've been using free probability to analyze the black hole information problem, whereas randomized constructions are in fact ubiquitous in quantum information. The rationale of injecting randomness dates back to Shannon's founding work of proving the channel coding theorem using a random coding~\cite{shannon1948mathematical}. Hastings' counterexample is also a random construction, and the existence of additivity violation is argued probabilistically. He didn't find an explicit channel that gives the violation, but rather $\mathcal{N}$ is a random unitary channel of form
\begin{equation}\label{eq:hastings}
    \mathcal{N}(\, \cdot\, ):=\sum_i^d p_i U_i (\, \cdot\, ) U_i^*
\end{equation}
with independently sampled Haar random unitaries $U_i$, weighted by some carefully chosen probability vector $p_i$ that is roughly uniform. Then Hastings showed that in the regime of $1\ll d\ll N$, some (or in fact any typical) instance of the random channel ensemble $\mathcal{N}$ violates additivity as in~\eqref{eq:additivity_violation}.

Hastings' work was built upon the earlier work of Hayden-Winter whose proved false the R\'enyi versions of the conjecture~\cite{hayden2008counterexamples}, where Hayden-Winter used a similar random construction in the Stinespring representation with random isometries. 

There are many follow-ups to the Hastings-Hayden-Winter (HHW) breakthrough~\cite{brandao2010hastings,fukuda2014revisiting,belinschi2016almost,collins2016haagerup,collins2016random,fukuda2022additivity}, but they almost all use the same family of random channels with minor variations. Unfortunately, no concrete counterexamples nor extensive violations are found. 

Let us mention a particular variant of the HHW-type construction that is been analyzed using free probability~\cite{fukuda2022additivity}.\footnote{I thank Tony Metger for pointing out this reference to me.} Consider the following random channel mapping from the $N$-dimensional quantum state space to itself,
\begin{equation}\label{eq:randomchannel}
   \mathcal{E}(\, \cdot\, ):= \frac1d\sum_i^d C_iT(\, \cdot\, )TC_i^*,\quad T:=\sqrt{d}\left(\sum_{i=1}^dC_i^*C_i\right)^{-\frac12}
\end{equation}
where we have $n$ independent $N\times N$ Ginibre matrices $C_i$, and $T$ is a \emph{rectification}  operator that ensures the channel is trace-preserving. Note that $T$ almost surely converge to identity at large $d$, so the rectification is a technicality that is only relevant for finite $d$ and can be approximately ignored at large $d$.

In the regime 
\begin{equation}\label{eq:violation_regime}
    1\ll d\ll N\ ,
\end{equation}
it's shown that typical instances of the random channel $\mathcal{E}$ (together with its conjugate channel $\mathcal{\bar E}$) violates the additivity the with a small gap~\cite{fukuda2022additivity}. The proof is built upon the earlier work by Collins~\cite{collins2016haagerup}. Note that \eqref{eq:randomchannel} is a variant of the HHW-type counterexample because Ginibre matrices can be thought of as partially-traced Haar random unitries/isometries (cf. \cite{collins2016haagerup} and the discussion in section 4.1 in \cite{fukuda2022additivity}). 

Note that the output of $\mathcal{E}$ has a nearly flat spectrum. (cf. Theorem 3.1 in \cite{collins2016haagerup} for a precise statement. ) This is a universal feature of the HHW-type construction. The output of the channel is very close to a maximally mixed state of the maximal possible rank.\footnote{It means that either the channel itself or its complement channel is close to a fully depolarizing channel. It is known that a channel is additivity-violating if and only if its complement channel is also additivity-violating~\cite{holevo2007complementary}. } It's somewhat ironic that the only family of channels that we know to have superadditive capacity has almost no practical use.\\

Consider now the spectrum of Hawking radiation in a microcanonical ensemble for a very old black hole. Namely, we are in the regime
\begin{equation}\label{eq:blackhole_regime}
    1\ll k/2^S\ll k\ .
\end{equation}
where as assume $k/2^S$ is an integer.

The microcanonical spectral distribution of the radiation reads~\eqref{eq:micro_spec_dist},
\begin{equation}
    \nu_{\tilde r}= (2^S/k)\mu^{\boxplus k/2^S}_{1,\nu_r}+(1-2^S/k)\delta_0\ .
\end{equation}

Now we look for a quantum channel that takes bulk radiation density operator $R$ as input and the output has its spectrum described by $\nu_{\tilde r}$ to a good approximation at large dimension $N$. Following the discussion in \ref{sec:freevariables}, a density operator with this spectrum is
\begin{equation}
    \tilde r=(k/2^S)pcrc^*p 
\end{equation}
where $c$ is a circular element and $p$ is a free compression of trace $\tau(p)=2^S/k$. However, when we model $\tilde r$ in random matrices, it doesn't take the form of a trace-preserving channel from $r$ to $\tilde r$. We need to consider something else.

Let's restrict to the non-zero part of $\nu_{\tilde r}$ and ask what channel has output spectrum described by $\mu^{\boxplus k/2^S}_{1,\nu_r}$ up to the normalization. This is the spectrum of $\sum_{i=1}^{k/2^S} c_i r c_i^*$ for $k/2^S$ free circular elements $\{c_i\}_{i\in [k/2^S]}$. We can put in a normalization factor to obtain a completely positive trace-preserving map on $(\W,\tau)$,
\begin{equation}
   E(r):=\frac{2^S}{k}\sum_{i=1}^{k/2^S} c_i r c_i^*\ .
\end{equation}

Using again the connection between random matrices and free probability, we can model $E(\,\cdot\,)$ by $\mathcal{E}(\,\cdot\,)$ in \eqref{eq:randomchannel} at large $N$. Then in the regime \eqref{eq:blackhole_regime}, which is identified with \eqref{eq:violation_regime} through $k/2^S=d, k=N$, one would find that this channel violates the additivity of minimum output entropy.  

Therefore, we conclude that the HHW-type random channel \eqref{eq:randomchannel}, which is originally constructed purely for technical convenience in approaching the additivity conjecture, can be physically encoded by an evaporating black hole at late time. However, it still begs the question what is the physical meaning of the channel $x\mapsto E(x)$. Could it be interpreted as a holographic map from bulk to boundary in some appropriate sense? 

It's important to remark that this violation is only visible here because we sum over all possible geometries, albeit that most of them are subleading and usually deemed unimportant at late time. If we only care about entropies, the island formula is good enough as the GPI at late time is dominated by the symmetric replica wormhole saddle alone. This is in accordance with the fact that the channel $\mathcal{E}$ is only slightly different from a depolarizing channel that outputs a maximally mixed state of rank $d=k/2^S$. However, it is exactly this subtle difference that is responsible for the violation in the HHW counterexamples. Therefore, in order to see the additivity violation from an old black hole, taking replica symmetry breaking into account is indeed indispensable. 

It's plausible that realistic black holes at late time encode information via a more sophisticated (random) channel such that one can hope for more counterexamples to additivity or even large extensive violations. This possibility is indeed contemplated in \cite{hayden2020black} by relating it to the existence of certain entanglement properties of black hole microstates. However, as we learnt in the PSSY model, one perhaps needs to wield the full GPI in order to extract such channels, which remains technically challenging for realistic black holes.

\section{More on random matrices}\label{app:random_matrices}

In this appendix, we introduce the notion of random matrices and their spectral distributions. We shall see how random matrices make contact with free probability.

\subsection{Preliminaries}
We have the probability space given by a Kolmogorov triple $(\Omega,\mathcal{F},P)$. The sample space $\Omega$ is equipped with some $\sigma$-algebra $\mathcal{F}$ and a probability measure $P$. We can simply take the $\mathcal{F}$ to be generated by the relevant random variables. Consider the matrix elements of an $N\times N$ matrix $A$ as random variables. They are defined as measurable functions, $A_{ij}:\omega\mapsto A_{ij}(\omega)$, from $\Omega$ to $\mathbb{C}$ equipped with the Borel $\sigma$-algebra. Then we define the random matrix $A$ as $A: \Omega\to M_N(\mathbb{C}), \omega\mapsto A(\omega)$. We can view it as the joint random variable of its matrix elements $(A_{ij})_{i,j=1}^N$ mapping from $\Omega$ to $\mathbb{C}^{N^2}$ equipped with the Borel $\sigma$-algebra. We can then pushforward $P$ to define the probability distribution of the matrix elements $\nu(A_{ij})$ on $\mathbb{C}$. 

We are particularly interested in the spectra of random matrices. For a $N\times N$ matrix $A$ with eigenvalues $\{\lambda_i(A)\}_{i=1}^N$, we can defined its \emph{spectral distribution}, which is a probability measure on $\mathbb{C}$, as an average over the Dirac mass on each eigenvalue,
\begin{equation}
    \nu_{A} (\lambda) := \frac1N\sum_{i=1}^N \delta (\lambda-\lambda_i(A)) \ .
\end{equation}

Often we need to consider a sequence of random matrices of size $N\in\mathbb{N}$, denoted as $\left(A_N\right)_N$. For technical convenience,\footnote{Generally we need to work with the probability space consisting of the product space $\prod_N\Omega_N$, with the product $\sigma$-algebra $\prod_N\mathcal{F}_N$ and the product measure $\prod_N P_N$.} we assume these random matrices are all defined on the same probability space $(\Omega,\mathcal{F},P)$.

Let the eigenvalues of $A_N(\omega)$ be $\{\lambda_{i}(A_N(\omega))\}_{i=1}^N$. Then we define the random variables $\{\lambda_i(A_N)\}_{i=1}^N$ to be measurable functions from $\Omega$ to $\mathbb{C}$, $\lambda_i(A_N): \omega\mapsto\lambda_{i}(A_N(\omega))$. Let the set of probability measures on $\mathbb{C}$ be $\mathcal{M}(\mathbb{C})$. Then we define the \emph{empirical spectral distribution} (ESD) of $A_N$ as a measurable function from $\Omega$ to $\mathcal{M}(\mathbb{C})$,
\begin{equation}
  \nu_{A_N}: \omega  \mapsto \nu_{A_N(\omega)}(\lambda) := \frac1N\sum_{i=1}^N \delta (\lambda-\lambda_i(A_N(w))) \ ,
\end{equation}
which is a \emph{probability measure-valued} random variable.\footnote{The notion of an ESD is perhaps confusing at the first encounter, essentially because we have two types of randomness put together. It would be conceptually more transparent when we work with a non-commutative probability space in section~\ref{sec:primer}.}  

For a random variable, we can discuss different notions of convergence as we send $N\to\infty$. In particular, we need the notion of almost sure convergence. For a sequence of classical (scalar) random variables $(X_N)_N$ on~$\mathbb{C}$, it converges almost surely to some random variable $X$ if 
\begin{equation}
    P\left(\omega\in\Omega: \lim_{N\to\infty} X_N(\omega)=X(\omega)\right)=1 \ .
\end{equation}

Let us restrict to Hermitian matrices. For a sequence of ESD $(\nu_{A_N})_N$ of hermitian matrices $(A_N)_N$, we are interested in the cases that the sequence converges to a deterministic limit. 

The strongest convergence is the almost sure convergence. We say a sequence of ESD $(\nu_{A_N})_N$ converges almost surely to a measure $\nu_a\in\mathcal{M}(\mathbb{R})$ if for any bounded continuous function $f:\mathbb{R}\mapsto\mathbb{R}$, $\int\nu_{A_N}(\lambda) f(\lambda)\dd\la$, which is a scalar random variable, converges almost surely to the value of $\int\nu_a(\lambda) f(\lambda)\dd\la$. In other words, we  almost surely have the ESD $(\nu_{A_N})_N$ converge to $\nu_a$ in \emph{weak topology}. Then we call $\nu_a$ the \emph{limiting spectral distribution} (LSD) of the random matrix ensemble $(A_N)_N$.

For classical random variables, the almost sure convergence implies convergence in probability, which in turn implies convergence in distribution.\footnote{For a sequence of classical (scalar) random variables $(X_N)_N$ on $\mathbb{R}$, it converges in probability if for any $\eps$, $\lim_{N\to\infty}P(|X_N(\omega)-X(\omega)|>\eps)=0$; and it converges in distribution if $\lim_{N\to\infty}\E f(X_N)=\E f(X)$ for any bounded continuous function $f:\mathbb{R}\mapsto \mathbb{R}$, i.e. the sequence of distributions $\nu_{X_N}$ converges weakly to $\mu_X$.} For a sequence of ESD, the convergence in probability is defined similarly as the almost sure convergence up to changing the sense of convergence, whereas the convergence in distribution should correspond to the notion of \emph{convergence in expectation}. It concerns the \emph{expected ESD} $\E\nu_{A_N}$ defined as
\begin{equation}
    \int\E\nu_{A_N}(\lambda)f(\lambda) \dd\la :=\E\int\nu_{A_N}(\lambda)f(\lambda) \dd\la
\end{equation}
for all bounded continuous function $f:\mathbb{R}\mapsto\mathbb{R}$. Then we say $(\nu_{A_N})_N$ converges in expectation to some probability measure $\nu_a$ on $\mathbb{R}$ if the sequence of \emph{deterministic} distributions $(\E\nu_{A_N})_N$ converges weakly to $\nu_a$, i.e.
\begin{equation}\label{eq:spectralmeasure}
    \lim_{N\to\infty}\int\E\nu_{A_N}(\lambda) f(\lambda)\dd\la =\int\nu_a(\lambda)f(\lambda)\dd\la
\end{equation}
for all bounded continuous function $f:\mathbb{R}\mapsto\mathbb{R}$. Of course, almost sure convergence implies convergence in expectation. In summary, for a sequence of  ESDs, almost sure convergence implies convergence in probability which implies convergence in expectation.  \\

\subsection{Making contact with free probability}
To see how asymptotic limit of large random matrices make contact with free probability, we would like to abstract away the Kolmogorov triple and resort to the non-commutative probability space $(L^{\infty-}(\Omega)\otimes M_N(\mathbb{C}),\E\otimes\ttr)$, which is a $*$-probability space (cf. Footnote~\ref{ft:noW*}). We now discuss the convergence of random matrices in terms of the convergence of non-commutative random variables.

Note that because ESD itself is a probability measure-valued random variable the convergence of ESD entails two notions convergences. For example, the almost sure convergence we've been discussing concerns the random variable itself, whereas we demand each instance to converge weakly as a probability measure. The latter is also known as the convergence in distribution for the random variables that carry these instances of probability measures. Since we've stashed the Kolmogorov triple and opted for $(L^{\infty-}(\Omega)\otimes M_N(\mathbb{C}),\E\otimes\ttr)$ to describe random matrices, we take the expectation and hence left the deterministic distributions $\E\nu_{A_N}$, the only notion of convergence available is the convergence in expectation, or convergence in distribution for the respective random variables in $L^{\infty-}(\Omega)\otimes M_N(\mathbb{C})$.

\begin{defn}[Convergence (in distribution) of non-commutative random variables]\label{def:convergence}
Let $(\A_N,\tau_N)_N$ be a sequence of non-commutative probability spaces, and let $(\A,\tau)$ be another non-commutative probability space. 
Let $(a_{N;1})_N,\ldots,(a_{N;d})_N\in (\A_N,\tau_N)_N$ be a collection sequences of random variables, and $a_1,\ldots,a_d\in(\A,\tau)$, we say $(a_{N;1})_N,\ldots,(a_{N;d})_N$ converges in distribution to $a_1,\ldots,a_d$ if 
\begin{equation}
    \lim_{N\to\infty}\tau_N(a_{N;i_1}\ldots a_{N;i_k}) = \tau(a_{i_1},\ldots,a_{i_k})
\end{equation}
for any $\{i_1,\cdots,i_k\}\in[d]$. We denote it as
\begin{equation}
   (a_{N;i})_N \to a_i,\,\, i\in[d]\ .
\end{equation}
\end{defn}

Concretely, for a sequence of random matrices $(A_N)_N\in(\A_N,\tau_N)_N$, we can now talk about its limit at $N\to\infty$ as some non-commutative random variable $a$ in $(\A,\tau)$ as long as the moments match asymptotically. Equivalently, when the ESD $(\nu_{A_N})_N$ converge in expectation to $\nu_a$, we can identify $\nu_a$ as the (deterministic) spectral distribution for some $a\in(\A,\tau)$.

With the appropriate notion of convergence, we can now discuss the when do large random matrices become free from each other.
\begin{defn}[Asymptotic freeness]
A collection of sequences of non-commutative random variables $(a_{N;1})_N,\ldots,(a_{N;d})_N\in(\A_N,\tau_N)_N$ is asymptotically free if they converge in distribution to some elements $a_1,\ldots,a_d$ respectively in some non-commutative probability space $(\A,\tau)$ such that the collection $a_1,\ldots,a_d$ is free.
\end{defn}

When we choose $(\A_N,\tau_N)_N$ to be $(L^{\infty-}(\Omega)\otimes M_N(\mathbb{C}),\E\otimes\ttr)_N$, asymptotic freeness implies that the ESD $\nu_{A_N}$ converge in expectation to $\nu_a$. Hence, asymptotic freeness for random matrices is a property defined at the level of expectations, capturing a relatively weak sense of asymptotic freeness. More explicitly, consider \eqref{eq:free} for any two sequences of random matrices $(A_N)_N$ and $(B_N)_N$. If for any finite collection of integers $\{m_i,n_i\}_i\ge 0$ such that,
\begin{equation}\label{eq:free2}
   \lim_{N\to\infty} \frac1N\E\tr\left(\prod_{i}\left((A_N)^{m_i}-\frac1N\E[\tr(A_N)^{m_i}]\cdot I_N\right)\cdot\left((B_N)^{n_i}-\frac1N\E[\tr(B_N)^{n_i}]\cdot I_N\right)\right)=0 \ ,
\end{equation}
then $(A_N)_N$ and $(B_N)_N$ are asymptotically free.

Asymptotic freeness does not only concern random matrices, due to the flexibility in choosing the non-commutative probability spaces $(\A_N,\tau_N)_N$. While one can choose $(\A_N,\tau_N)_N$ to be $(L^{\infty-}(\Omega)\otimes M_N(\mathbb{C}),\E\otimes\ttr)_N$ for $N\times N$ random matrices, we can also discuss asymptotic freeness for sequences of deterministic matrices in $(M_N(\mathbb{C}),\ttr)_N$ when they converge in distribution to elements in $(\A,\tau)$. We can exploit this to define the stronger notion of \emph{almost sure asymptotic freeness} for random matrices.

\begin{defn}[Almost sure asymptotic freeness]
A collection of sequences of random $N\times N$ matrices $(A_{N;1})_N,\ldots,(A_{N;d})_N$ in $(L^{\infty-}(\Omega)\otimes M_N(\mathbb{C}),\E\otimes\ttr)_N$ is almost surely asymptotically free if, for almost all $\omega\in\Omega$, the sequences $(A_{N;1}(\omega))_N,\ldots,(A_{N;d}(\omega))_N\in(M_N(\mathbb{C}),\ttr)_N$ are asymptotically free, i.e., they converge in distribution to some elements $a_1,\ldots,a_d$ respectively in some non-commutative probability space $(\A,\tau)$ such that the collection $a_1,\ldots,a_d$ is free.
\end{defn}
Note that here we demand the elements in $(M_N(\mathbb{C}),\ttr)$ under fixed $\omega\in\Omega$ to converge in distribution for almost all $\omega$, which implies the almost sure convergence of the ESDs. Therefore, almost sure asymptotic freeness is stronger than asymptotic freeness. 

More explicitly, consider again \eqref{eq:free} for any two sequences of random matrices $(A_N)_N$ and $(B_N)_N$ with each admitting an asymptotic spectral distribution $\nu_a$ and $\nu_b$ respectively. If for any finite collection of integers $\{m_i,n_i\}_i\ge 0$ such that the following random variable converges almost surely
\begin{equation}\label{eq:free3}
   \left(\frac1N\tr\prod_{i}\left((A_N)^{m_i}-\nu_a(\lambda^{m_i})\cdot I_N\right)\cdot\left((B_N)^{n_i}-\nu_b(\lambda^{n_i})\cdot I_N\right)\right)_N\stackrel{a.s.}{\longrightarrow} 0 \ ,
\end{equation}
then $(A_N)_N$ and $(B_N)_N$ are almost surely asymptotically free. 

Lastly, for two sequences of random matrices $(A_N)_N$ and $(B_N)_N$ in $(L^{\infty-}(\Omega)\otimes M_N(\mathbb{C}), \E\otimes\ttr)_N$, asymptotic freeness is equivalent to that any sequence of mixed cumulants, as defined via \eqref{eq:m-c0}, vanish in the limit $N\to\infty$; and almost sure asymptotic freeness is equivalent to that the mixed cumulants vanish almost surely in the limit $N\to\infty$.\\

Now we can now state Voiculescu's theorem, that is central in bridging random matrix theory and free probability.
\begin{thm}[Voiculescu~\cite{voiculescu1991limit}]\label{thm:Voiculescu}
Let $(A_N)_N$ and $(B_N)_N$ be two sequences of $N\times N$ random matrix ensembles in $(L^{\infty-}(\Omega)\otimes M_N(\mathbb{C}),\E\otimes\ttr)_N$ that admit asymptotic spectral distributions such that for any $N$, their matrix elements are independently distributed and $A_N$ (or $B_N$) is a unitarily invariant ensemble,\footnote{A unitarily invariant ensemble means the joint distribution of the matrix elements is invariant under any unitary conjugation.} then $(A_N)_N$ and $(B_N)_N$ are almost surely asymptotically free.
\end{thm}

Let us revisit the matrix matrix model proposed in Section~\ref{sec:matrixmodel}. We claimed that $\tilde R_N$ and $\tilde R$ model $\tilde r_N$ and $\tilde r$, in the sense that the ESD of the former is well approximated by the spectral distribution of the later at large fixed values of $N$ and $k$. We want to argue it by leveraging Voiculescu's theorem~\ref{thm:Voiculescu}.  However, an obstacle is that we are not given a sequence of random matrices in \eqref{eq:matrixmodel1} but rather a random matrix with fixed size. We do not have a sequence of random matrices so Voiculescu's theorem doesn't seem to be applicable here. 

In order to make the contact with free probability, a standard solution is to ``blow up'' a finite matrix, say the deterministic $k\times k$ radiation density matrix $R$ given to us, by tensoring an identity factor of dimension $L$, therefore creating a sequence of $kL\times kL$ matrices $(R_{kL})_L:= R\otimes I_{L}$ by adjusting $L$.\footnote{\label{ft:freeequivalent} More generally, in a (random) matrix polynomial involving standard independent random matrix ensembles, such as GUE, GE, CUE, and some deterministic matrices. One simply replaces the deterministic ones by some random variables ($\{d_i\}$) that share the same spectral distribution. Then the corresponding polynomial in $\{d_i\}$ and the free semi-circular, circular or Haar elements is called the \emph{free deterministic equivalent}, and it serves as a good approximation at large $N$.  This procedure is often used to deal with (random) matrix polynomials involving single deterministic matrices or non-converging sequence of deterministic matrices. In our case, for example, $r$ is a free deterministic equivalent of $(R_{kL})_L$. For a precise formulation of free deterministic equivalents, see Chapter 10.1 in Mingo-Speicher~\cite{mingo2017free} and \cite{speicher2012free,belinschi2017analytic,vargas2015free}.} Note that the ESD of $(R_{kL})_L$ remains the same and thus trivially converges to $\nu_r$ as $L\to\infty$. 

Consider now the matrix model \eqref{eq:matrixmodel1}. We can shift this factor $N$ from $C_N$ to $R_N$ without changing the overall expression. Let's rewrite it as $\sqrt{NR_N}C_NB_NC_N\sqrt{NR_N}$ where we have $\E\ttr B_N=1, \ttr NR_N=1$ and $C_N$ has variance $1/N$ instead of $1$ as in \eqref{eq:matrixmodel1}. Then we blow up $NR_N$ to sequences and $R_{NL} := NR_N\otimes I_L$. The Ginibre ensemble can be chosen to be of any shape so we can have the sequence of $NL\times NL$ Ginibre matrices $C_{NL}$.

Since any unitary from left or right can be absorbed by the Ginibre ensemble, we can conjugate $(B_{NL})_L$ by a Haar random unitary for free. Then it follows that $(C_{NL})_L$, $(B_{NL})_L$ and $(R_{NL})_L$ are all independently distributed and any pair consists of one ensemble being unitarily invariant. Also, each ensemble admits a LSD. All the conditions of Voiculescu's theorem are thus satisfied and the almost sure asymptotic freeness follows. 

As sequences of non-commutative random variables in $(L^{\infty-}(\Omega)\otimes M_{NL}(\mathbb{C}),\E\otimes\ttr)_L$, we then have the convergence in the sense of Definition~\ref{def:convergence}, $(C_{NL})_L\to c$, $(B_{NL})_L\to b_N$ and $(R_{NL})_L\to r_N$ such that $c,b_N,r_N$ are free from each other. This gives a concrete random matrix realization of the abstract non-commutative random variables we invoke in Section~\ref{sec:freevariables}. In the next subsection, we discuss why the random matrix models we proposed capture the desired limiting spectral distributions.

\subsection{Sample covariance matrix and separable sample covariance matrix}

It now makes sense to ask for the LSD for any polynomial of sequences a collection of asymptotically free random matrix ensembles. When each random matrix ensemble admits a LSD, the answer is given by the distribution of the corresponding non-commutative random variables at the large $N$ limit. If the random matrices are almost surely asymptotically free, then the ESD of any polynomial almost surely converges weakly to the LSD. For the precise statement and its proof, see Lemma~II.4.1 in~\cite{mai2017analytic}. 

For example, we are mostly interested in the product of two asymptotically free random matrices. Given two almost surely asymptotically free sequences of random positive\footnote{In fact, we only demand one of them, say $A_N$, to be positive.} Hermitian matrices $(A_N)_N$ and $(B_N)_N$, which have the LSD of $\nu_a$ and $\nu_b$ respectively, we almost surely have the sequence of ESDs $\left(\nu_{\sqrt{A_N}B_N\sqrt{A_N}}\right)_N$ converges weakly to the free multiplicative convolution $\nu_a\boxtimes\nu_b$,
\begin{equation}\label{eq:xconvo_a.s.}
    \left(\nu_{\sqrt{A_N}B_N\sqrt{A_N}}\right)_N \stackrel{a.s.}{\longrightarrow} \nu_a\boxtimes\nu_b \ .
\end{equation}
The same also holds for the free additive convolution.\\

We now look at the sample covariance matrix ensemble that models the free compound Poisson distribution, and the separable sample covariance matrix that models our free convolution formula~\eqref{eq:mainresult}.

The free Poisson distribution is modeled by the \emph{Wishart ensemble}. Let $X_N$ to be $N\times M$ rectangular random matrices with i.i.d. complex Gaussian entries of zero mean and variance $\alpha/M$.  Consider now the Wishart matrix
\begin{equation}\label{eq:wishart}
    W_N=X_NX_N^*\in (L^{\infty-}(\Omega)\otimes M_N(\mathbb{C}),\E\otimes\ttr)\ .
\end{equation}
We demand that the ratio $M/N \to \lambda$ as $N,M\to\infty$.

The resulting LSD is called the \emph{Marchenko–Pastur (MP) distribution}, denoted as $\mu^\mathrm{MP}_{\la,\al}$.
\begin{equation}\label{eq:mp}
\mu^\mathrm{MP}_{\la,\al}:=
    \begin{cases}
      (1-\lambda)\delta_0 + \mu , &\mathrm{if}\quad 0\le\lambda\le 1 \\ 
      \mu, &\mathrm{if}\quad \lambda>1
    \end{cases}
\end{equation}
where $\mu(x):=\frac{\lambda}{2\pi\alpha x}\sqrt{(x_+ - x)(x - x_-) }\mathbf{1}_{\left[x_-, x_+\right]} \ , \quad x_\pm:=\alpha(1\pm1/\sqrt{\lambda})^2$. \\

Its moments reads
\begin{equation}\label{eq:mp_mom}
    m_n(\mu^\mathrm{MP}_{\la,\al}) =\lim_{N\to\infty}\ttr\ W^n_N= \alpha^n\sum_{k=1}^n \lambda^{-k} N(n,k)
\end{equation}
where $N(n,k):=\binom{n}{k}\binom{n}{k-1}/n$ are the Narayana numbers. The free cumulants read
\begin{equation}\label{eq:mp_cumu}
    \kappa_n(\mu^\mathrm{MP}_{\la,\al}) = \lambda^{1-n}\alpha^n\ .
\end{equation}
We can write the MP distribution in terms of a free Poisson distribution
\begin{equation}
\mu^\mathrm{MP}_{\la,\al}=\mu_{\la,\al/\la}\ . 
\end{equation}


We now show that a compound Poisson distribution is the LSD of the following random matrix ensemble,
\begin{equation}\label{eq:sam_cov}
    X_N^* A_NX_N\in(L^{\infty-}(\Omega)\otimes M_N(\mathbb{C}),\E\otimes\ttr)\ ,
\end{equation}
which consists of the same Ginibre ensemble as above and a sequence of independently distributed Hermitian matrices $(A_N)_N$ that admit a LSD $\nu$.

It shares the same non-zero eigenvalues with a \emph{sample covariance matrix}, $\sqrt{A_N}X_NX_N^*\sqrt{A_N}$. Their spectra are different when $X_N$ is rectangular ($N\times M$ with variance $1/M$). Nonetheless, one can easily obtain one from the other via\footnote{This formula holds in general for the ESDs $\nu_{AB}$ and $\nu_{BA}$ of two rectangular matrices $A$ and $B$ with sizes $N\times M$ and $M\times N$ respectively. }
\begin{equation}\label{eq:twospectrum}
    \nu_{X_N^* A_NX_N}=(N/M)\nu_{\sqrt{A_N}X_NX_N^*\sqrt{A_N}}+\left(1-N/M\right)\delta_0\ .
\end{equation}

If $M/N \to \lambda$ as $N,M\to\infty$ and $\nu_{A_N}\to \nu$, we claim that\footnote{This result was established by Silverstein and Bai for $X_N$ being the more general Wigner ensemble~\cite{silverstein1995empirical}.}
 \begin{equation}\label{eq:convergence_sample_cov}
\begin{aligned}
    \nu_{X_N^* A_NX_N}\ &\stackrel{a.s.}{\longrightarrow} \  \mu_{1/\lambda,\nu}\ ,\\
    \nu_{\sqrt{A_N}X_NX_N^*\sqrt{A_N}}\ &\stackrel{a.s.}{\longrightarrow} \  \la\mu_{1/\lambda,\nu} + (1-\la)\delta_0\ .
\end{aligned}
 \end{equation}

We can use free probability to obtain this result. Since $A_N$ is independent from $X_N$, $X_N$ is unitarily invariant and so is $W_N=X_NX_N^*$ and both $W_N$ and $A_N$ admits the LSD's $\mu^\mathrm{MP}_{\la,1}$ and $\nu$ respectively, they are almost surely asymptotically free as implied by Voiculescu's theorem~\ref{thm:Voiculescu}.  The LSD of $(\sqrt{A_N}X_NX_N^*\sqrt{A_N})_N$ is given by the convolution $\mu^\mathrm{MP}_{\la,1}\boxtimes\nu$. 

Consider now the LSD of $X_N^* A_NX_N$ and using \eqref{eq:twospectrum} gives,
\begin{equation}
    \nu_{X_N^* A_NX_N}\ \stackrel{a.s.}{\longrightarrow} \ \lambda^{-1}\cdot\mu^\mathrm{MP}_{\la,\al}\boxtimes\nu + (1-\lambda^{-1})\delta_0\ .
\end{equation}

Let's evaluate the moments of the above LSD,
\begin{equation}
\begin{aligned}
  m_n(\lambda^{-1}\cdot\mu^\mathrm{MP}_{\la,1}\boxtimes\nu+ (1-\lambda^{-1})\delta_0)&=\lambda^{-1}\sum_{\pi\in \nc_n}\left(\prod_{V\in\pi}\kappa_{|V|}(\mu^\mathrm{MP}_{\la,1})\cdot \prod_{\bar V\in\bar\pi} m_{|\bar V|}(\nu)\right)\\
   &=\lambda^{-1}\sum_{\pi\in \nc_n}\prod_{V\in\pi}\lambda^{1-|V|} \cdot\prod_{\bar V\in\bar\pi} m_{|\bar V|}(\nu)\\
   &=\sum_{\pi\in \nc_n}\prod_{V\in\pi}\lambda^{-1} m_{|V|}(\nu) = m_n(\mu_{1/\la,\nu})\ ,
\end{aligned}
\end{equation}
where we used the formula \eqref{eq:freemulti1} in the first line; in the second line we used the formula of free cumulants \eqref{eq:mp_cumu} with $\alpha=1$; and in the last step we used $|\pi|+|\bar\pi|=1+n$. Hence, the cumulants of the LSD is given by $\kappa_n = \lambda^{-1} m_{n}(\nu)$, which yields \eqref{eq:convergence_sample_cov}. \\

As for the sample covariance matrix \eqref{eq:sample_cov}, we can also consider the matrix ensemble that shares the same non-zero spectrum as \eqref{eq:sep_sample_cov}, known as the \emph{separable sample covariance matrix}
\begin{equation}\label{eq:sep_sample_cov2}
    \sqrt{D_N}C_N^* A_N C_N\sqrt{D_N}\in(L^{\infty-}(\Omega)\otimes M_M(\mathbb{C}),\E\otimes\ttr)\ .
\end{equation}
where we have the a sequence of additional independently distributed random $N\times N$ matrix $(D_N)_N$ that admits a LSD $\xi$ . $C_N$ is an $N\times N$ Ginibre matrix. We assume w.l.o.g. that $M=N$ by allowing $D_N$ to be rank-deficient. $(D_N)_N$ is almost surely asymptotically free from the sample covariance matrix $C_N^* A_N C_N$ according to Voiculescu's theorem. It follows from the above result on sample covariance matrix that 
\begin{equation}
    \nu_{\sqrt{D_N}C_N^* A_N C_N\sqrt{D_N}} \stackrel{a.s.}{\longrightarrow} \xi\boxtimes\mu_{1,\nu}\ .
\end{equation}
In terms of the non-commutative random variables, we have the following convergence in distribution, 
\begin{equation}\label{eq:convergence_sep_sam_cov}
    (\sqrt{D_N}C_N^* A_N C_N\sqrt{D_N})_N\longrightarrow \sqrt{d}c^*ac\sqrt{d}
\end{equation}
where $(D_N)_N\to d$, $(A_N)_N\to a$ and $c$ is a circular element in some $W^*$-probability space $(\W,\tau)$.\\

Now we can verify that our matrix models in Section~\ref{sec:matrixmodel} correctly model the random variables in Section~\ref{sec:freevariables}. We let $D_N$ be the rescaled $N\times N$ embedded radiation density matrix $NR_N$, and $A_N$ be the black hole (random) thermal density matrix $B_N$. Then the ensemble $\eqref{eq:sep_sample_cov2}$ matches exactly with $\tilde R_N$ in \eqref{eq:matrixmodel1}. 
 
Based on the discussion in the end of last subsection, we fix $N$ to be some large value and we use $L$ for the varying parameter for the matrix size. We $NR_N$ to a sequence, and define a sequence that extends \eqref{eq:matrixmodel1} to the large $L$ limit.  
\begin{equation}
   (\tilde R_{NL})_L :=(N\sqrt{R_{NL}}C_{NL}^*B_{NL} C_{NL}\sqrt{R_{NL}})_L\in (L^{\infty-}(\Omega)\otimes M_{NL}(\mathbb{C}),\E\otimes\ttr)
\end{equation}
where at $L=1$ we have the $N\tilde R_N$. We need to put in the factor $N$ to adjust the normalization.

Let the free deterministic equivalent of $NR_N$  be $r_N$ (cf. Footnote~\ref{ft:freeequivalent}), in the sense that $(R_{NL})_L\to r_N$. Then \eqref{eq:convergence_sep_sam_cov} implies that $\tilde R_N$ \eqref{eq:matrixmodel1} models $\tilde r_N$ \eqref{eq:operatororder1}, in the sense that $(\tilde R_{NL})_L\to \tilde r_N$. More explicitly, we have almost surely the sequence of ESD $(\nu_{\tilde R_{NL}})_L$ converges weakly to $\nu_{\tilde r_N}$. Similarly, the arguments apply for $\tilde B_N$ and $\tilde b_N$. 

Lastly, we consider the large $N$ limit. We have $\tilde R_N\to \tilde R$ and $p_N\tilde r_N p_N\to \tilde r$. Because almost surely the sequence of ESD $(\nu_{P_{NL}\tilde R_{NL}P_{NL}})_L$ converges weakly to $\nu_{p_N\tilde r_N p_N}$. If for any bounded continuous function $f$ on $\mathbb{R}$, the convergence of the expectation value of $f$ evaluated with the distribution $(\nu_{P_{NL}\tilde R_{NL}P_{NL}})_L$ is uniform on $N$, we can then swap the limits to obtain $\tilde R\to \tilde r$.


\bibliographystyle{jhep}
\bibliography{free}

\end{document}